\documentclass[a4paper,11pt]{article}
\usepackage{jheppub} 
\usepackage{lineno}
\usepackage[caption=false]{subfig}

\def\rar{\rightarrow}

\def\ra{\rangle}

\def\bea{\begin{eqnarray}}
\def\eea{\end{eqnarray}}
\def\be{\begin{equation}}
\def\ee{\end{equation}}

\usepackage{mathrsfs}
\usepackage{physics}
\usepackage{amsmath}
\usepackage{dsfont}
\usepackage{extarrows}
\DeclareMathOperator{\arcsinh}{arcsinh}
\newcommand{\dbar}{d\hspace*{-0.08em}\bar{}\hspace*{0.1em}}

\title{\boldmath Quantum thermodynamics in a rotating BTZ black hole spacetime}

\author[a]{Wenjing Chen,}
\author[a]{Yixuan Ma,}
\author[a]{Si-Wei Han,}
\author[a]{Zihao Wang,}
\author[a,b,c]{Jun Feng\footnote{Corresponding author}}
\affiliation[a]{School of Physics, Xian Jiaotong University, Xi'an 710049, Shaanxi, P.R. China}
\affiliation[b]{Institute of Modern Physics, Xian Jiaotong University, Xi'an 710049, Shaanxi, P.R. China}
\affiliation[c]{Hefei National Laboratory, Hefei 230088, Anhui, P.R. China}

\emailAdd{j.feng@xjtu.edu.cn}

\abstract{ We address the problem of the thermalization process for an Unruh-DeWitt (UDW) detector outside a BTZ black hole, from a perspective of quantum thermodynamics. In the context of an open quantum system, we derive the complete dynamics of the detector, which encodes a complicated response to scalar background fields. Using various information theory tools, such as quantum relative entropy, quantum heat, coherence, quantum Fisher information, and quantum speed of evolution, we examined three quantum thermodynamic laws for the UDW detector, where the influences from BTZ angular momentum and Hawking radiation are investigated. In particular, based on information geometry theory, we find an intrinsic asymmetry in the detector's thermolization process as it undergoes Hawking radiation from the BTZ black hole. In particular, we find that the detector consistently heats faster than it cools, analogous to the quantum Mpemba effect for nonequilibrium systems. Moreover, we demonstrate that the spin of a black hole significantly influences the magnitude of the asymmetry, while preserving the dominance of heating over cooling.
}

\begin{document}
\maketitle
\flushbottom
\section{Introduction}
\label{1}

The fundamental laws of thermodynamics dictate the average behavior of heat and work in macroscopic systems at thermal equilibrium. Over the past two centuries, thermodynamics has consistently revealed the underlying nature of various physical mechanisms. Alongside the bulk processes of a particle ensemble and the thermal history of the early universe, one of its most surprising applications is seen in self-gravitating systems, particularly black holes. Since the original work of Bekenstein \cite{sec1-1} and Hawking \cite{sec1-2-1,sec1-2-2} more than half a century ago, black hole thermodynamics has raised many challenging questions regarding the fundamental nature of quantum mechanics and gravity.

The thermodynamic nature of black holes is evident from two perspectives \cite{sec1-3}. From a classical standpoint, black holes behave like thermal objects, as analogous thermodynamic laws can be identified using their physical parameters \cite{sec1-4}, such as mass $M$, electric charge $Q$, and angular momentum $J$. On the other hand, the intrinsic temperature of black holes and their entropy, which is related to the area of the horizon, suggest that black holes must be seen as a quantum system. In particular, the quantum fluctuations near the horizon are stretched by the spacetime geometry, causing the black hole to radiate like a black body. Unfortunately, the conjectured evaporation process is inconsistent with quantum unitary evolution, which states that pure states always evolve to other pure states, leading to the black hole information paradox. It is widely believed that this paradox can only be resolved with a full understanding of the quantum nature of gravity \cite{sec1-5}, thereby promoting it as a touchstone for any candidate of a consistent quantum gravity theory.

The thermal nature of Hawking radiation can be manifested by the spectral behavior of Unruh-deWitt (UDW) detector \cite{sec1-6,sec1-7}, an idealized particle with energy gap $\omega$, interacting with fluctuating quantum fields. Outside a static black hole, regardless of its initial state, a hovering detector would be excited and asymptotically be driven to a Gibbs state 
\be
\rho_{\text{detector}}\left(t \rightarrow \infty\right)=\frac{e^{-\mathfrak{h}_{\text{detector}}/T_{\text{eff}}}}{\operatorname{tr} e^{- \mathfrak{h}_{\text{detector}}/T_{\text{eff}}}}\equiv \sigma_{\text{thermal}},
\label{eq1.1}
\ee
which is irrelevant to the detector's initial states and in equilibrium with the perceived radiation quanta, indicating the semiclassical Planckian blackbody law of Hawking radiation. In practice, the effect temperature of Hawking radiation $T_{\text{eff}}$ can be recognized from the detailed-balance condition of the thermalization \emph{end} (\ref{eq1.1}) of the detector, justified by an excitation-to-deexcitation ratio \cite{sec1-7-1,sec1-7-2} 
\be
\mathcal{F}(\omega)=\mathcal{F}(-\omega)e^{-\omega/T_{\text{eff}}},
\label{eq1.2}
\ee
where $\mathcal{F}(\omega)$ is the response function of the UDW detector, and $T_{\text{eff}}$ would approach the Hawking temperature $T_H$ in the spatial infinity. The same analysis can be extended to various curved scenarios \cite{sec1-8}, especially the analog Hawking-Unruh radiation perceived by an accelerating linear detector in flat spacetime \cite{sec1-9}. Very recently, people realized that black holes are fast scramblers \cite{sec1-10} and shockwave interactions are a universal feature of quantum gravity. Therefore, quantum black holes are expected to be chaotic quantum systems \cite{sec1-11} whose chaotic level statistics modify the spectral properties of Hawking-Unruh radiation and were shown to be experimentally accessible by probing the Unruh heat bath with a UDW detector \cite{sec1-12}.

Nevertheless, such a "conventional" field approach is incompetent for fully exploring the dynamical influence of the Hawking effect, as it cannot capture detector density matrix decoherence, i.e., how the off-diagonal terms evolve toward the Gibbs state (\ref{eq1.1}) undergoing the Hawking-Unruh effect. This indicates that the detailed-balance condition (\ref{eq1.2}) can only be \emph{necessary but not sufficient} condition for detector thermalization. In other words, once arriving at the same thermalization end $\sigma_{\text{thermal}}$, one cannot distinguish the specific thermalization \emph{trajectory} followed by the detector in its Hilbert space. In recent years, the dynamics of UDW detectors in various curved spacetimes have been fully addressed through an open quantum system approach \cite{sec1-13}. With resolved complete open dynamics, the understanding of the detector thermalization \emph{process} can be refined by many feature functions \cite{sec1-14,sec1-15,sec1-16,sec1-17} from quantum information theoretical or metrological perspectives.

Essentially, the thermalization process of the UDW detector is an irreversible process that fully includes quantum effects. This naturally suggests that the non-equilibrium thermodynamic dynamics of a UDW detector can be re-investigated in the context of quantum thermodynamics, which is a rapidly growing field that has presented rich and exotic insights 
\cite{QT0,QT1}. The combination is expected to inspire new insights into our understanding of the Hawking-Unruh effect from an alternative perspective. In particular, the \emph{thermodynamic process functions}, such as quantum heat and work, become trajectory-dependent quantities in the quantum regime and exhibit quantum thermodynamic laws \cite{QT2,QT3}. This aligns with the required process functions needed to distinguish different thermalization trajectories in the state space of the UDW detector \cite{sec1-18}. Recent interesting developments have begun to address such problems in an engineering way. For instance, some authors \cite{QT4,QT5,QT6,QT7} have designed a quantum heat engine using the thermal properties of the quantum vacuum, where the thermal nature of the Unruh effect plays a central role. 

In this work, we explore the thermalization process of a UDW detector outside a rotating BTZ black hole from a quantum thermodynamic perspective. We solve the dynamics of the UDW detector within an open quantum system framework \cite{sec1-13}, where the detector is considered a local open system governed by a quantum master equation that encodes dissipation and decoherence due to quantum field fluctuations in the BTZ spacetime. 

The main structure of the paper is outlined as follows.

In Section \ref{2}, we consider a co-rotating detector in the exterior region of the BTZ black hole. Beyond previous studies on spinless BTZ geometry \cite{QT7,sec1-19,sec1-20}, we are particularly interested in the open dynamics of a rotating BTZ spacetime \cite{sec1-21,sec1-22}, where the angular momentum of the black hole is expected to enrich the possible thermalization trajectories of the UDW detector. Employing a sharp switching-on/off in the asymptotic past/future, we are able to analytically calculate the response function of the detector to a massless scalar field, which determines the complete open dynamics of the detector, encoding the influences from Hawking radiation and classical angular momentum degrees.

In Section \ref{3}, we examine quantum thermodynamic laws applicable to the UDW detector. Firstly, by using quantum relative entropy (QRE) to quantify the distance between distinct states, i.e., $S(\rho \| \sigma) \equiv \operatorname{tr}(\rho \log \rho)-\operatorname{tr}(\rho \log \sigma)$, we formulate the \emph{Zeroth Law} of quantum thermodynamics as a vanishing QRE \cite{QT8} between the detector state at a specific time and its thermalization end:
\be
\mathcal{D} \equiv S\left(\rho_{\text {detector }}(t) \| \sigma_{\text {thermal}}\right)=0,
\label{eq1.3}
\ee
indicating that the detector has reached its equilibrium with the environment. The \emph{First Law} of quantum thermodynamics captures the detailed time evolution of the detector's quantum heat and work during the thermalization process. In particular, the change rate of detector internal energy, i.e., $U_t:= \operatorname{Tr}\left[\rho_{\text {detector }}(t) H_{\text{detector}}\right]$, can be decomposed into \cite{QT9}
\be
\dot{U_t}=\dot{\mathbb{W}}+\dot{\mathbb{Q}}+\dot{\mathbb{C}},
\label{eq1.4}
\ee
where, besides quantum work $\mathbb{W}$ and heat $\mathbb{Q}$, the evolution of quantum coherence $\mathbb{C}$ plays a significant role. Finally, for an irreversible open process, the time derivative of the detector's von Neumann entropy can be separated into entropic flow $\Phi$ from the environment and entropy production rate $\Pi$ as $\dot{S}(t)=\Pi-\Phi$. For any time interval, the nonnegative entropy production rate determines a nonnegative entropy production as $\Pi\geqslant 0$ (or equivalently $\int  \Pi dt\geqslant 0$), which is the context of the \emph{Second Law} of quantum thermodynamics \cite{QT10}. 

Since the irreversible open dynamics of the co-rotating detector in the BTZ are significantly influenced by its perceived Hawking radiation, as well as black hole angular momentum, we expect that particular quantum thermodynamic laws can manifest these influences during the thermalization process. For each quantum thermodynamic law, besides the local effective Hawking radiation, we demonstrate how the non-vanishing black hole angular momentum modifies every thermodynamic process function. Furthermore, in BTZ spacetime, the background scalar field admits different boundary conditions at spatial infinity, which drastically affect the thermodynamic laws. Specifically, we find that for the scalar field with Dirichlet boundary conditions, a significantly longer timescale for the detector thermalization process is required compared to other Neumann or transparent boundary cases. This is attributed to the complicated behavior of detector response in rotating BTZ spacetime, e.g., the discrepancy arising from field boundary choices reflects the fact that the Dirichlet condition provides a transition rate that varies least rapidly.  

In Section \ref{4}, we analyze an intriguing effect of the nonequilibrium UDW detector immersed in Hawking radiation: the asymmetry of the time evolution of heating and cooling trajectories during the process of detector thermalization \cite{QT12,QT13}. Using information geometric measures such as fidelity, quantum speed, and quantum degree of completion \cite{QT11}, we examine this phenomenon for two UDW detectors co-rotating with the black hole at different spatial orbits, and find that detector heating is always faster than its cooling. We show that the angular momentum of the BTZ black hole may significantly affect this thermal asymmetry in heating/cooling processes, implying that the classical degrees of black hole can also be encoded in the quantum thermodynamics of a local quantum probe, analogous to the recently proposed quantum Mpemba effect \cite{QT14} for nonequilibrium systems.

For simplicity, throughout the analysis, we use natural units $\hbar=c=k_{B}=1$.

\section{Open dynamics in black hole spacetime}
\label{2}

We model the UDW detector as a two-level system (TLS) interacting with a massless field in BTZ spacetime. The time evolution of the TLS as a local open system is then governed by a {quantum Markovian master equation} (QMME), where its interaction with background quantum fluctuations acts as the "bath", inducing dissipation and decoherence terms. In the seminal works \cite{sec1-23,sec1-24}, a rigorous mathematical proof demonstrates that, once the bath meets general conditions, the QMME for open dynamics becomes accurate in the van Hove limit (i.e., $g^2 t \sim \mathcal{O}(1)$ while taking $g \rightarrow 0$ and $t \rightarrow \infty$ simultaneously), as all higher-order terms tend toward zero.

\subsection{Quantum Markov master equation for a UDW detector}
\label{2.1}

The Hamiltonian of the detector-field system takes the form:
\begin{equation}
H_{\text{tot}}=H_{\text{detector}}+H_\Phi+g H_{\text{int}},
\label{eq2.1}
\end{equation}
where the Hamiltonian $H_{\Phi}$ of the massless quantum field is defined in a specific spacetime geometry, and the interaction Hamiltonian $H_{\text{int}}$ encapsulates the system-environment interaction with the coupling constant $g$. For the linear coupling, the interaction Hamiltonian takes the form $\sum_{\mu=0}^3 \mathfrak{m}_\mu \otimes \Phi_\mu[x(t)]$, where $\mathfrak{m}_\mu$ are system monopole momentum operator and $x(t)$ denotes the classical trajectory of the detector.

While the detector-field total system is governed by a Liouville-von Neumann equation
\be
\frac{d}{dt}\rho_{\text{tot}}(t)=-i\left[g {H}_{\text{int}}(t), \rho_{\text{tot}}(t)\right],
\label{eq2.2}
\ee
we are interested in the dynamics of the detector density matrix $\rho(t)=\operatorname{Tr}_{\mathrm{\Phi}}\left[\rho_{\mathrm{tot}}\right]$ obtained by tracing over all field degrees of freedom from the total state. In a "bath" (quantum field) with significantly larger degrees than the detector, it is reasonable to assume that the detector evolves while the field remains largely unaffected. Thus, Born’s approximation at time $t$ can be employed
\begin{equation}
\rho_{\text{tot}}(t) \approx \rho(t) \otimes \rho_{\Phi},
\label{eq2.3}
\end{equation}
where $\rho_\Phi$ is the time-independent, stationary field state. 

Working in the interaction picture, the time evolution of the detector density matrix is determined by an integro-differential equation
\begin{equation}
\frac{d}{dt}{\rho}(t)=g^2 \sum_{\mu,\nu =0}^3 \int_0^t d s\left\{\mathcal{W}_{\mu\nu}(s)\left[\mathfrak{M}_{\nu}(t-s) \rho(t-s),\mathfrak{M}_{\mu}^{\dagger}(t)\right]+\text {h.c.}\right\}, 
\label{eq2.4}
\end{equation}
where $\mathfrak{M}_{\mu}(t)\equiv \exp\left(i H_{\text{detector}} t \right) \mathfrak{m}_{\mu} \exp\left(-i H_{\text{detector}} t \right)$ is the momentum operator in the interaction picture, and $\mathcal{W}_{\mu\nu}(s)$ denotes the Wightman function of the quantum field as
\begin{equation}
\mathcal{W}_{\mu\nu}\left(s_1-s_2\right):=\langle \Phi^\dagger_\mu\left[x\left(s_1\right)\right] \Phi_\nu\left[x\left(s_2\right)\right]\rangle=\operatorname{Tr}_\Phi\left\{\rho_\Phi \Phi^\dagger_\mu\left[x\left(s_1\right)\right] \Phi_\nu\left[x\left(s_2\right)\right]\right\},
\label{eq2.5}
\end{equation}
which describes the correlation of the quantum field at various times along the detector trajectory. The detector dynamics is usually hard to be resolved from \eqref{eq2.4} as it is a time-nonlcal differential equation for $\rho$, which means to know $\rho$ at a specific time $t$, the complete evolution history of $\rho(s)$ from $0$ to $t$ in the integral of \eqref{eq2.4} is necessary, reflecting the \textit{memory effect}. 

Nevertheless, we can assume that the background field has a short correlation time $t_B$, meaning the corresponding Wightman function decays rapidly (e.g., $\left|\mathcal{W}(s)\right| \sim e^{-s / t_B}$) and becomes zero for $s \gg t_B$. Under this Markov approximation, the main contribution to the integral in Eq.(\ref{eq2.4}) comes from the vicinity of $s\approx 0$. The short "memory" of the bath correlation function allows us to replace $\rho(t-s)\approx\rho(t)$ without introducing significant errors, which results in the \textit{Redfield equation}
\begin{equation}
\dot{\rho}(t)=-g^2 \sum_{\mu,\nu =0}^3 \int_0^\infty d s\left\{\mathcal{W}_{\mu\nu}(s)\left[\mathfrak{M}_{\mu}^{\dagger}(t), \mathfrak{M}_{\nu}(t-s) \rho(t)\right]+h . c .\right\}.
\label{eq2.6}
\end{equation}
an expected time-local (i.e., memoryless) differential equation for $\rho$, which depends solely on $t$ and not on the detector's history. The Markov approximation could become increasingly accurate for longer relaxation timescales of the detector, says $s \gg t_B$.

Unfortunately, the Redfield equation is unreliable as it fails to maintain the CP (completely positive) property of quantum dynamics, risking a non-positive density matrix \cite{sec1-26}. To ensure physical consistency, the rotating wave approximation (RWA) is required, averaging out "fast-oscillating" terms, yielding a reduced dynamics that is both trace-preserving and completely positive. Eventually, we arrive at the so-called Gorini-Kossakowski-Sudarshan-Lindblad (GKSL) form of QMME as \cite{sec1-27,sec1-26}:
\be
\frac{d}{dt} \rho(t) = -i [H_{\text{LS}},\rho(t)] + g^2 \mathcal{L}\left[\rho\right]     \label{eq2.7-1}
\ee
where $H_{\text{LS}}$ is known as the Lamb shift Hamiltonian arising from the renormalization of the unperturbed energy levels induced by the system-reservoir coupling. The $\mathcal{L}[\rho(t)]$ is called a Lindblad dissipator of the master equation, as it controls the dissipation and decoherence in the system:
\be
\mathcal{L}\left[\rho\right] = \sum_{\alpha,\beta} \gamma_{\alpha \beta} \left( \mathfrak{M}_{\beta}\hspace{1pt}\rho\hspace{1pt} \mathfrak{M}_{\alpha}^{\dagger} - \frac{1}{2} \{\mathfrak{M}_{\alpha}^{\dagger} \mathfrak{M}_{\beta}, \rho \} \right) \label{eq2.7-2}
 \ee
where the Kossakowski matrix $\gamma_{\alpha\beta}$ is determined by the Wightman function of the background field.

For a UDW detector modeled by a two-level atom, the detector Hamiltonian can be written as $H_{\text{detector}}=\frac{1}{2} \omega \sigma_3$ with its energy gap $\omega$. The monopole momentum operator has a form $\mathfrak{m}_{\mu}=\sigma_{\mu}$, which  $\sigma_0$ is $2\times 2$ identity matrix and $\sigma_i$ are Pauli matrices. The Lindblad dissipator \eqref{eq2.7-2} reduces to \cite{sec1-25}
\be
\mathcal{L}\left[\rho\right]  = \sum_{i, j=1}^3 \gamma_{i j}\left[\sigma_j \rho \sigma_i-\frac{1}{2}\left\{\sigma_i \sigma_j, \rho\right\}\right]  \label{eq2.8}
\ee
where the Kossakowski matrix is decomposed into
\be
\gamma_{ij}=\frac{\gamma_{+}}{2} \delta_{i j}-i \frac{\gamma_{-}}{2} \epsilon_{i j 3} +\gamma_0 \delta_{3, i} \delta_{3, j}
\label{eq2.9}
\ee
with coefficients
\be
\gamma_{\pm}=\mathcal{C}(\omega) \pm \mathcal{C}(-\omega), \quad \gamma_0=\mathcal{C}(0)-\gamma_{+} / 2
\label{eq2.10}
\ee
defined in terms of the two-sided Fourier transformation on the Wightman function of the background field
\be
\mathcal{C}(\omega)=\int_{-\infty}^{\infty} d t~ e^{i \omega t}\left\langle  \Phi[x(t)] \Phi[x(0)]\right\rangle=\int_{-\infty}^{\infty} d t ~e^{i \omega t}~\mathcal{W}(t)
\label{eq2.11}
\ee

For the two-level UDW detector, its density matrix can be expressed in a Bloch form
\be
\rho(t)=\frac{1}{2}\left(1+\sum_{i=1}^3 n_i(t) \sigma_i\right)
\label{eq2.12}
\ee
where the Bloch vector $\boldsymbol{n}(t)$ has its length $l=\sqrt{\sum_{i=1}^3 n_i^2}$ to characterize its purity. Substituting it into the master equation \eqref{eq2.7-1} and specific Lindblad dissipator \eqref{eq2.8}, assuming a general initial state $\rho_0$ with Bloch vector length $l_0$ (for mixed state $l_0<1$, for pure state $l_0=1$) and angle $\theta_0$ to the $z-$axis, the density matrix (13) can be explicitly resolved in terms of time-dependent elements
\be
\boldsymbol{n}(t)=\bigg(\mathcal{E}_1(t)l_0 \sin \theta_0 \cos \Omega t,~\mathcal{E}_1(t)l_0 \sin \theta_0 \sin \Omega t,~\mathcal{E}_2(t)(l_0\cos \theta_0+\gamma)-\gamma \bigg)
\label{eq2.13}
\ee
where the decay rate $\mathcal{E}_1(t) \equiv \exp \left(-2g^2(\gamma_{+} +\gamma_{0})t \right)$, $\mathcal{E}_2(t) \equiv \exp \left(-2 g^2 \gamma_{+} t \right)$ and $\gamma \equiv \gamma_{-} / \gamma_{+}$ have been introduced. The effective Hamiltonian of the detector is $H_{\text{LS}}=\Omega\sigma_z/2$, with a Lamb-type renormalized energy levels, i.e., $\Omega=\omega+$ $i[\mathcal{K}(-\omega)-\mathcal{K}(\omega)]$ \cite{sec1-25}, where $\mathcal{K}(\lambda)=\frac{1}{i \pi} \mathrm{P} \int_{-\infty}^{\infty} d \omega(\omega-\lambda)^{-1}{\mathcal{G}(\omega)}$ with P denoting the principal value.

The length of the Bloch vector $\boldsymbol{n}(t)$ at an arbitrary time is
\be
l(t)=\sqrt{\left[\mathcal{E}_2(l_0\cos \theta_0+\gamma)-\gamma\right]^2+\mathcal{E}_1^2l_0^2 \sin ^2 \theta_0}.
\label{eq2.14}
\ee
Note that asymptotically, the detector achieves a unique thermalization end with $l(t\rar\infty)=\gamma$, which has nothing to do with its initial state and is solely determined by the quantum field correlation in specific spacetime geometry.

\begin{figure}[hbtp]
\centering  
\includegraphics[width=0.7\textwidth]{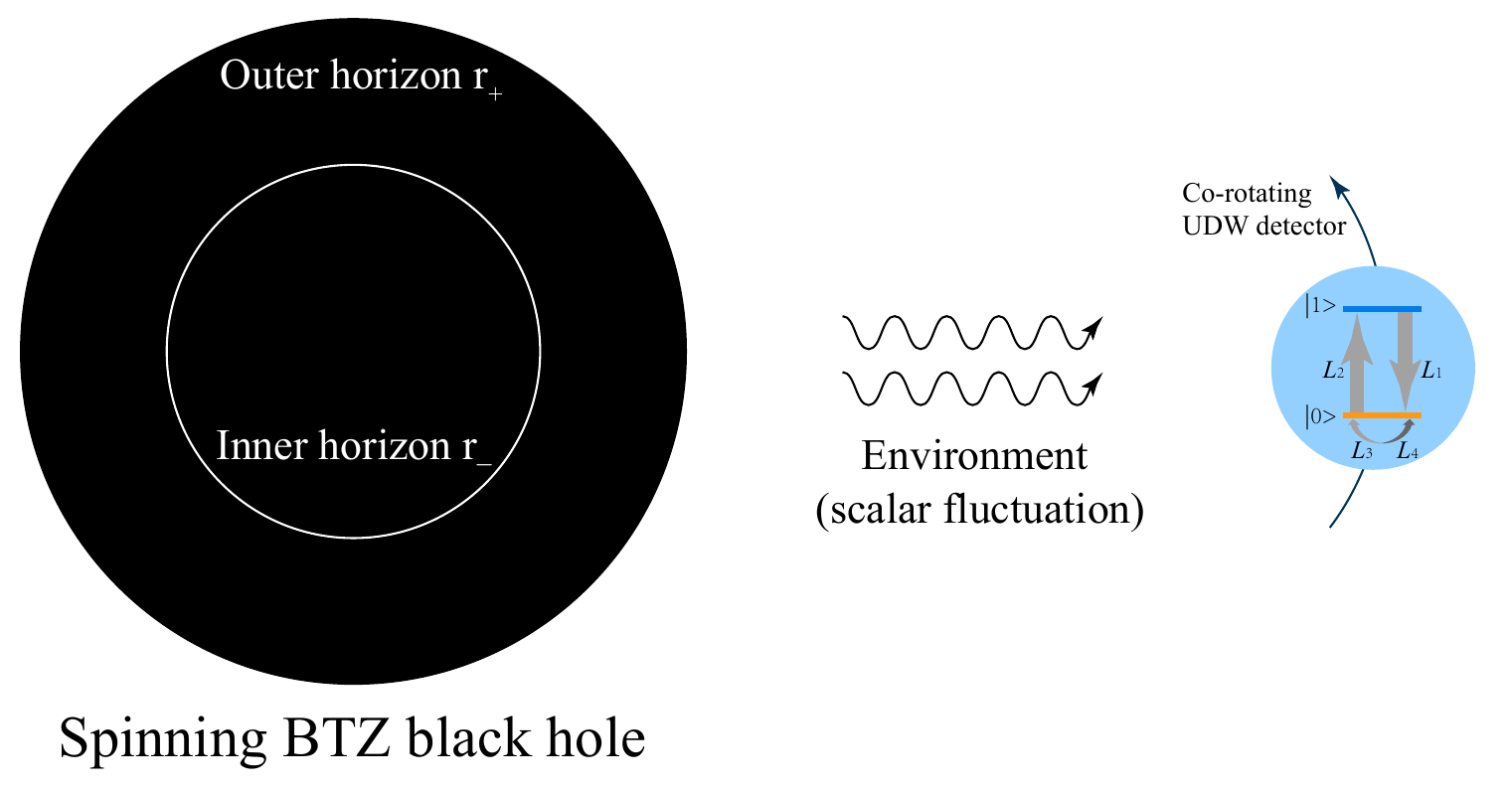}
\caption{Detector-field model in a spinning BTZ spacetime. Jump operators $L_1$ and $L_2$ relate two energy levels of the detector, while operators $L_3$ and $L_4$ keep the detector in the ground state.} 
\label{model}
\end{figure}

For the later construction of quantum thermodynamics for the UDW detector, we need to further derive the instantaneous spectrum of the open dynamics from the QMME \eqref{eq2.7-1} and \eqref{eq2.7-2}. The key point is to reinterpret the open dynamics of the detector as a sequence of Markov jump processes, relating the probabilities of forward and reverse jumps, which can be linked to the entropy production in the environment (see Fig.\ref{model}). 

In terms of diagonalized density matrix $\rho(t) = \sum_x p_x (t) \ket{x(t)}\bra{x(t)}$ with its instantaneous eigenbasis $\bra{x(t)}$, we can rewrite the QMME \eqref{eq2.7-1} into \cite{IG1,IG2}:
\be
\frac{d}{dt} \rho(t) = -i [H_{\text{LS}},\rho(t)] + g^2\sum_k \left(2L_k\rho(t)L_k^\dagger-\left\{L_k^\dagger L_k,\rho(t)\right\}\right),
\label{eq-IG1}
\ee
where the dissipative superoperator is written with so-called jump operators $L_k$ that appears in pairs $(k,k')$ satisfy the local detailed balance condition
\be
L_k = e^{\phi^{k}/2} L_{k^{\prime}}^{\dagger}
\label{eq-IG2}
\ee
Here, $\phi^{k} = -\phi^{k^{\prime}}$ is the entropy change in the environment upon the action of the jump operator $L_k$. Crucially, we may then define the transition rate \cite{IG3}
\be
w_k^{x y}(t):=\left|\bra{x(t)} L_k \ket{y(t)}\right|^2.
\label{eq-IG3}
\ee
Taking the derivative of $p_x(t)=\langle x(t)| \hat{\rho}|x(t)\rangle$, one finds \cite{IG4} 
\be
\frac{{d} p_x(t)}{{d} t}=\sum_{k, y(y \neq x)}\left[w_k^{x y}(t) p_y(t)-w_k^{y x}(t) p_x(t)\right],
\label{eq-IG4}
\ee
which is a master equation for the instantaneous spectrum of the detector state. It should be clarified that, although \eqref{eq-IG4} appears the same in classical stochastic thermodynamics, it is quantum because $p_x(t)$ is meaningful only for the instantaneous basis. Non-diagonal terms of the detector density matrix would emerge if the eigenbasis were fixed. 

For the two-level UDW detector, we know its time-dependent density matrix \eqref{eq2.12} (with \eqref{eq2.13}) can then be diagonalized as $\rho(t) = p_+ \ket{+}\bra{+} + p_- \ket{-}\bra{-}$, where the eigenvalues $p_{\pm}(t) = \frac{1}{2}(1 \pm l(t))$ and eigenvectors are (up to a phase $\varphi_t$):
\be
\ket{+} = \left( e^{-i \varphi_t} \cos{\frac{\Theta}{2}} , ~ \sin{\frac{\Theta}{2}} \right)^{\intercal} ,~~~~~~~
\ket{-} = \left( - \sin{\frac{\Theta}{2}}, ~ e^{i \varphi_t} \cos{\frac{\Theta}{2}} \right)^{\intercal}
\label{eq-IG5}
\ee
with $\cos{\Theta} = {n_3}/l(t)$ and $\tan{\Theta}=n_2/n_1$ are time-dependent parameters.

Similar to diagonalizing the Kossakowski matrix $\gamma_{\alpha\beta}$, we can identify the jump operators for the UDW detector by comparing \eqref{eq2.6} and \eqref{eq-IG1}, and give:
\be
L_1 = \sqrt{\mathcal{C}(\omega)}\sigma_{-},~~~~ L_2 = \sqrt{\mathcal{C}(-\omega)}\sigma_{-}^{\dagger},~~~~L_3 = \sqrt{\mathcal{C}(0)}\pi_{-},~~~~L_4 = \sqrt{\mathcal{C}(0)}\pi_{-}^{\dagger},
\label{eq-IG6}
\ee
with $\sigma_{-} = \ket{0} \bra{1}$ and $\pi_{-} = \ket{1} \bra{1}$. 

For different instantaneous states of the UDW detector, the related transition rate \eqref{eq-IG3} is now calculated by taking $k=1,\cdots,4$ and $x,y=+,-$ and using eigenstates \eqref{eq-IG5}, then explicitly gives
\be
\left\{\begin{aligned} 
w_1^{++} & = w_1^{--} = \frac{\mathcal{C}(\omega)\sin^2{\Theta}}{2}, ~~~ w_1^{+-} = 2\mathcal{C}(\omega)\sin^4{\frac{\Theta}{2}}, ~~~ w_1^{-+} = w_1^{+-} \cot^4{\frac{\Theta}{2}}, \\ 
w_2^{++} & = w_2^{--} = \frac{\mathcal{C}(-\omega)\sin^2{\Theta}}{2},~~~w_2^{+-} = 2\mathcal{C}(-\omega)\cos^4{\frac{\Theta}{2}}, ~~~w_2^{-+} = w_2^{+-} \tan^4{\frac{\Theta}{2}} , \\ 
w_3^{++} & = w_4^{++} = 2\mathcal{C}(0)\cos^4{\frac{\Theta}{2}}, ~~~w_3^{--} = w_4^{--} = 2\mathcal{C}(0)\sin^4{\frac{\Theta}{2}}, \\
w_3^{+-} & = w_3^{-+} = w_4^{+-} = w_4^{-+} = \frac{\mathcal{C}(0)\sin^2{\Theta}}{2}
\end{aligned}
\right.
\label{eq-IG7}
\ee

These results, together with \eqref{eq2.13} and \eqref{eq2.14}, provide two approaches to the complete open dynamics of the two-level UDW detector, once the response function $\mathcal{C}(\omega)$ is determined for a specific spacetime geometry, as discussed later.

\subsection{Quantum fields in BTZ spacetime}

We are interested in a UDW detector in a rotating BTZ black hole spacetime, conformally coupling to a massless scalar field. In this section, we briefly recall (see Ref. \cite{sec1-29} for a review) the BTZ spacetime geometry and relevant properties of the scalar Wightman function in this curved background.

\subsubsection{Boundary conditions in BTZ geometry}

The $(2+1)$-dimensional black hole solution found by Ba\~nados, Teitelboim, and Zanelli is most easily described as $\mathrm{AdS}_3$ identified under a discrete subgroup $\simeq\mathbb{Z}$ of its isometry group. Recall that $\text{AdS}_3$ is hyperboloid with cosmological constant $\Lambda=-1/\ell^2$:
\be
X_{1}^{2}+X_{2}^{2}-T_{1}^{2}-T_{2}^{2}=-\ell^2,
\label{eq2.16}
\ee
embedded in $\mathbb{R}^{2,2}$ with metric 
\be
ds^{2}=dX_{1}^{2}+dX_{2}^{2}-dT_{1}^{2}-dT_{2}^{2},
\label{eq2.17}
\ee
The black hole solution is constructed by identifying the parameters describing boosts in the ( $T_1, X_1$ ) and $\left(T_2, X_2\right)$ planes. Using a set of coordinates $(t,r,\phi)$ that are adapted to the relevant isometries and cover the exterior region of the black hole, the hyperboloid can be parameterized as:
\be
\begin{aligned}
T_1=\ell\sqrt{\alpha}\cosh{\left(\frac{r_{+}}{\ell}\phi - \frac{r_{-}}{\ell^2}t \right)}~~~~~&,~~~~~T_2=\ell\sqrt{\alpha - 1}\sinh{\left(\frac{r_{+}}{\ell^2}t - \frac{r_{-}}{\ell}\phi\right)}, \\
X_1=\ell\sqrt{\alpha}\sinh{\left(\frac{r_{+}}{\ell}\phi - \frac{r_{-}}{\ell^2}t\right)}~~~~~&,~~~~~X_2=\ell\sqrt{\alpha - 1}\cosh{\left(\frac{r_{+}}{\ell^2}t - \frac{r_{-}}{\ell}\phi\right)}
\label{eq2.18}
\end{aligned}
\ee
where 
\be
\alpha(r) = \frac{r^2-r_{-}^2}{r_{+}^2-r_{-}^2}
\label{eq2.19}
\ee
with $r_{\pm}$ the outer/inner horizon radius. 

The $\mathbb{Z}$ quotient is through the identification $(t,r,\phi) \sim (t,r,\phi + 2\pi)$, resulting in a BTZ metric that takes the form:
\be
ds^{2}=-(N^{\perp})^2 dt^2+f^{-2} dr^2+r^2(d\phi + N^{\phi}dt)^2,
\label{eq2.20}
\ee
where the lapse and shift functions
\be
N^{\perp} = f = \left(-M + \frac{r^2}{\ell^2} + \frac{J^2}{4r^2} \right), ~~~ N^{\phi} = -\frac{J}{2r^2},
\label{eq2.21}
\ee
defined by the mass $M$ and the angular momentum $J$. The metric is singular at the location of the inner and outer horizons, now defined by 
\be
r_{ \pm}^2=\frac{M \ell^2}{2}\left(1 \pm \sqrt{1-\frac{J^2}{M^2 \ell^2}}\right)
\ee
Thus, we can express the mass and angular momentum in terms of $r_\pm$ as
\be
M = \frac{r_{+}^2+r_{-}^2}{\ell^2}, ~~ J=\frac{2 r_{+}r_{-}}{\ell},
\label{eq2.22}
\ee
which satisfy $|J|<M\ell$. One notes that the coordinate ranges covering the black hole exterior are $r_{+} < r < \infty, -\infty < t < \infty, -\infty < \phi < \infty$, and the black hole solution asymptotically approaches the $\text{AdS}_3$.

To resolve the explicit open dynamics of the UDW detector \eqref{eq2.12} and \eqref{eq2.13} in BTZ, we need to calculate the response of the detector \eqref{eq2.11} to the conformally coupled massless scalar field, particularly in the Hartle-Hawking vacuum.

Since the BTZ black hole is given by $\mathbb{Z}$ quotient relation with $\text{AdS}_3$, we can use the method of images to derive the Wightman function on the black hole spacetime. Given the Wightman function $G_{\text{A}}(x, x')$ on $\mathrm{AdS}_3$ space, we have the BTZ Wightman function as a summation
\be
G_{\text{BTZ}}(x,x') = \sum_{n=-\infty}^{n=\infty} G_{\text{A}}(x, \Lambda^n x') 
\label{eq2.23}
\ee
where $\Lambda x$ takes $(t, r, \phi) \rightarrow(t, r, \phi+2 \pi)$.  

The quantum fields in BTZ spacetime are complicated by the fact that $\mathrm{AdS}_3$ is not globally hyperbolic. Therefore, it is crucial to establish the proper boundary conditions for quantum fields at infinity \cite{sec1-30}. Otherwise, information can escape or leak in through this surface in a finite coordinate time, spoiling the composition law property of the propagator, and has the most disruptive effect on the Cauchy problem. To reflect the information at the spatial infinity, one can impose either Dirichlet or Neumann boundary conditions, which means that the quantum field or its normal derivative vanishes on the boundary \cite{sec1-31}. Another field quantization scheme on $\mathrm{AdS}_3$ space involves quantization in a transparent box; however, such transparent boundary condition requires a two-time effective Cauchy surface to recirculate the energy and angular momentum, lost at timelike infinity, whose physical interpretation is vague \cite{sec1-32}. 

Different boundary condition choices for the quantum fields at infinity would modify the correlation functions on $\mathrm{AdS}_3$. For example, the Wightman function $G_{\text{A}}(x, x')$ of a massless, conformally coupled field becomes \cite{sec1-29}
\be
G_{\text{A}}(x,x';\zeta) = \frac{1}{4\pi} \left( \frac{1}{\sqrt{\Delta X^2(x,x')}} - \frac{\zeta}{\sqrt{\Delta X^2(x,x') + 4\ell^2}} \right)
\label{eq2.24}
\ee
where parameter $\zeta$ specifies respectively the Neumann ( $\zeta=-1$ ), transparent ( $\zeta=0$ ), or Dirichlet ( $\zeta=1$ ) boundary conditions imposed on the field at infinity. The squared geodesic distance $\Delta X^2(x,x')$ between $x$ and $x'$ in the flat embedding spacetime $\mathbb{R}^{2,2}$ is given by
\be
\Delta {X}^2\left({x}, {x}^{\prime}\right):=  -\left(T_1-T_1^{\prime}\right)^2-\left(T_2-T_2^{\prime}\right)^2 +\left(X_1-X_1^{\prime}\right)^2+\left(X_2-X_2^{\prime}\right)^2.
\label{eq2.25}
\ee 

Naturally, for a UDW detector, we can anticipate that the boundary conditions of the fields at infinity significantly affect its local dynamics. For example, in the spinless BTZ spacetime, the presence of local extrema in the quantum Fisher information has been noted \cite{sec1-19} for the transparent boundary condition, which is absent in AdS spacetimes. In the BTZ with a spinning degree, the choices of boundary conditions heavily impact the transition rate of the detector. An intriguing observation is that the Dirichlet condition provides a transition rate that varies least rapidly \cite{sec1-33}, unlike the Neumann or transparent boundary conditions. Further evidence shows a qualitative discrepancy in the Fisher information for the Dirichlet boundary compared to that for the Neumann or transparent cases \cite{sec1-34}. Indeed, this distinctive property of the Dirichlet boundary, which separates it from the Neumann or transparent choices, has also been reported by studies on entanglement harvesting in BTZ \cite{sec1-35,sec1-36}.

However, beyond previous studies that highlight quantitative differences in the transition rate or Fisher information of the detector, we prefer to identify an observable that is insensitive to detector design (e.g., its energy gap), thus providing incontrovertible discrimination regarding the choice of Dirichlet boundary over Neumann or transparent boundaries. We find that such an approach could be relevant to quantum thermodynamics, as thermodynamic laws or asymmetry in the detector thermalization process are significantly modified by the choices of Neumann or transparent boundaries, but are intrinsically unaffected by the Dirichlet boundary.

\subsubsection{Analytic response of co-rotating observer}
\label{Fourier-correlator}

We consider a co-rotating UDW detector in the exterior region of the BTZ black hole, with a timelike worldline reads \cite{sec1-33}:
\be
r=\text{constant}, ~~~ t=\frac{\ell}{\mathfrak{R}(r)}\tau, ~~~ \phi=\frac{ r_-}{  r_+}\frac{\tau}{\mathfrak{R}(r)};~~~~~\mathfrak{R}(r):=\frac{\sqrt{r^2-r_{+}^2}\sqrt{r_{+}^2-r_{-}^2} }{ r_{+}}.
\label{eq2.26}
\ee
The detector has the same angular velocity as the BTZ black hole, given by
\be
\Omega_{\text{BTZ}} =\frac{J}{2 r_{+}^2}= \frac{r_{-}}{r_{+}\ell}.
\label{eq2.27}
\ee
The Hawking temperature of the BTZ black hole at infinity can be obtained in many approaches \cite{sec1-29}, which is 
\be
T_{\text{BTZ}}=\frac{\kappa_H}{2\pi}=\frac{1}{2\pi\ell^2}\left(\frac{r^2_+-r^2_-}{r_+}\right),
\label{eq2.30}
\ee
where $\kappa_{H}$ is the surface gravity of the black hole with respect to the horizon-generating Killing vector $\partial_t+\Omega_H \partial_\phi$.

For a co-rotating detector, it is important to note that $\mathfrak{R}(r)$ relating its proper time $\tau$ at radial location $r$  to the coordinate time $t$ is a Tolman redshift factor \cite{sec1-37,sec1-38}, i.e., $\mathfrak{R}(r):=\left(-g_{00}\right)^{1 / 2}$, which determines a local temperature for the co-rotating detector as:
\be
T_{\text{KMS}}:=\frac{T_{\text{BTZ}}}{\mathfrak{R}(r)}=\frac{1}{2\pi\ell^2\sqrt{\alpha(r)-1}}.
\label{eq2.29}
\ee 
This is also known as the Kubo-Martin-Schwinger (KMS) temperature \cite{sec1-39,sec1-40}, as it exhibits the periodic property of the Hartle-Hawking state through an appropriately defined imaginary time coordinate \cite{sec1-32}. 

We are now ready to calculate the response $\mathcal{C}(\omega)$ defined in \eqref{eq2.11} for the co-rotating detector. To avoid a possible regularization issue, we adjust the integral \eqref{eq2.11} by applying the sharp switching function and taking the switch-on to be in the asymptotic past \cite{sec1-41}:
\be
\mathcal{C}(\omega) := \frac{1}{\sigma} \int d \tau d \tau^{\prime} \chi_D(\tau) \chi_D\left(\tau^{\prime}\right) e^{i \omega\left(\tau-\tau^{\prime}\right)} G_{\text{BTZ}}\left(x(\tau), x\left(\tau^{\prime}\right)\right). 
\label{eq2.31}
\ee
This integral can be analytically calculated (for details see Appendix \ref{appendixA}) and gives:
\be
\mathcal{C}(\omega) = \frac{1}{2(e^{-\beta \omega}+1)} \sum_{n=-\infty}^{n=\infty}\cos{\left(\frac{n \beta \omega r_-}{\ell}\right)}\left[P_{-\frac{1}{2}+\frac{i \beta \omega}{2 \pi}}\left(2K_{n}+1\right)-\zeta P_{-\frac{1}{2}+\frac{i \beta \omega}{2 \pi}}\left(2Q_{n}+1\right)\right],
\label{eq2.32}
\ee
where $P_{\nu}$  is the associated Legendre function of the first kind, satisfying $P_{-1/2+i\lambda} = P_{-1/2-i\lambda}$, and $K_n = (1-\alpha^{-1})^{-1} \sinh^2(n \pi r_{+}/\ell)$, $Q_n = K_n + (\alpha-1)^{-1}$ are determined by the BTZ black hole parameters. For $\omega=0$, we induce \eqref{eq2.32} to
\be
\mathcal{C}(0)= \frac{1}{4} \sum_{n=-\infty}^{n=\infty}\left[P_{-\frac{1}{2}}\left(2K_{n}+1\right)-\zeta P_{-\frac{1}{2}}\left(2Q_{n}+1\right)\right],
\label{eq2.35}
\ee
which together with \eqref{eq2.32} can straightforwardly give the transition rate $w^{xy}_k(t)$ between different eigenstates $|x(t)\ra$ and $|y(t)\ra$ of the co-rotating detector by \eqref{eq-IG7}, following the Markov jump approach to the open dynamics \eqref{eq-IG4}.

It is obvious that the the response $\mathcal{C}(\omega)$ satisfies:
\be
\mathcal{C}(\omega) = \mathrm{e}^{\beta \omega} \mathcal{C}(-\omega),
\label{eq2.33}
\ee
which is the KMS condition in frequency space, where the parameter $\beta^{-1}:= T_{\text{KMS}}$ is precisely identified as the KMS temperature of Hawking radiation, locally perceived by the detector. Additionally, from the response functions \eqref{eq2.32}, we note that, unlike for a four-dimensional black hole, the Fermi-Dirac type distribution is displayed \cite{sec1-32}. This can't help but remind us that a similar situation exists for the Unruh effect, which exhibits a "statistics inversion" phenomenon \cite{Unruh1,Unruh2,Unruh3} between scenarios in odd and even spacetime dimensions.

Substituting \eqref{eq2.32} back to \eqref{eq2.10}, we obtain the coefficients in the Kossakowski matrix for the co-rotating UDW detector:
\be
\begin{aligned}
& \gamma_{+} = \frac{1}{2} \sum_{n=-\infty}^{n=\infty}\cos{\left(\frac{n \beta \omega r_-}{\ell}\right)}\left[P_{-\frac{1}{2}+\frac{i \beta \omega}{2 \pi}}\left(2K_{n}+1\right)-\zeta P_{-\frac{1}{2}+\frac{i \beta \omega}{2 \pi}}\left(2Q_{n}+1\right)\right], \\
& \gamma_{-} =  \frac{1}{2}\tanh{\left(\frac{\beta\omega}{2}\right)} \sum_{n=-\infty}^{n=\infty}\cos{\left(\frac{n \beta \omega r_-}{\ell}\right)}\left[P_{-\frac{1}{2}-\frac{i \beta \omega}{2 \pi}}\left(2K_{n}+1\right)-\zeta P_{-\frac{1}{2}-\frac{i \beta \omega}{2 \pi}}\left(2Q_{n}+1\right)\right],\\
& \gamma_0 = \frac{1}{4} \sum_{n=-\infty}^{n=\infty}\left[P_{-\frac{1}{2}}\left(2K_{n}+1\right)-\zeta P_{-\frac{1}{2}}\left(2Q_{n}+1\right)\right] - \frac{\gamma_{+}}{2},
\label{eq2.34}
\end{aligned}
\ee
In rotating BTZ spacetime, these functions are considerably complicated. They depend not only on the black hole's angular momentum and effective Hawking temperature $\beta=1/T_{\text{KMS}}$, but also on the choice of boundary conditions of the background field, which can lead to drastic differences \cite{sec1-33}. 

\begin{figure}[htbp]
\centering
\subfloat[]{\includegraphics[width=.33\textwidth]{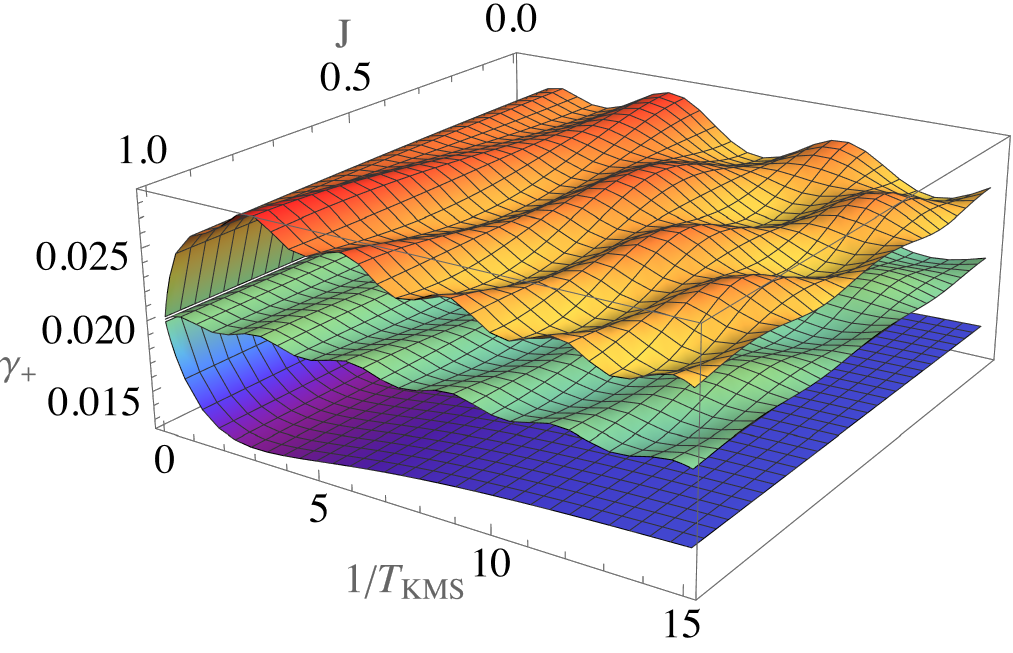}}~
\subfloat[]{\includegraphics[width=.33\textwidth]{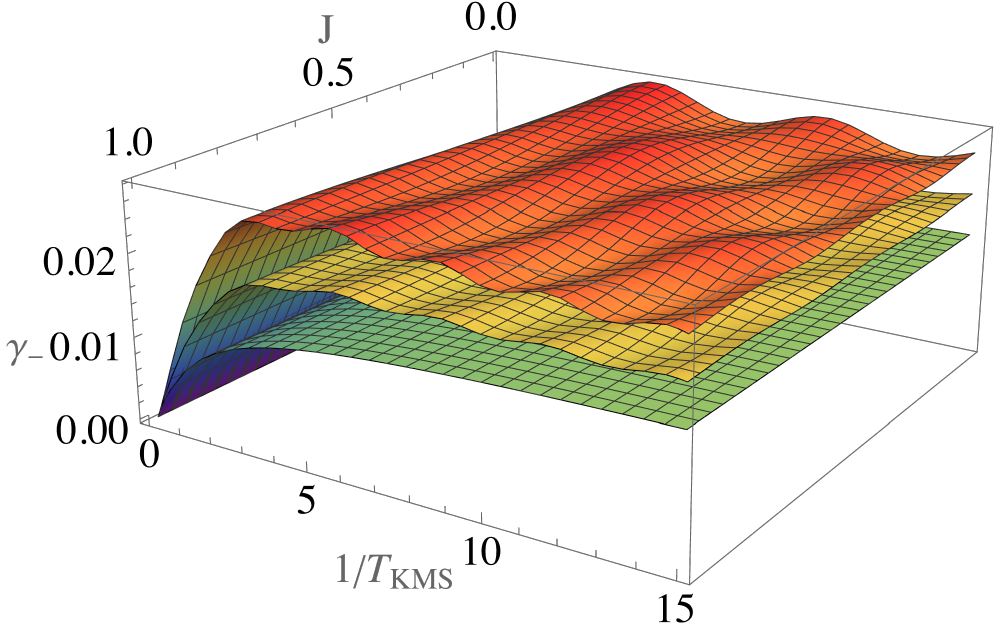}}~
\subfloat[]{\includegraphics[width=.33\textwidth]{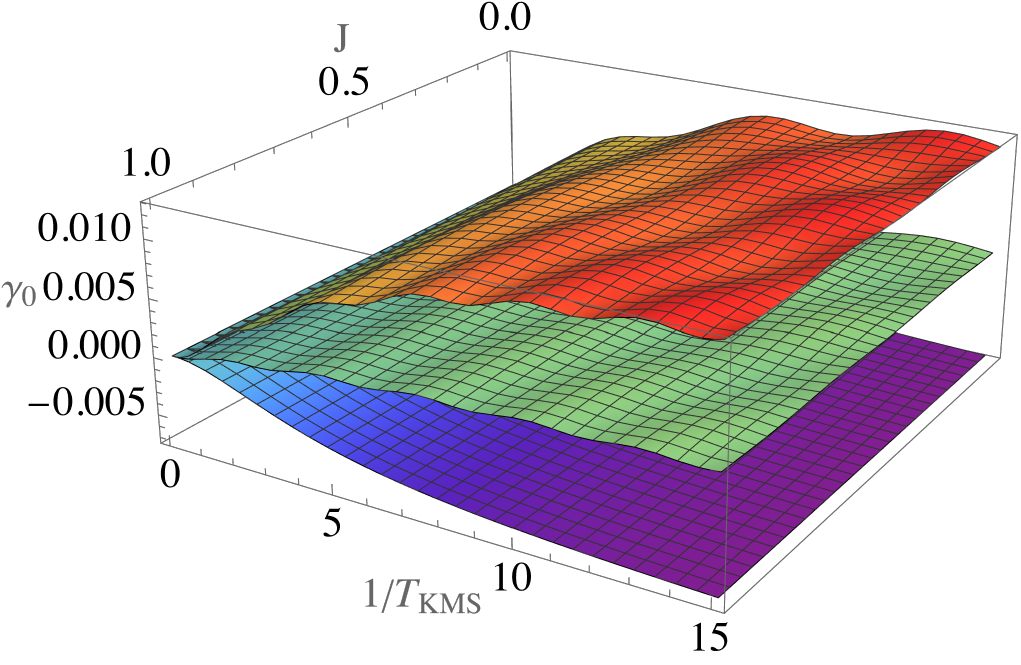}}
\caption{The Kossakowski coefficients as a function of black hole angular momentum $J$ and effective Hawking temperature while Neumann ($\zeta=-1$, top), transparent ($\zeta=0$, middle), and Dirichlet ($\zeta=1$, bottom) boundary conditions are imposed for the scalar field at infinity. Through the estimation, we assume the BTZ black hole with a fixed outer horizon given by $r_+=1.2$.
%
%
 }
\label{fig2}
\end{figure}

To illustrate, we numerically estimate the Kossakowski coefficients in Fig.\ref{fig2} for the black hole with a fixed outer horizon. For a sufficiently high effective temperature, we see that all the Kossakowski coefficients exhibit monotonicity. However, as $T_{\text{KMS}}$ degrades, the oscillatory dependence of the Kossakowski coefficients on $J$ and $T_{\text{KMS}}$ emerges for the Neumann and transparent boundary conditions, but is absent for the Dirichlet case. In the following sections, we will demonstrate that this distinctive behavior of the Kossakowski coefficients for the Dirichlet case always leads to the unique quantum thermodynamics of the UDW detector, compared to the other two boundary conditions.

.

\section{Quantum thermodynamics in BTZ spacetime}
\label{3}

Quantum thermodynamics examines thermodynamic quantities, including temperature, heat, work, and entropy, in microscopic quantum systems, even those involving a single particle \cite{QT0,QT1}. In irreversible processes, for example, those experienced by an open quantum system, these thermodynamic quantities become trajectory-dependent and may serve as \emph{process probes} constrained by quantum thermodynamic laws.

We are interested in applying general quantum thermodynamic laws to the UDW detector system and expect that the evolution of those quantum thermodynamic probes in the Hilbert space could unravel the thermalization dynamics subjected to a specific BTZ geometry. 

%

\subsection{Warm-up: time-evolution of Bloch vector}

As it undergoes thermalization, the detector state should evolve toward equilibrium $\sigma_{\text{thermal}}$, which can be derived from \eqref{eq2.13} in the limit $t\rar\infty$:
\be
\sigma_{\text{thermal}}:=\frac{1}{2}\left(\begin{array}{cc}
1-\gamma & 0 \\
0 & 1+\gamma
\end{array}\right) \equiv \frac{e^{-H_{\text {detector }} / T_{\mathrm{eff}}}}{\operatorname{Tr}\left[e^{-H_{\text {detector }} / T_{\mathrm{eff}}}\right]},
\label{eq2.38}
\ee
with an effective temperature read as
\be
T_{\mathrm{eff}}=\frac{\Omega}{2 \tanh ^{-1}(\gamma)}.
\label{eq2.39}
\ee

\begin{figure}[htbp]
\centering
\subfloat[Neumann: $\zeta=-1$]{\includegraphics[width=.3\textwidth]{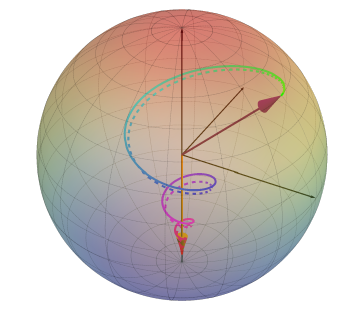}}~
\subfloat[Transparent: $\zeta=0$]{\includegraphics[width=.3\textwidth]{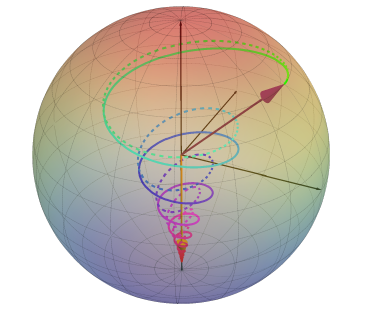}}~
\subfloat[Dirichlet: $\zeta=1$]{\includegraphics[width=.3\textwidth]{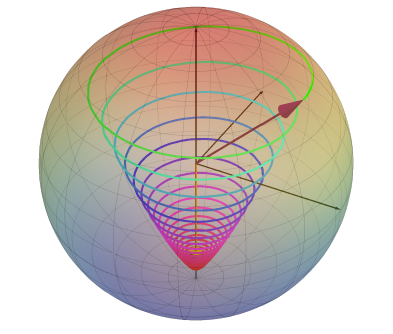}}
\caption{The trajectories in the Bloch sphere are swept out by a co-rotating UDW detector located outside a BTZ black hole. The trajectories are distinct for spinless (solid curves) and spinning (dashed curves) BTZ geometry, as well as for the scalar field with (a) Neumann b.c. ($\zeta=-1$), (b) transparent b.c. ($\zeta=0$), and (c) Dirichlet b.c. ($\zeta=1$). The detector is assumed to start from a pure state with $l_0=1$ and $\theta=\pi/4$. Its Bloch vector then traces a spiral trajectory with a continuously contracting radius until it coincides with the Bloch vector of the thermal state. The effective Hawking temperature is fixed by $\alpha=2$ and $\beta\omega=2$. The black hole mass and angular momentum are determined from \eqref{eq2.22}, where we assume fixed mass $M\ell^2=1$, and $J=0$ for solid curves, and $J=\sqrt{3}/2$ for dashed curves, respectively.
%
%
 }
\label{fig3}
\end{figure}

Intuitively, we anticipate that this process can be characterized by the time-evolution of the Bloch vector \eqref{eq2.13}, which determines the UDW detector state through \eqref{eq2.12}. Depicted in a Bloch sphere (see Fig.\ref{fig3}), we find that the origin-anchored Bloch vector of a UDW detector sweeps out a spiral trajectory with a continuously contracting radius over time, moving toward the specific location on the $z$-axis that labels the thermal state.

We note that the classical spinning degree of a BTZ black hole directly influences the state evolution of the detector. In Fig.\ref{fig3}, we illustrate the evolving trajectories for the co-rotating UDW detector outside a spinless (solid curves) and spinning (dashed curves) BTZ black hole, both with the same mass $M$ and effective temperature $T_{\text{KMS}}$. These distinctive trajectories arise from the non-vanishing angular momentum of the black hole. Furthermore, they may depend on the boundary conditions imposed on the scalar field at spatial infinity. Typically, among the three boundary conditions, we observe a unique evolving trajectory for the Dirichlet case, with no significant difference resulting from the black hole spinning. This phenomenon may be ascribed to the least rapidly varying detector response \eqref{eq2.32} under the Dirichlet condition, due to partial cancellations between terms in the Dirichlet Wightman function \eqref{eq2.24}. We note that, compared to the transparent and Neumann boundary conditions, a similar specialty arising from the Dirichlet condition on the background field has been reported on several occasions \cite{sec1-33}, such as in entanglement harvesting \cite{sec1-34,sec1-35}. 

In the following section, we will revise the open dynamics of the UDW detector using a quantum thermodynamic approach, where the trajectories of the Bloch vector are mapped to various thermal process functions.

\subsection{The quantum Zeroth Law}

In classical thermodynamics, the zeroth law of thermodynamics states that a system can reach an equilibrium state after a sufficiently long interaction, thereby acting like a thermometer that indicates the temperature of the bath. However, in the quantum regime, the “in equilibrium” condition is not necessarily equal to “equal temperature” \cite{QT15,QT16}. Otherwise, the detailed balance condition is superior, which should be guaranteed once $\mathcal{L}\left[\rho_f\right] =0$ at thermalization end, defined from Eq. \eqref{eq2.7-1}. 

One can use quantum relative entropy (QRE) to quantify the distance between the state $\rho(t)$ of a quantum system and its stationary state $\rho_{f}=\mathds{B}(\rho)$ after a certain process $\mathds{B}$ (interacting with the bath)
\be
S[\rho \Vert \mathds{B}] = \Tr{\rho \log{\rho}} - \Tr{\rho \log{\rho_{f}}}.
\label{eq2.36}
\ee
Then the Zeroth Law of quantum thermodynamics can be manifested as: a state $\rho$ is in quantum thermodynamic equilibrium with a bath $\mathds{B}$ when $S[\rho \Vert \mathds{B}]=0$ \cite{QT8}. 

Specifying a UDW detector in BTZ spacetime, we translate the law into 
\be
\mathcal{D} \equiv S\left(\rho(t) \| \sigma_{\text{thermal}}\right)=0  ,
\label{eq2.37}
\ee
where we have identified $\rho_f$ as the thermalization end state.

The QRE at the left-hand side of \eqref{eq2.37} can be explicitly given as \cite{sec1-18}
\be
\mathcal{D}(\tau)
=  \frac{1}{2} \log \left(\frac{1-\ell^2}{1-\ell_{\text {thermal }}^2}\right)+\frac{\ell}{2} \log \left(\frac{1+\ell}{1-\ell}\right) 
+\frac{\ell \cos \alpha}{2} \log \left(\frac{1+\ell_{\text {thermal }}}{1-\ell_{\text {thermal }}}\right)
\ee
where $\cos \alpha=n_3 / \ell$ is the angle between the Bloch vector and the $z$-axis. The Bloch vector of the thermal state is aligned with the negative $z$-axis and has a length $\ell_{\text {thermal }}=\gamma$, as observed from \eqref{eq2.38}.

\begin{figure}[htbp]
\centering
\subfloat[Neumann: $\zeta=-1$]{\includegraphics[width=.31\textwidth]{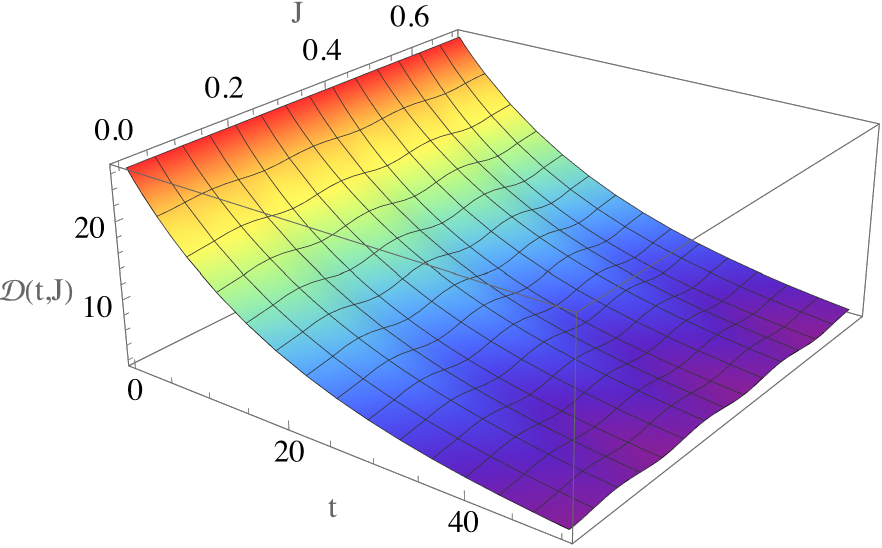}}
\subfloat[Transparent: $\zeta=0$]{\includegraphics[width=.35\textwidth]{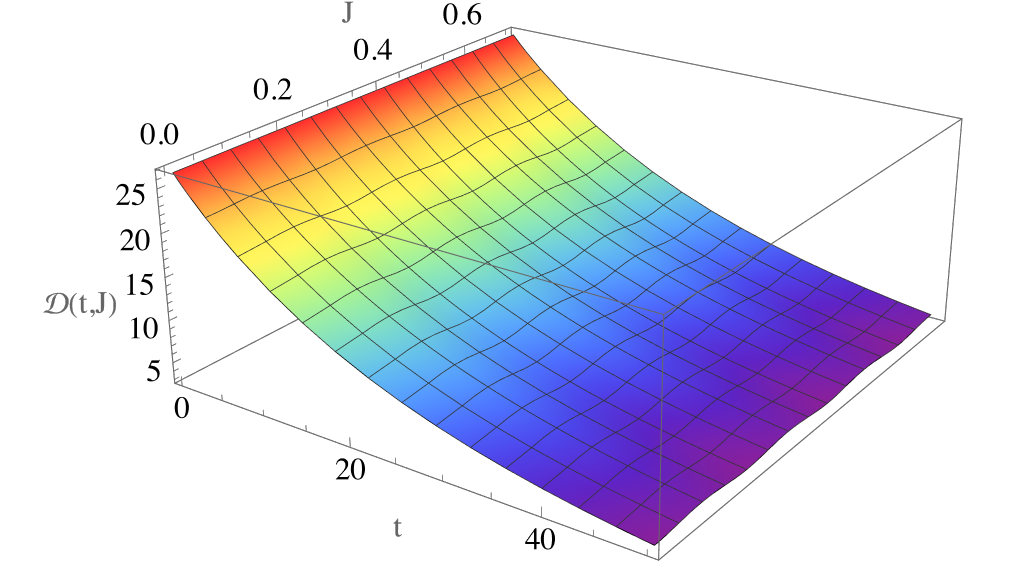}}
\subfloat[Dirichlet: $\zeta=1$]{\includegraphics[width=.30\textwidth]{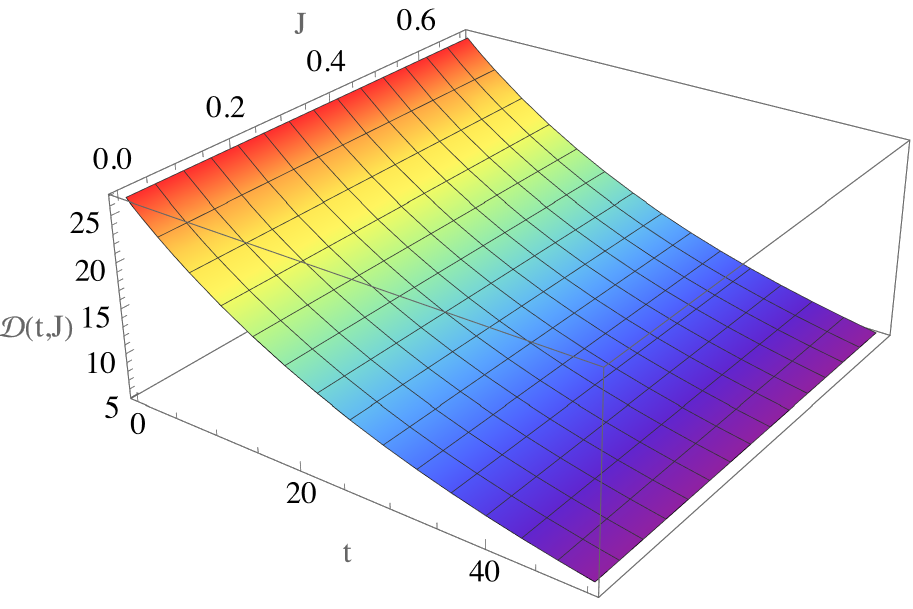}}
\caption{The QRE of the co-rotating detector as a function of black hole angular momentum $J$ and time. As time passed, the QRE monotonously degrades and eventually vanishes, indicating the unique thermalization end. At an arbitrary time, the QRE has a periodic dependence on the black hole's angular momentum, which is obvious for (a) Neumann b.c. ($\zeta=-1$), (b) transparent b.c. ($\zeta=0$), but hardly observed for (c) Dirichlet b.c. ($\zeta=1$). The detector is assumed to start from a pure state with $l_0=1$ and $\theta=\pi/4$. The effective Hawking temperature is fixed by $\beta\omega=10\pi$.
l
 }
\label{fig4}
\end{figure}

We depict the QRE of the detector as a function of black hole angular momentum $J$ and time in Fig.\ref{fig4}, with other physical parameters fixed, such as effective Hawking temperature $T_{\text{KMS}}$. As time passes, we observe that for arbitrary $J$, the QRE monotonously degrades to zero, which confirms the intuitive depiction in Fig.\ref{fig3}, where the detector Bloch vector continuously approaches the unique thermalization end $\sigma_{\text{thermal}}$. Outside the black hole with different angular momentum, we recognize that the QRE degrades along numerically distinct trajectories, reflecting the fact that the spinning degree of the black hole may be encoded in the detector's open dynamics, as seen in Fig.\ref{fig3}. Nevertheless, from a quantum thermodynamic perspective, we note a novel periodic dependence of QRE on the black hole's angular momentum. Such periodic behavior is evident for a scalar background with von Neumann and transparent boundary conditions (see Fig.\ref{fig4}(a)(b)), but is hardly observed for Dirichlet boundary conditions (see Fig.\ref{fig4}(c)).

%
\begin{figure}[htbp]
\centering
\subfloat[Neumann: $\zeta=-1$]{\includegraphics[width=.33\textwidth]{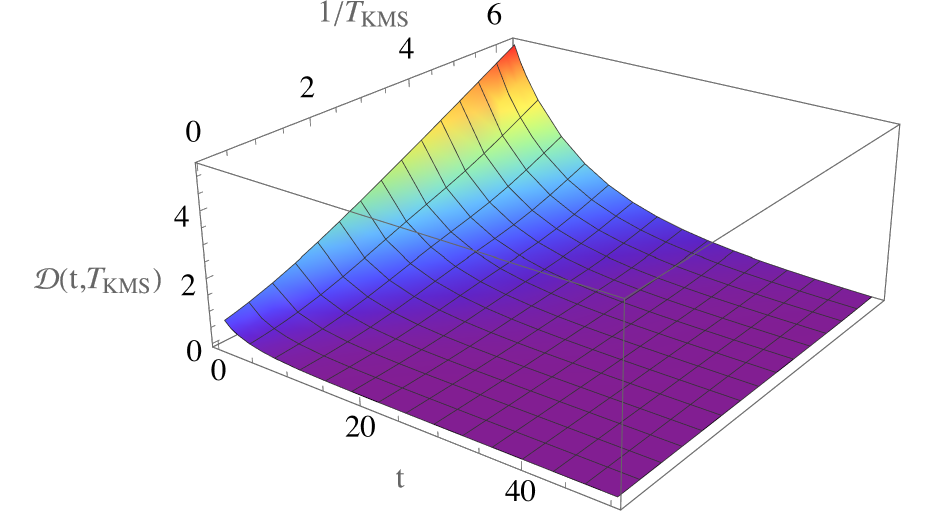}}
\subfloat[Transparent: $\zeta=0$]{\includegraphics[width=.31\textwidth]{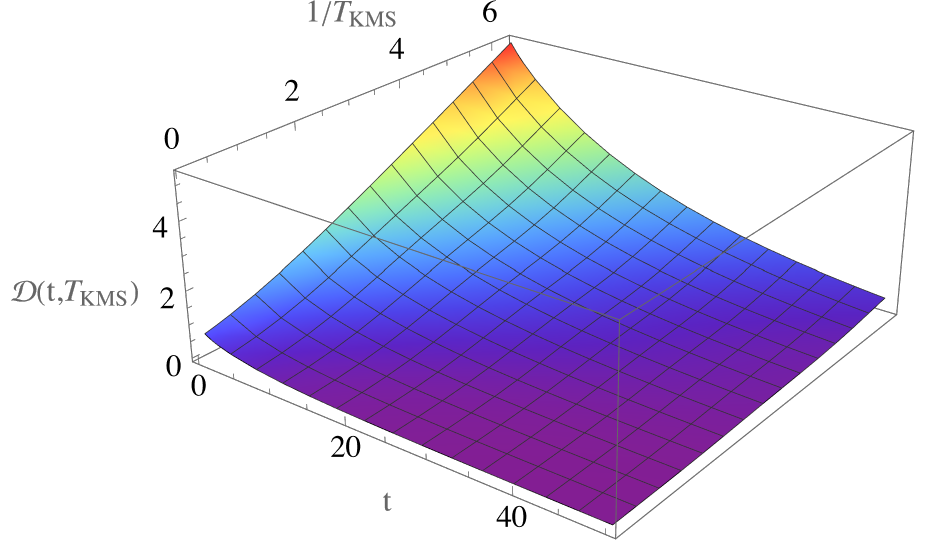}}
\subfloat[Dirichlet: $\zeta=1$]{\includegraphics[width=.32\textwidth]{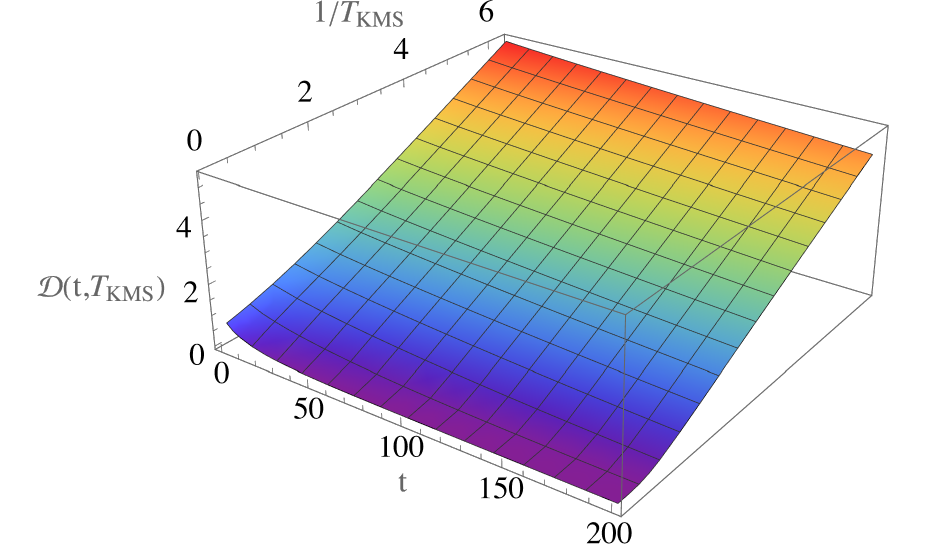}}
\caption{The QRE of the co-rotating detector as a function of the effective Hawking temperature $T_{\text{KMS}}$ and time. The QRE starts from different initial values determined by the distance of its initial state to the final equilibrium set by $T_{\text{KMS}}$, but eventually approaches zero. The detector is assumed to begin from a pure state with $l_0=1$ and $\theta=\pi/4$. We estimate the QRE using the fixed BTZ geometry, specifically $r_+=1$ and $r_-=0.5$. The scalar field is specified with (a) Neumann b.c. ($\zeta=-1$), (b) transparent b.c. ($\zeta=0$), and (c) Dirichlet b.c. ($\zeta=1$). One can observe that the thermalization timescale for the Dirichlet boundary is significantly longer than the other two boundary condition choices.
 }
\label{fig5}
\end{figure}

The co-rotating detector in orbit at different radial distances from the black hole would perceive radiation at effective temperatures $T_{\text{KMS}}$. In Fig.\ref{fig5}, we illustrate the detector's QRE as a function of effective Hawking temperature $T_{\text{KMS}}$ and time outside a BTZ black hole with fixed inner and outer horizons $r_\pm$. The unique thermalization end is indicated by vanishing QRE after a sufficiently long time. For different effective temperatures, we note that the time evolution of QRE starts from different initial values. This is because the QRE measures the distance between the detector's initial state, which we choose to be a fixed pure state, and the final equilibrium state determined by a specific effective temperature through \eqref{eq2.38}. On the other hand, the specialty of the Dirichlet boundary condition choice is again demonstrated, as shown in Fig.\ref{fig5}(c), where the thermalization process sustains a significantly longer time than the von Neumann and transparent boundary cases (see Fig.\ref{fig5}(a)(b)).


\subsection{The quantum First Law}

For a classical system that cannot exchange any matter with the surrounding medium, work $W$ and heat $Q$ are the only two forms of energy transfer. This is the core of the First Law in a classical context, stated as the conservation of energy, $dU = \dbar {W} + \dbar {Q}$. Extending to the quantum domain, the internal energy of a quantum system is defined as $U\equiv\langle H\rangle=\text{tr}\left\{\rho H\right\}$ \cite{QT2}, and the energetic contribution of the dynamics of coherence must be taken into account, which is an element absent in both classical and stochastic thermodynamics. 

For an open quantum system undergoing a relaxation process, its density matrix $\rho(t) = \sum_x p_x (t) \ket{x(t)}\bra{x(t)}$ is not commutative with its Hamiltonian $H=\sum_{n=1}^D E_n \ket{n} \bra{n}$, since $c_{nx}:= \bra{n}\ket{x}$ is in genenral not a constant during the evolution. The internal energy is then $U(t)= \sum_{n,x} E_n p_x c_{nx} $, whose time evolution can be divided into three parts:
\be
\dot{U}(t) = \sum_{x,n} \left( \dot{E}_n p_x \lvert c_{nx}\rvert^2 +E_n \dot{p}_x \lvert c_{nx}\rvert^2 + E_n p_x \lvert \dot{c}_{nx}\rvert^2 \right)  \label{eq2.40}
\ee
The first term measures the variation of the quantum work $\mathbb{W}$, which arises from alterations in the energy-level configuration $E_n$ \cite{QT17}. The second term is determined by the variation of $p_x$, which appears in the von Neumann entropy change of the system $dS=-k_B \sum_x\left[\log \left(p_x\right) d p_x\right]$. Recall the classical interpretation of heat exchanged between the working substance and the external environment that is accompanied by an entropy change, we can identify the term with $\dot{p}_x$ as a variation of quantum heat $\mathbb{Q}$ \cite{QT3}. Finally, note that the third term in \eqref{eq2.40} does not depend on either $dE_n$ or $d p_x$, which directly classifies it as work or heat, respectively. Instead, it depends on the variation of the quantity $\left|c_{nx}(t)\right|^2=|\langle n(t) \mid x(t)\rangle|^2$, which in a quantum process varies only if the directions of the basis vectors $|x\rangle$ of the density operator change with respect to the basis vectors $|n\rangle$ of the Hamiltonian. Physically, such a variation, like the third term in \eqref{eq2.40} denoted as $\dot{\mathbb{C}}$ \cite{QT8}, occurs if the quantum coherence of the system changes with time \cite{QT18}. In conclusion, the First law of quantum thermodynamics can now be written as \cite{QT9}
\be
dU = \dbar \mathbb{W} + \dbar \mathbb{Q}+ \dbar \mathbb{C}
 \label{eq2.41}
\ee
which claims that in the quantum regime the quantum work, heat, and coherence determine the variation of system internal energy\footnote{Utilizing some specific protocols, the quantum coherence can be used to extract work \cite{QT19,QT20}, which confirms the role of a quantum resource played by coherence.}. We note that for a closed quantum system, the unitary evolution means $[H, \rho]=0$, leading to a constant $c_{nx}$. The First Law then reduces to the standard form $dU = \dbar \mathbb{W} + \dbar \mathbb{Q}$ as in stochastic thermodynamics, since no coherence dynamics exists.

\subsubsection{Trade-off between quantum heat and coherence}
\label{3.3.1}

We now apply the quantum First Law \eqref{eq2.41} to a UDW detector dynamics in BTZ spacetime, whose Hamiltonian is time-independent $H_{\text{detector}}=\frac{1}{2}\omega\sigma_3$ giving the eigenvalues and eigenstates are $E_1=\frac{\omega}{2}$ with $\ket{0}=(1,0)^{\intercal}$ and $E_2=-\frac{\omega_t}{2}$ with $\ket{1}=(0,1)^{\intercal}$, repectively. As time passes, the density matrix of the detector evolves as $\rho(t) = p_+ \ket{+}\bra{+} + p_- \ket{-}\bra{-}$, with its time-varying eigenvalues $p_{\pm}(t)$ and eigenvectors $\ket{\pm}$ have been given in \eqref{eq-IG5}.

Based on the definition, we obtain the internal energy of the detector as\footnote{Strictly speaking, the detector internal energy should be defined using the effective Hamiltonian $H_{\text{eff}}=\Omega\sigma_z/2$ with a renormalized frequency, as done in Appendix \ref{AppendixB}. However, since the contribution from the Lamb shift is always far less than the energy spacing $\omega$ of the detector, we usually ignore this part in estimations.}
\be
U_t := \Tr[\rho(t)H_{\text{detector}}]= \frac{\omega}{2} l(t)\cos{\Theta}
 \label{eq2.42}
\ee 
which is determined by the energy spacing of the detector, the norm of its Bloch vector \eqref{eq2.14}, as well as the angle $\cos\Theta=n_3/l(t)$ between the Bloch vector and the $z-$axis. Substituting \eqref{eq2.42} back into the decomposition \eqref{eq2.40}, we can identify the contributions to the detector internal energy from quantum work, heat, and coherence, respectively (for details see Appendix \ref{AppendixB}):
\be
\left\{\begin{aligned}
\dot{\mathds{W}}(t) &= \frac{{l}(t) \cos{\Theta}}{2}\frac{d\omega}{dt}=0  \\
\dot{\mathds{Q}}(t) &= \frac{\omega \cos{\Theta}}{2}\frac{d {l}(t)}{dt}= -g^2 \Omega \left[\gamma_{-} + \gamma_{+} n_3 + \left( \gamma_0 n_3 - \gamma_{-} \right) \sin^2 \Theta \right] \\
\dot{\mathds{C}}(t) &= \frac{\omega {l}(t)}{2} \frac{d\cos{\Theta}}{dt}
=g^2 \Omega \left( \gamma_0 n_3 - \gamma_{-} \right) \sin^2 \Theta 
\label{eq2.43}
\end{aligned}
\right.
\ee
We observe that for the standard UDW detector, the rate of change of quantum work is zero due to its time-independent Hamiltonian. This indicates that the quantum work performed by the detector remains constant, even as it undergoes thermalization. On the other hand, the varying $\mathds{Q}$ and $\mathds{C}$ suggest that an interchange between quantum heat and coherence undergoes Hawking radiation, while both of them depend on the measurement basis. It is interesting to note that a closely related trade-off between quantum coherence and heat (in terms of the classical Kullback-Leibler divergence) is observed \cite{sec1-18} in a thermalization process in Schwarzschild spacetime.

\begin{figure}[htbp]
\centering
\subfloat[Neumann: $\zeta=-1$]{\includegraphics[width=.33\textwidth]{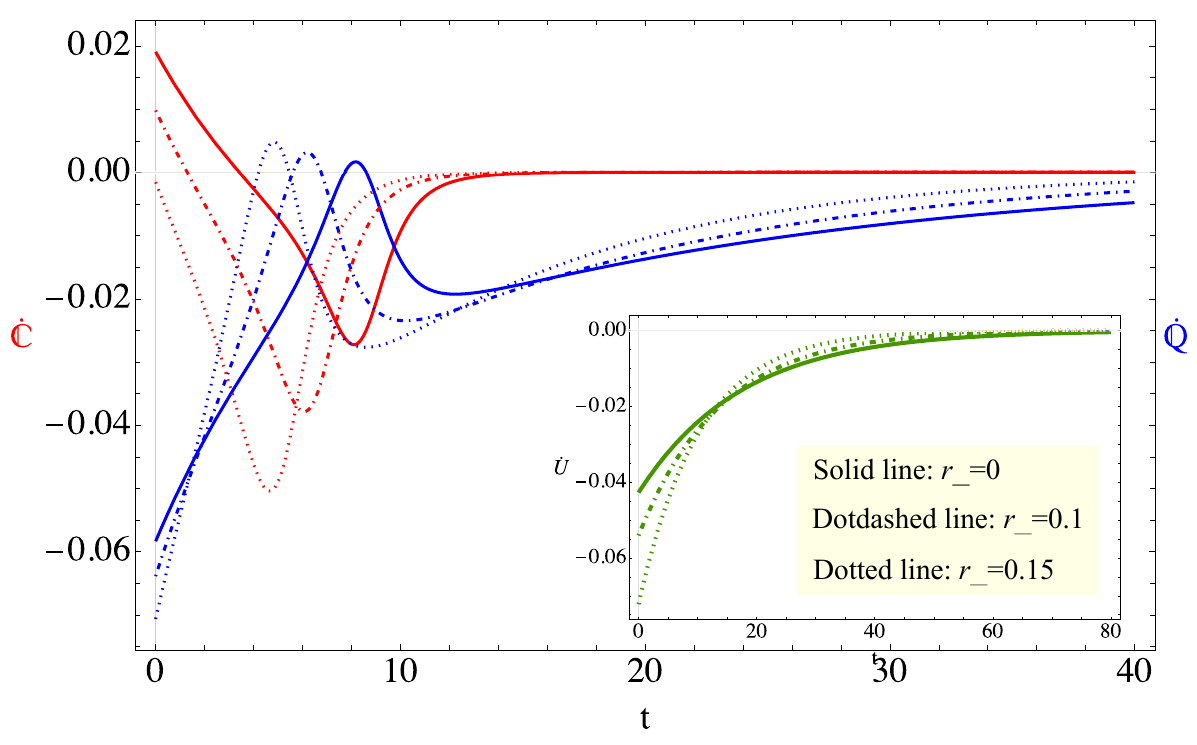}}
\subfloat[Transparent: $\zeta=0$]{\includegraphics[width=.34\textwidth]{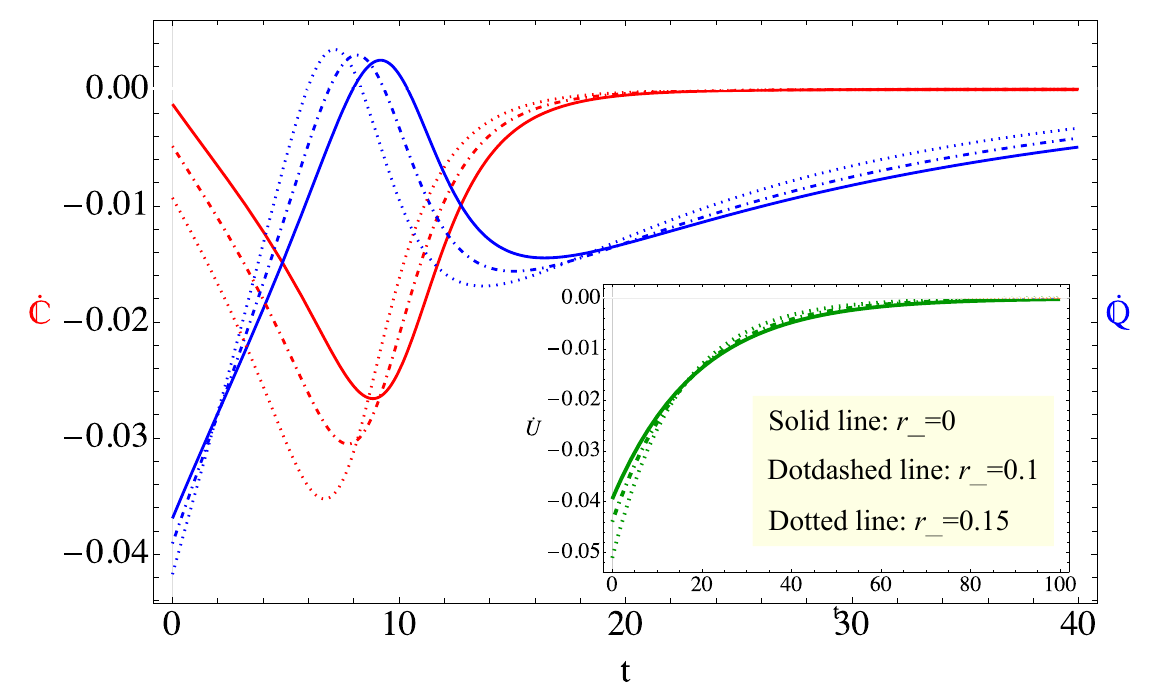}}
\subfloat[Dirichlet: $\zeta=1$]{\includegraphics[width=.33\textwidth]{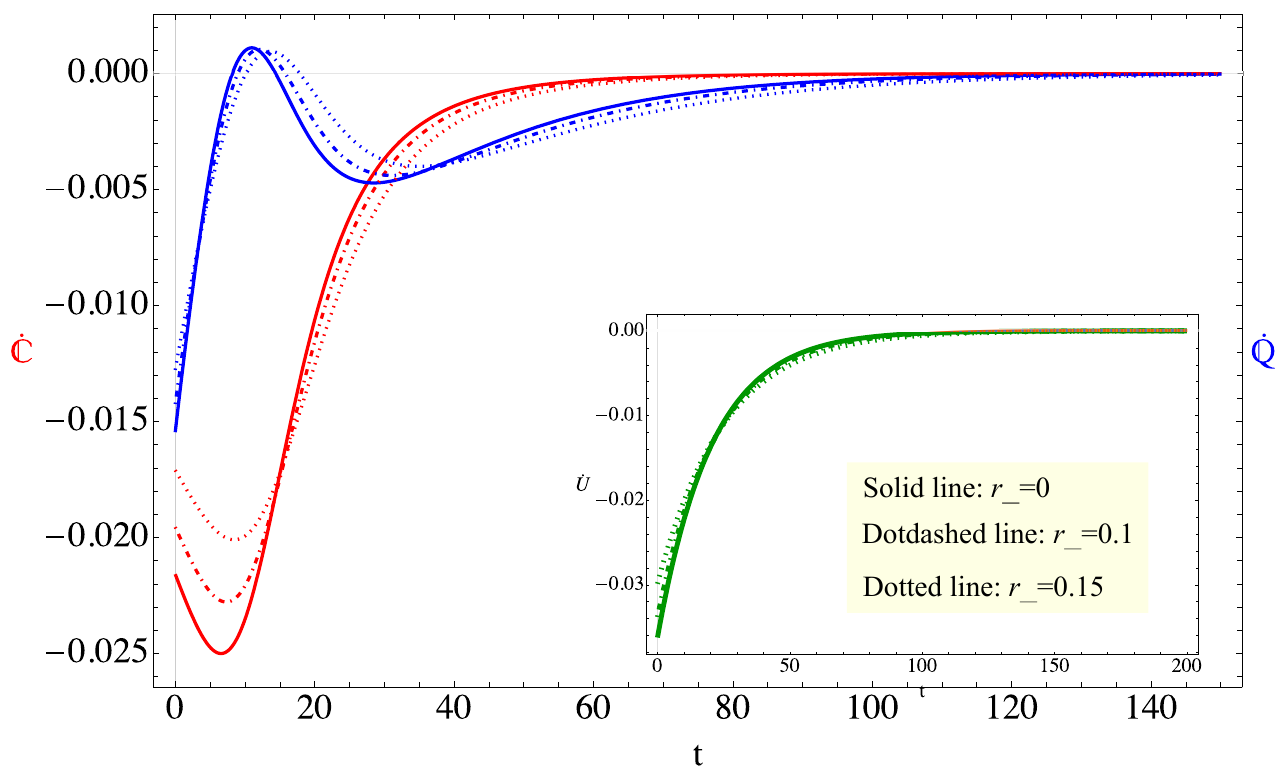}}
\caption{The change rate of quantum coherence and quantum heat of a co-rotating detector as a function of black hole angular momentum $J$ and time. The detector is prepared in an initial pure state with $l_0=1$ and $\theta_0=\pi/3$. The BTZ black hole is assumed to have a fixed outer horizon $r_+=0.2$, while the effective Hawking temperature perceived by the detector is set by $\beta\omega=2\pi$. The red curves denote the change rate of quantum coherence $\dot{\mathds{C}}$ and the blue curves denote the quantum heat change rate $\dot{\mathds{Q}}$. The nonvanishing angular momentum (represented by a larger $r_-$) has a significant impact on the evolution of quantum heat and coherence. Moreover, the estimation within (c) Dirichlet b.c. ($\zeta=1$) shows a significantly longer time for thermalization than that evolving within (a) Neumann b.c. ($\zeta=-1$) and (b) transparent b.c. ($\zeta=0$). In the insets of (a), (b), (c), the monotonously varying rate of the detector's internal energy is depicted.
 }
\label{fig6}
\end{figure}

We consider a co-rotating detector outside the BTZ black hole that perceives the Hawking radiation at a fixed effective temperature. Then, we can depict the \emph{typical} change rates of quantum heat, coherence, and internal energy as functions of time evolving with black hole angular momentum (see Fig.\ref{fig6}). 

From the insets of Fig.\ref{fig6}, we observe that the varying rate of the detector's internal energy (green curves) is a monotonous function of time, regardless of the choice of quantum field boundary condition. After a sufficiently long time has passed, $\dot{U}$ approaches zero, indicating that no energy transfer occurs between the detector and the background, i.e., a thermalization endpoint has been reached. We note that different boundary condition choices of the background field may affect the timescale of the detector's thermalization process. In particular, the Dirichlet case (inset of Fig.\ref{fig6}(c)) has a significantly longer timescale than the Neumann (inset of Fig.\ref{fig6}(a)) and transparent (inset of Fig.\ref{fig6}(b)) boundary choices.

Switching to a perspective from quantum First Law \eqref{eq2.41}, a complementary evolution between the change rates of quantum heat and coherence is clearly shown in Fig.\ref{fig6}. The detector's unique thermalization endpoint is characterized by a refined formula as $dU=\dbar\mathbb{Q}=\dbar\mathbb{C}=0$. For the rotating BTZ black hole with fixed outer horizon $r_+$, we note that the nonvanishing angular momentum $J$, represented by varying $r_-$, has a significant impact on the change rate of quantum heat (blue curves) and coherence (red curves). For example, for the background field under Neumann or transparent boundary conditions (Fig.\ref{fig6}(a), (b)), the larger black hole angular momentum leads to enhanced non-monotonic variations of $\dot{\mathbb{Q}}$ and $\dot{\mathbb{C}}$, whose extrema occur at progressively earlier times, indicating that a more violent thermalization process could happen\footnote{This is also consistant with the evolution of internal energy change rate in the insets of Fig.\ref{fig6}(a), (b), where the larger angular momentum corresponding a more drastic increase of $\dot{U}$.}. For the scalar field with the Dirichlet boundary condition, Fig.\ref{fig6}(c) reveal significantly extended timescales for $\dot{\mathbb{Q}}$ and $\dot{\mathbb{C}}$ approaching asymptotic zero, which means the thermalization process could persist for an anomalously long duration in this case.

\begin{figure}[htbp]
\centering
\subfloat[Neumann: $\zeta=-1$]{\includegraphics[width=.33\textwidth]{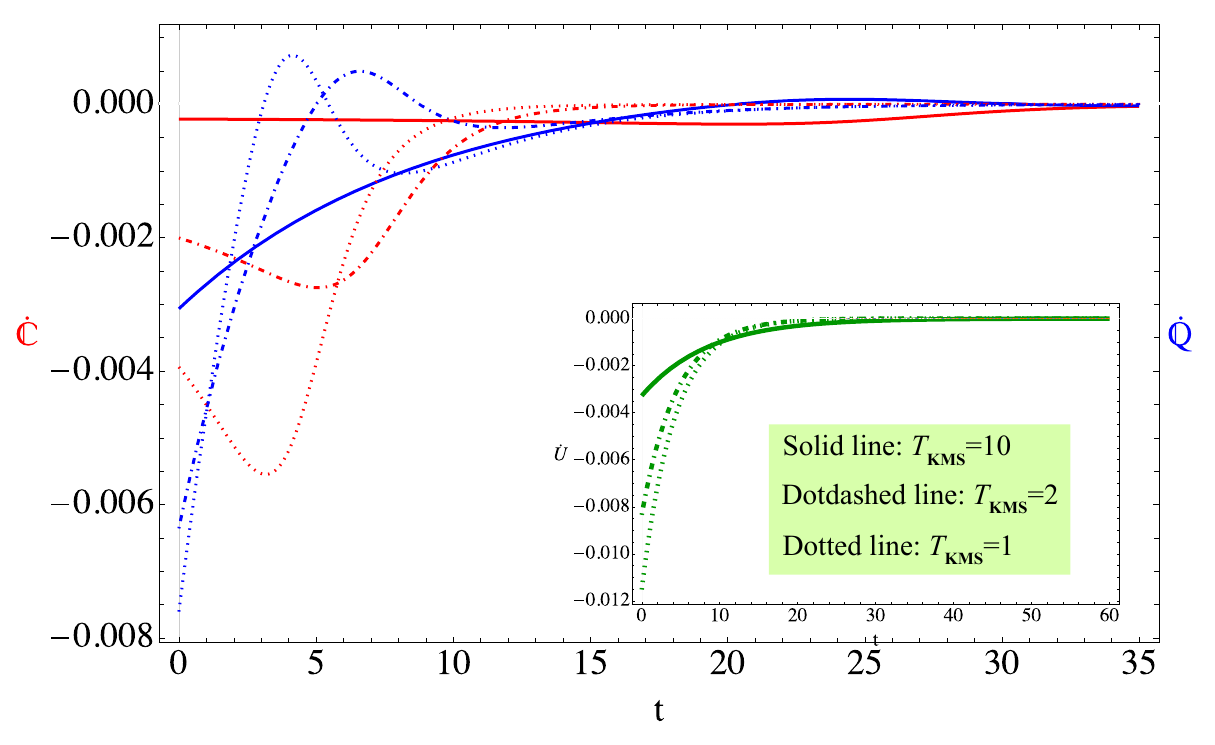}}
\subfloat[Transparent: $\zeta=0$]{\includegraphics[width=.33\textwidth]{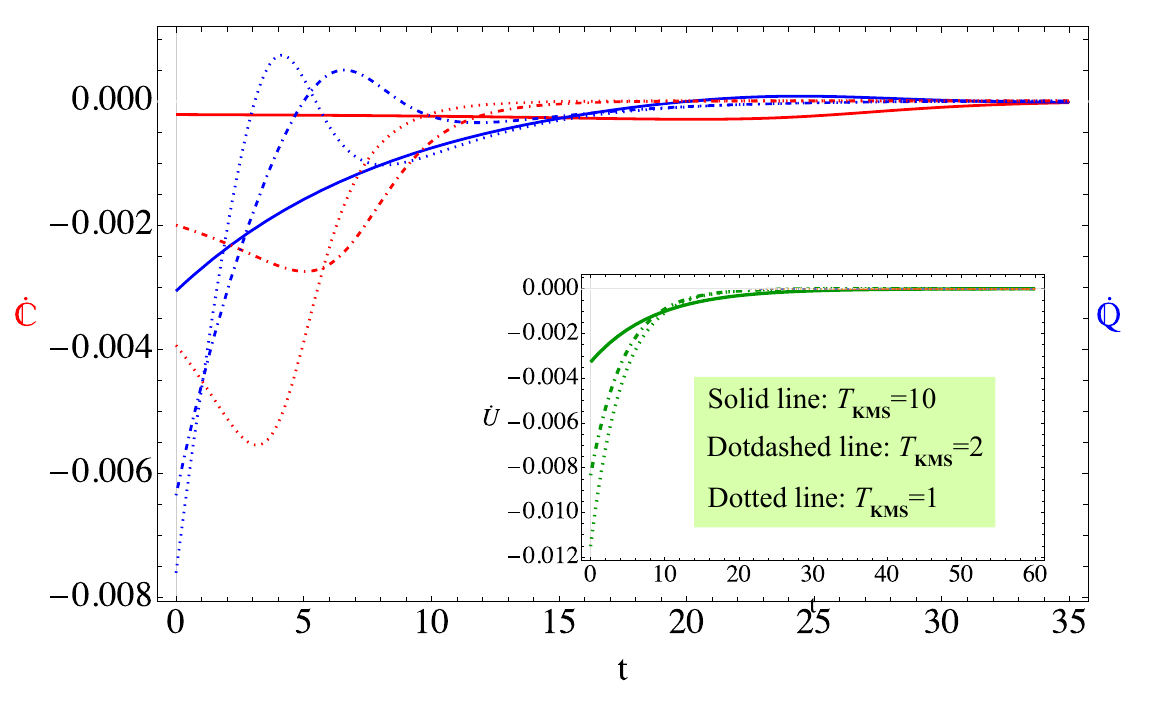}}
\subfloat[Dirichlet: $\zeta=1$]{\includegraphics[width=.33\textwidth]{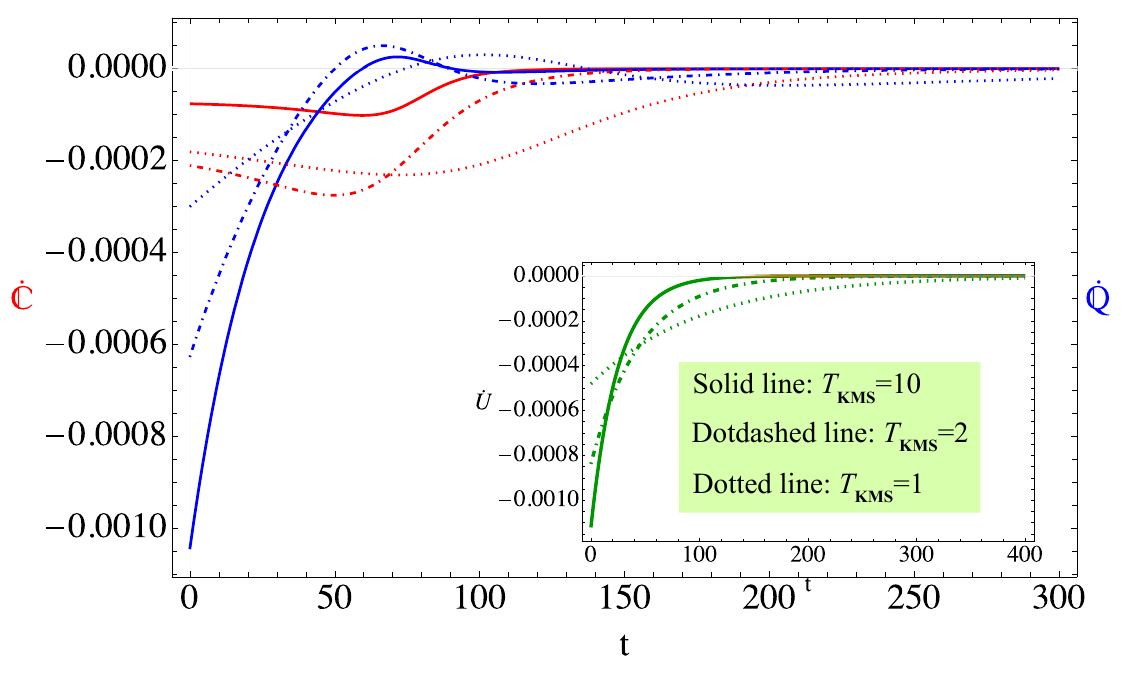}}
\caption{The change rate of quantum coherence and quantum heat of a co-rotating detector as a function of effective Hawking temperature $T_{\text{KMS}}$ and time. The detector is prepared in an initial pure state with $l_0=1$ and $\theta_0=\pi/3$. The BTZ black hole is assumed to be spinless and have a fixed outer horizon $r_+=0.2$. The red curves denote the change rate of quantum coherence $\dot{\mathds{C}}$ and the blue curves denote the quantum heat change rate $\dot{\mathds{Q}}$. Compared to (a) Neumann ($\zeta=-1$) and (b) transparent ($\zeta=0$) b.c., the thermalization process is significantly stretched to a very long timescale for (c) Dirichlet b.c. ($\zeta=1$).}
\label{fig7}
\end{figure}

We now study the influence of Hawking radiation on the rate of change of quantum heat and coherence of the detector, specifically focusing on a non-rotating BTZ black hole (set $r_-=0$) for simplicity. For selection of detector and black hole parameters, we depict $\dot{\mathbb{Q}}$ and $\dot{\mathbb{C}}$ as time-depedent functions of effective temperature $T_{\text{KMS}}$ in Fig.\ref{fig7}, and the detector's internal energy in the insets.

We observe that the monotonicity over time for the detector's internal energy variation rate $\dot{U}$, which is consistent for any boundary conditions on the background field, as shown in the insets of Fig.\ref{fig7}. Compared to the Neumann or transparent cases, the Dirichlet boundary condition again exhibits distinguishability (see Fig.\ref{fig7}(c)), as the timescale of the detector thermalization process (i.e., $\dot{U}$ approaching zero) has been significantly stretched in this case.

Refining our study with the First Law formula, we find that the relative evolution between $\dot{\mathbb{Q}}$ and $\dot{\mathbb{C}}$ can be significantly affected by Hawking radiation. For example, in Fig.\ref{fig7}(a),(b), for the detector exposed to Hawking radiation at higher $T_{\text{KMS}}$, the complementarity between the evolution of $\dot{\mathbb{Q}}$ and $\dot{\mathbb{C}}$ becomes more suppressed, as shown by the decreasing separation between their extrema. This leads to an interesting conclusion: with properly chosen parameters, the higher-temperature bath does not necessarily intensify the detector's thermalization process and may, in fact, do the opposite. This counterintuitive phenomenon, rarely observed with a classical thermal bath in flat spacetime \cite{sec1-7-1}, is closely linked to the complicated dependence of the Kossakowski coefficients on the effective temperature in BTZ black hole geometry.

%

\subsubsection{A remark on quantum work}

For the standard UDW detector \cite{sec1-6,sec1-7}, we show that its quantum work remains zero during the thermalization process (see \eqref{eq2.43}), as it is assumed to have fixed energy spacing. To enable the detector to perform nontrivial quantum work, the simplest way is to allow the detector Hamiltonian to be driven $H_{\text{detector}}(t)$, for example, in the context of quantum Floquet engineering \cite{Floquet1}. 
 
In this subsection, we analyze a type of time-dependent detector Hamiltonians, which, under certain adiabatic conditions, can naturally leverage the previous analyses for time-independent UDW detectors. We consider the detector with a time-dependent energy level $\widetilde{\omega}(t)$, which leads to the Hamiltonian as
\be
\widetilde{H}_{\text{detector}}(t)=\frac{\widetilde{\omega}(t)}{2} \sigma_3 
\label{eqT1}
\ee
To derive its related quantum master equation, we follow the same procedure outlined in Section \ref{2.1}, which accounts for the time dependence of the detector Hamiltonian. In particular, the operator $\mathfrak{m}$ in the interaction picture is now given by
\begin{equation}
\widetilde{\mathfrak{M}}(t) = \mathfrak{U}_S^{\dagger}(t,0) \mathfrak{m}(0) \mathfrak{U}_S(t,0)
\end{equation} \label{eqT2}
where the unitary operator is time-ordered as
\be
\mathfrak{U}_S(t,0) = \mathcal{T} \exp\left(-i\int_0^t \widetilde{H}_{\text{detector}}(t') \, dt'\right)
\ee

Since only the energy levels are time-dependent in \eqref{eqT1}, the eigenbasis of $H_{\text{detector}}(t)$ remains the same as in the time-independent case. Therefore, the unitary operator can be simplified to
\begin{equation}
\mathfrak{U}_S(t,0) = \begin{pmatrix}
\exp\left(-\frac{i}{2}\displaystyle \int_0^t \widetilde{\omega}(t') \, dt'\right) & 0 \\
0 & \exp\left(\frac{i}{2}\displaystyle \int_0^t \widetilde{\omega}(t') \, dt'\right)
\end{pmatrix}
\label{eqT3}
\end{equation}

Specified to the UDW detector with dipole momentum operator $\mathfrak{m}=\sigma_{+}+\sigma_{-}$, in the interaction picture, one has a decomposition like
\begin{equation}
\widetilde{\mathfrak{M}}(t) = e^{i \int_0^t \widetilde{\omega}(t') \, dt'} \, \sigma_{+} + e^{-i \int_0^t \widetilde{\omega}(t') \, dt'} \, \sigma_{-}
\label{eqT4}
\end{equation}
which, when plugged into the Redfield equation \eqref{eq2.6}\footnote{To obtain the Redfield equation for the time-dependent detector, we still need to employ the Markov approximation \cite{MasterEq}, which assumes that the Wightman function $\mathcal{W}(s)$ decays rapidly with a characteristic time scale $t_B$, much shorter than the system's relaxation time scale $t_S := 1/\widetilde{\omega}(t)$ characterized by a varying energy spacing $\widetilde{\omega}(t)$.}, induces the Kossakowski coefficients like in \eqref{eq2.9}. However, in the present scenario, the two-sided Fourier transform of the Wightman function should now be generalized to a more complicated form
\begin{equation}
\begin{aligned}
\widetilde{\mathcal{C}}\left(\widetilde{\omega}(t)\right)&=\int_{0}^{\infty} d s~ \exp\left(i\int_0^{t}\widetilde{\omega}(t')dt'\right) \exp\left(-i\int_0^{t-s}\widetilde{\omega}(t')dt'\right) \left\langle  \Phi[x(t)] \Phi[x(t-s)]\right\rangle + h.c.\\
&=\int_{0}^{\infty} d s ~\exp\left(i\int_{t-s}^{t}\widetilde{\omega}(t')dt'\right)~\mathcal{W}(s) + \int_{-\infty}^{0} d s ~\exp\left(i\int_{t}^{t+s}\widetilde{\omega}(t')dt'\right)~\mathcal{W}(s)
\label{eqT5}
\end{aligned}
\end{equation}
which resists direct computation.

To overcome this difficulty, we assume that the energy level of the detector varies adiabatically, meaning that $\widetilde{\omega}(t)$ changes sufficiently slowly and smoothly, $\lvert\dot{\widetilde{\omega}}(t)\rvert ~t_S \ll \widetilde{\omega}(t)$, such that $\lvert\widetilde{\omega}^{(n)}\rvert t_S^n \ll \lvert\widetilde{\omega}^{(n-1)}\rvert ~t_S^{n-1}\ll\cdots \ll\lvert\dot{\widetilde{\omega}}(t)\rvert~ t_S \ll \widetilde{\omega}(t)$. Together with $t_B \ll t_S$, this yields the useful hierarchy
\be
\lvert\dot{\widetilde{\omega}}(t)\rvert ~ t_B \ll \lvert\dot{\widetilde{\omega}}(t)\rvert ~t_S \ll \widetilde{\omega}(t)
\label{eqT6}
\ee

Under these assumptions, we can expand the exponentials in the integral \eqref{eqT5} as:
\begin{equation}
\begin{aligned}
\int_{0}^{\infty} ds~ \exp\left(i \int_{t-s}^{t} \widetilde{\omega}(t') dt'\right) \, \mathcal{W}(s) &= \int_{0}^{\infty} ds~ \exp\left(i \widetilde{\omega}(t) s + \frac{i}{2} \dot{\widetilde{\omega}}(t) s^2 + \mathcal{O}(\ddot{\widetilde{\omega}} s^3)\right) \, \mathcal{W}(s)\\
& \approx \int_{0}^{\infty} ds~ e^{i \widetilde{\omega}(t) s} \, \mathcal{W}(s)
\label{eqT7}
\end{aligned}
\end{equation}
and similarly
\be
\int_{-\infty}^{0} d s ~\exp\left(i\int_{t}^{t+s}\widetilde{\omega}(t')dt'\right)~\mathcal{W}(s) \approx \int_{-\infty}^0{ds ~ e^{i\widetilde{\omega}(t) s }~\mathcal{W}(s)}
\label{eqT8}
\ee
Substituting these estimations back into \eqref{eqT5}, we eventually have
\be
\widetilde{\mathcal{C}}\left(\widetilde{\omega}(t)\right)\approx \int_{-\infty}^\infty{ds ~ e^{i\widetilde{\omega}(t) s }~\mathcal{W}(s)}=\mathcal{C}\left(\widetilde{\omega}(t)\right)
\label{eqT9}
\ee
This indicates that for a UDW detector with adiabatically varying energy level $\widetilde{\omega}(t)$, the related transition rate $\widetilde{\mathcal{C}}\left(\widetilde{\omega}(t)\right)$ can be accurately approximated by the time-independent detector transition rate \eqref{eq2.11}, by replacing the constant frequency $\omega$ with a time-dependent one $\omega(t)$. We note that our result is consistent with the adiabatic limit of the quantum master equation, as discussed in \cite{TimeDependentMasterEq}.

To illustrate our claim, we consider a UDW detector with a time-varying frequency
\be
\widetilde{\omega}(t)=\omega_0+\omega_1 \sin (a t),~~~~ (\omega_1 \ll \omega_0)
\label{eqT10}
\ee
The driving parameter $a$ should be chosen at the order $\mathcal{O}(\omega_0)$ to ensure the validity of the adiabatic approximation:
\be
\left|\frac{\dot{\widetilde{\omega}}(t)}{(\widetilde{\omega}(t))^2}\right|=\frac{\omega_1 a|\cos (a t)|}{\left(\omega_0+\omega_1 \sin (a t)\right)^2}\simeq \frac{\omega_1 a}{\omega_0^2}=\left(\frac{\omega_1}{\omega_0}\right) \frac{a}{\omega_0} \ll 1.
\label{eqT11}
\ee

Once the detector co-rotates with a BTZ black hole, we substitute \eqref{eqT10} back into \eqref{eq2.32}, we obtain the transition rate $\mathcal{C}\left(\widetilde{\omega}(t)\right)$ for the time-dependent detector in BTZ spacetime. Following the same procedure in Section \ref{3.3.1}, we immediately derive the change rate of the detector's quantum work as
\be
\dot{\widetilde{\mathds{W}}}(t)= \frac{{l}(t) \cos{\Theta}}{2}\frac{d\widetilde{\omega}(t)}{dt}=\frac{a\omega_1{l}(t) \cos{\Theta}\cos(at)}{2} 
\label{eqT12}
\ee
Because of the size of parameters $\omega_1$ and $a$, this is a small but nonvanishing quantity at the order $\mathcal{O}(\omega_1)$, oscillating over time. Therefore, for a co-rotating UDW detector with a time-dependent Hamiltonian in BTZ spacetime, a trade-off between its quantum heat, quantum work, and coherence could exist during the thermalization process. This may open up the possibility of designing a quantum thermal machine in spinning spacetime in future studies\footnote{The Floquet driving UDW detector may be a more optimal choice for high-performance quantum thermal machines. However, the related quantum master equation is more complicated, leading to significantly increased solution difficulty, which remains a future task. }.

\subsection{The Second Law}

For the open quantum system, its quantum entropy $S(t)$ may not always increase positively due to the entropy flux $\Phi(t)$ from the system to the bath. Once introducing the quantum entropy production rate $\Pi(t)$ of the system, which quantifies the irreversibility of the thermodynamic process and satisfies the exact fluctuation theorem \cite{QT22}, the entropy change can be split into two contributions:
\be
\dot{S}(t)=\Pi(t)-\Phi(t)
\label{eqT13}
\ee
Then, the quantum Second Law claims that the entropy production should be nonnegative: $\Pi\geqslant 0$, with the equality holding if and only if the system is at equilibrium.

\subsubsection{Information geometry}
\label{3.4.1}

We use the language of information geometry to analyze the quantum Second Law in detail. In this framework \cite{QT23}, the system states are considered as points within a geometric space, mapped to statistical distance through a quantum Fisher information (QFI) with respect to time\footnote{The widely known QFI in quantum metrology is a measure to discriminate quantum states $\rho(X)$ and $\rho(X+\delta X)$ with an infinitesimal change in the parameter $X$. Many studies utilize QFI to probe sensitive parameters in quantum gravity, such as Hawking-Unruh temperature \cite{sec1-19,sec1-34,Unruh3} or spacetime structure \cite{ds1,ds2,ds3}. Alternatively, we may also consider time itself as a parameter, thus endow the QFI with the ability to probe the details of quantum evolution, especially be insightful for quantum thermodynamic processes.}, which is particularly significant in thermodynamics since it can be identified as the metric of the quantum space \cite{QT24} and sets the bounds on quantum speed limits of the evolution \cite{QT25,QT26}.

We begin with the master equation \eqref{eq-IG4} for the instantaneous spectrum of the detector state. By defining the probability current $j_k^{x y}(t)$ between instantaneous eigenstates $|y(t)\rangle$ and $|x(t)\rangle$ as:
\be
j_k^{x y}(t):=w_k^{x y}(t) p_y(t)-w_{k^{\prime}}^{y x}(t) p_x(t),
\label{eqT14}
\ee
the master equation \eqref{eq-IG4} can be rewritten into $\dot{p}_x(t)=\sum_{k, y(y \neq x)} j_k^{x y}(t)$. This just enables us to calculate directly the time-derivative of quantum entropy $S(t)=-\sum_xp_x(t)\log~p_x(t)$ as
\be
\dot{S}(t)= \frac{1}{2} \sum_{x, y, k} j_k^{x y}(t) \log \left[\frac{w_k^{x y}(t) p_y(t)}{w_{k^{\prime}}^{y x}(t) p_x(t)}\right]  -\frac{1}{2} \sum_{x, y, k} j_k^{x y}(t) \log \left[\frac{w_k^{x y}(t)}{w_{k^{\prime}}^{y x}(t)}\right] .
\label{eqT15}
\ee
By comparing the above result with \eqref{eqT13}, we can identify the entropy production rate and entropy flux to the environment as \cite{QT27}
\be
\begin{aligned}
\Pi(t) &=\frac{1}{2} \sum_{x, y, k} j_k^{x y}(t) \log \left[\frac{w_k^{x y}(t) p_y(t)}{w_{k^{\prime}}^{y x}(t) p_x(t)}\right]:=\langle\langle f(t)\rangle\rangle \\
\Phi(t) &=\frac{1}{2} \sum_{x, y, k} j_k^{x y}(t) \log \left[\frac{w_k^{x y}(t)}{w_{k^{\prime}}^{y x}(t)}\right]:=\langle\langle\phi(t)\rangle\rangle
\end{aligned}
\label{eqT16}
\ee
where $\langle\langle \star \rangle\rangle:=\frac{1}{2} \sum_{x y} \left(j^{x y} ~\star\right)$ means a current-weighted average of quantity. The split \eqref{eqT16} shows that the entropy production rate $\Pi(t)$ is a current-weighted average of the so-called thermodynamic force
\be
f_k^{x y}(t)=\log \left[\frac{w_k^{x y}(t) p_y(t)}{w_{k^{\prime}}^{y x}(t) p_x(t)}\right],
\label{eqT17}
\ee
while the entropy flux to the environment $\Phi(t)$ is the current-weighted average of the entropy flow
\be
\phi_k^{x y}(t)=\log \left[\frac{w_k^{x y}(t)}{w_{k^{\prime}}^{y x}(t)}\right],
\label{eqT18}
\ee
associated with the jump ${L}_k$ at time $t$ resulting in a probability current $j_k^{x y}(t)$ from state $|y(t)\rangle$ to $|x(t)\rangle$. 

By applying the inequality $(a-b) \log \frac{a}{b} \geq 0$ to the first line of \eqref{eqT16}, it automatically proves the non-negativity of the entropy production rate, i.e., $\Pi(t)\geq 0$, which corresponds to the Second Law of quantum thermodynamics. On the other hand, the entropy flux determines the heat exchange rate by $\dot{Q} =-\beta^{-1}\Phi(t)$, from the perspective of the entropy change split.

We now turn to the information geometry approach, which provides a more refined description of the time-derivative of quantum entropy. The key concept is the QFI with respect to the time parameter \cite{QFI1}
\be
\mathcal{F}_Q(t)=\sum_{x, y} \frac{\left|\partial_t {\rho}_{x y}(t)\right|^2}{p_x(t) f\left(p_y(t) / p_x(t)\right)},
\label{eqT19}
\ee
where $\partial_t {\rho}_{x y}:=\langle x(t)| \partial_t {\rho}|y(t)\rangle$ and the function $f(x)$ admits several choices, e.g., for the symmetric logarithmic derivative (SLD) QFI one has $f_{\text{SLD}}(x)=(x+1)/2$. 

In quantum state space, the QFI \eqref{eqT19} provides a metric contractive under quantum stochastic maps \cite{QT28}, by which a squared line element of the form $ds^2=\frac{1}{4}\mathcal{F}_Qdt^2$ can be defined. Along the system evolution path $\Gamma$ in quantum state space, the path length is
\be
\mathcal{L}:=\int_\Gamma ds=\frac{1}{2}\int_0^T dt\sqrt{\mathcal{F}_Q(t)},
\label{eqT20}
\ee
representing a statistical distance that faithfully measures the distinguishability between the initial state $\rho(0)$ and the final state $\rho(t)$. In an optimal case, there is a geodesic path with constant speed, that
connects the initial state and the final state, so that $\mathcal{L}_{\text{geo}}\leqslant \mathcal{L}$. 

It is important to note that any QFI with respect to the parameter time $t$ \eqref{eqT19} can be split into an incoherent contribution and a coherent contribution $\mathcal{F}^{\text{C}}_Q(t)$, so that
\be
\mathcal{F}_Q(\rho(t))=\mathcal{F}^{\text{IC}}_Q(\rho(t))+\mathcal{F}^{\text{C}}_Q(\rho(t)),
\label{eqT22}
\ee
where
\be
\begin{aligned}
\mathcal{F}^{\text{IC}}_Q(\rho(t))&:=\sum_x p_x(t)\left(\frac{d}{dt}\log p_x(t)\right)^2\\
\mathcal{F}^{\text{C}}_Q(\rho(t))&:=\sum_{x\neq y} \frac{\left|\partial_t {\rho}_{x y}(t)\right|^2}{p_x(t) f\left(p_y(t) / p_x(t)\right)}
\end{aligned}
\label{eqT23}
\ee
denoting the contribution from the changes in the spectrum of the state via $\mathcal{F}^{\text{IC}}_Q(t)$, and the nondiagonal contribution from the time-evolving eigenbasis of the quantum systems via $\mathcal{F}^{\text{C}}_Q(t)$.

In a quantum thermodynamics regime, the QFI is closely connected to the second derivative of quantum entropy, which is entropy acceleration, given by \cite{QT11}
\be
\ddot{S}(t)=\mathcal{B}-\mathcal{F}^{\text{IC}}_Q,
\label{eqT24}
\ee 
where $\mathcal{B}:=-\sum_x \ddot{p}_x(t) \log p_x(t)$ measures the difference between the average local information rate and the bulk information rate. Combining this with definitions \eqref{eqT16},  one has the alternative formula for the incoherent contribution to the QFI as
\be
\mathcal{F}^{\text{IC}}_Q(t)=-\left\langle\left\langle\frac{\mathrm{d} f(t)}{\mathrm{d} t}\right\rangle\right\rangle+\left\langle\left\langle\frac{\mathrm{d} \phi}{\mathrm{~d} t}\right\rangle\right\rangle .
\label{eqT25}
\ee 

On the other hand, the coherent part of QFI plays a significant role in the geometric bound on the von Neumann entropy rate. Specifically, the time-averaged variance of the QFI, known as the geometric uncertainty, satisfies
\be
\delta:=\mathbb{E}\left[\mathcal{F}_Q\right]-\mathbb{E}[\sqrt{\mathcal{F}_Q}]^2 \geqslant 0.
\label{eqT26}
\ee 
Here, $\mathbb{E}[\star]$ represents a time-averaged operation. For example, a time-averaged rate of information change is defined as $\mathcal{I}:=\mathbb{E}\left[\mathcal{F}_Q\right]=\frac{1}{T} \int_0^T dt \mathcal{F}_Q$. Due to the Cauchy-Schwarz inequality, a quantum thermodynamic uncertainty relation can be expressed as \cite{QT23}
\be
\frac{\mathcal{I}}{\delta} \geqslant 1,
\label{eqT27}
\ee 
indicating that lower uncertainty in the path comes at the expense of a lower time-averaged rate of information change. A direct consequence of this relation is that the difference in the entropy rate between the final and initial states over a duration $T$ is bounded from above
\be
\Delta \dot{S}=\dot{S}(T)-\dot{S}(0)\leqslant \int_0^T \mathcal{B}(t) \mathrm{d} t-T \delta+\int_0^T dt \mathcal{F}_Q^{\text{C}}.
\label{eqT28}
\ee
Unlike the classical case, an additional nonnegative quantum contribution appears in the quantum bound \cite{QT11}, which is related to the coherent dynamics of the system.

\subsubsection{Entropy production rate in BTZ spacetime}

We now apply the results from Section \ref{3.4.1} to the co-rotating UDW detector outside a BTZ black hole and examine its quantum Second Law by evaluating the QFI with respect to time in the context of quantum geometry.

Based on the spectral decomposition of the detector density matrix $\rho(t)=p_{+}|+\rangle\langle+|+p_{-}|-\rangle\langle-|$, with eigenvalues and eigenstates determined from \eqref{eq2.14} and \eqref{eq-IG5}, the first time derivative of von Neumann entropy:
\be
\dot{S}(\rho(t)) = - \sum_{x=\pm} \dot{p}_x \log{p_x} 
= \left[ \gamma_{-}\cos{\Theta}+l(t)\gamma_{+}+l(t)\gamma_{0}\sin^2{\Theta} \right]\log{\left[\frac{1+l(t)}{1-l(t)}\right]} ,
\label{eqT29}
\ee
In the context of \eqref{eqT13}, the entropy rate can be divided into the contributions of the entropy production rate $\Pi$ and the entropy flux $\Phi$ from the scalar field background, which can be calculated directly from \eqref{eqT16} as:
\be
\begin{aligned} 
\Pi (t)& = \langle\langle f(t)\rangle\rangle = \dot{S}(t) + \Phi(t),\\
\Phi(t)& = \langle\langle \phi(t)\rangle\rangle= -\frac{\omega}{2T_{\text{KMS}}} \frac{dn_3}{dt}  = \frac{\omega\left[ \gamma_{-}+l(t)\gamma_{+}\cos{\Theta} \right] }{T_{\text{KMS}}}.
\end{aligned}
\label{eqT30}
\ee
Recall that the above entropy flux determines the heat exchange rate of the UDW detector to the environment as
\be
\dot{Q}:=-\Phi(t)/\beta=-\frac{\omega}{2} \dot{n_3}.
\label{eqT31}
\ee
Comparing it with \eqref{eq2.43}, one immediately observes that the quantum First Law is properly linked to the heat exchange rate through $\dot{Q}=\dot{\mathbb{Q}} + \dot{\mathbb{C}}$. Note that both quantum heat change $\dot{\mathbb{Q}}$ and (de)coherence change $\dot{\mathbb{C}}$ are sorts of dissipative energy changes for detector thermalization, thus could be identified in the quantum Second Law as the heat exchange rate $\dot{Q}$ to the environment.

\begin{figure}[htbp]
\centering
\subfloat[Neumann: $\zeta=-1$]{\includegraphics[width=.33\textwidth]{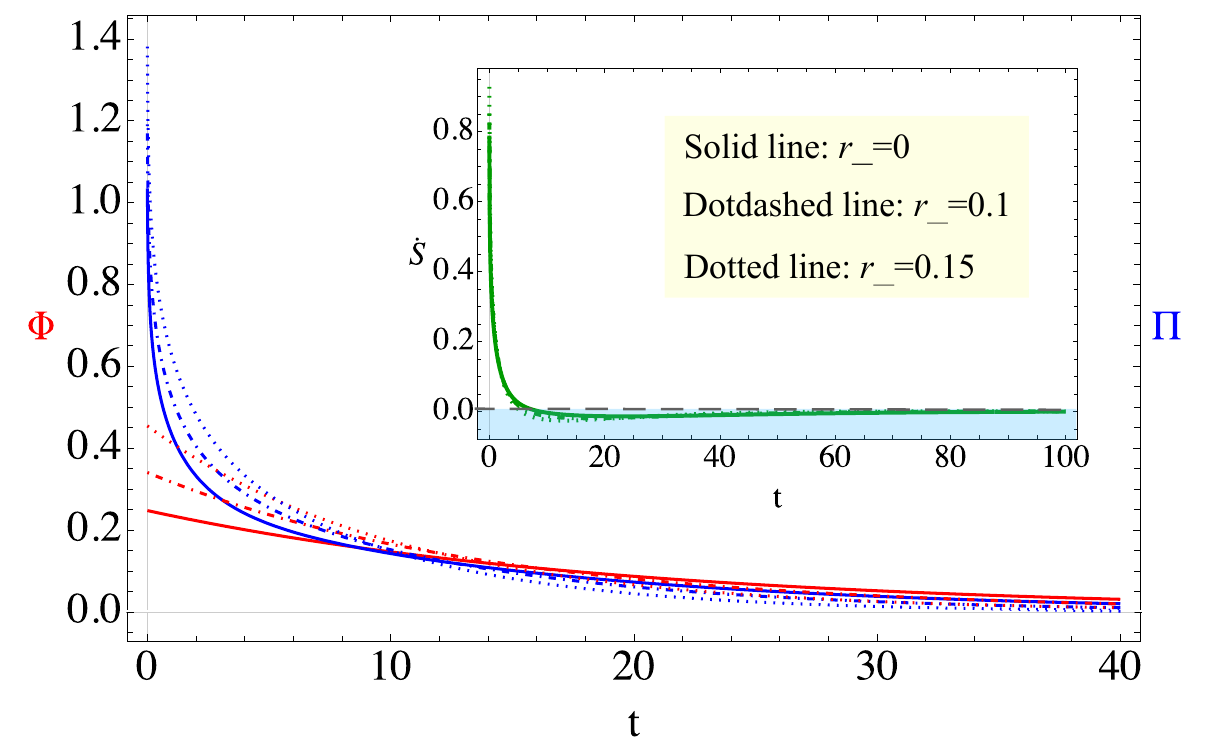}}
\subfloat[Transparent: $\zeta=0$]{\includegraphics[width=.33\textwidth]{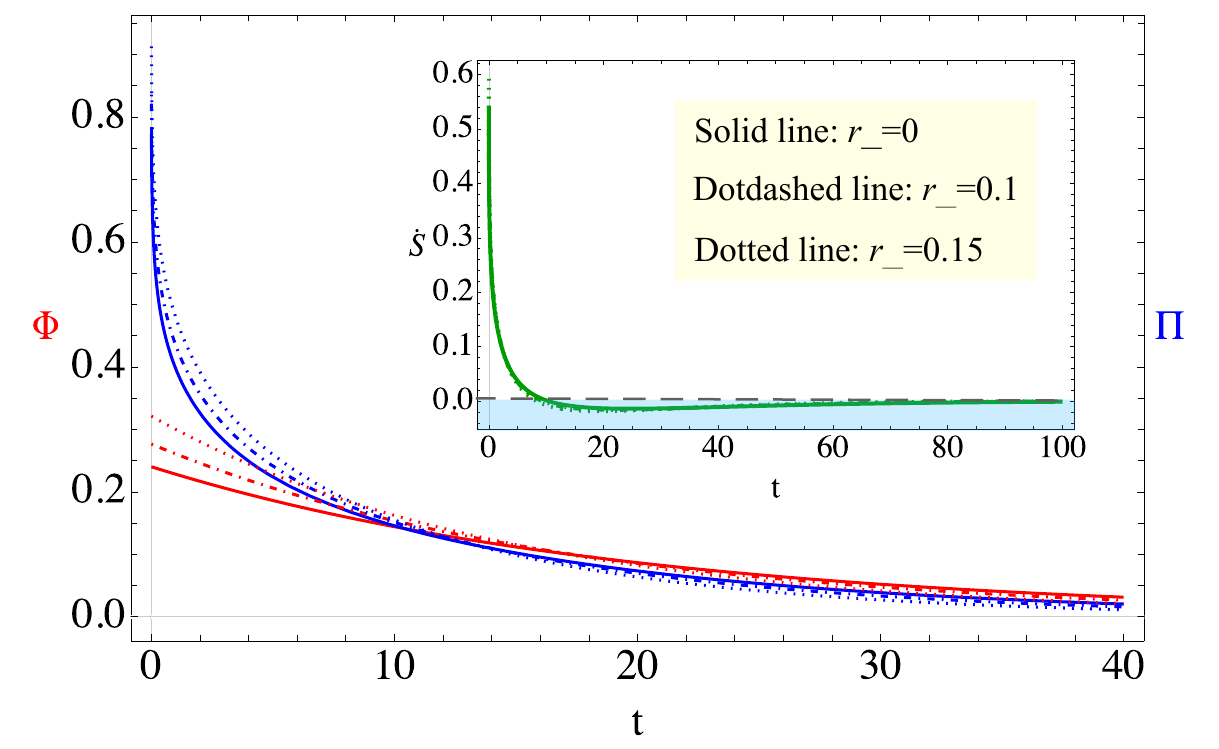}}
\subfloat[Dirichlet: $\zeta=1$]{\includegraphics[width=.33\textwidth]{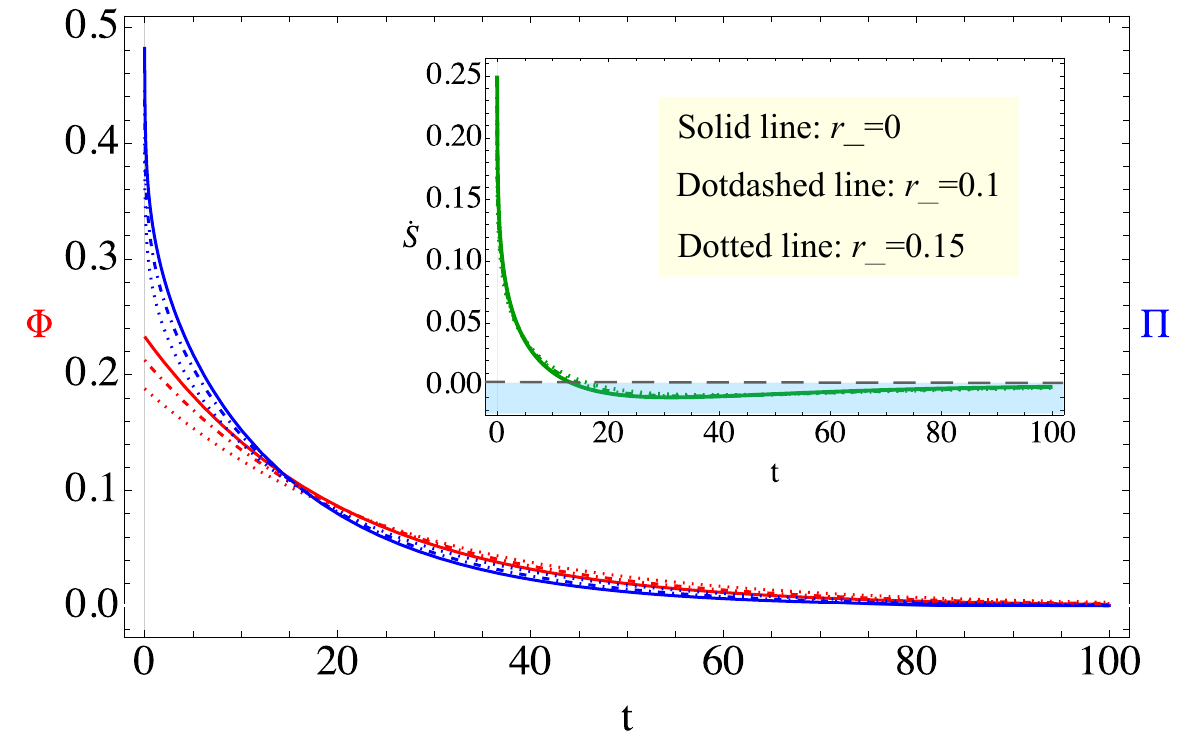}}
\caption{The quantum Second Law of a co-rotating detector outside a BTZ black hole with different angular momentum. The detector is prepared in an initial pure state with $l_0=1$ and $\theta_0=\pi/3$. The BTZ black hole is assumed to have a fixed outer horizon $r_+=0.2$, while the effective Hawking temperature perceived by the detector is set by $\beta\omega=2\pi$. The red curves denote the entropy flux $\Phi$ of the detector to the environment, and the blue curves denote the entropy production rate $\Pi$. The estimations are taken for (a) Neumann b.c. ($\zeta=-1$), (b) transparent b.c. ($\zeta=0$), and (c) Dirichlet b.c. ($\zeta=1$), respectively. The derivative of the quantum entropy $\dot{S}$ for three boundary conditions is illustrated in the insets, with the blue shadow regions denoting the negative values.}
\label{fig8}
\end{figure}

In Fig.\ref{fig8}, we show the time derivative of quantum entropy $\dot{S}$, the entropy flux $\Phi(t)$, and the entropy production rate $\Pi(t)$, respectively, for a detector outside a rotating BTZ black hole. First, the insets reveal that the quantum entropy change rate $\dot{S}$ varies non-monotonically as described by \eqref{eqT29} and can become negative (indicated by the blue-shaded areas in the insets) during certain periods. This highlights the non-equilibrium nature of the detector's evolution, which leads to the failure of the standard thermodynamic Second Law, i.e., the non-decreasing entropy over time. Although the influence of the black hole rotation on the evolution of $\dot{S}$ appears subtle in the insets, it is more evident when examining the split \eqref{eqT30}, by noting the change in the monotonically evolving behaviors of $\Phi(t)$ and $\Pi(t)$ due to black hole angular momentum (represented by different values of $r_-$). Finally, similar as before, different boundary conditions for the scalar background affect the thermalization time: the Dirichlet case in Fig.\ref{fig8}(c) requires a longer time for thermalization than the Neumann (Fig.\ref{fig8}(a)) and transparent (Fig.\ref{fig8}(b)) cases. 

We now turn to an information geometry description of the quantum Second Law, which requires an explicit form of the QFI with respect to time. Substituting \eqref{eqT30} into \eqref{eqT25}, we obtain the incoherent contribution of QFI for the UDW detector, which is
\be
\mathcal{F}_Q^{\text{IC}}(t) = -\left\langle\left\langle\frac{ {d} f(t)}{ {d} t}\right\rangle\right\rangle+\left\langle\left\langle\frac{ {d} \phi}{ {d} t}\right\rangle\right\rangle = \frac{4\left[ \gamma_{-}\cos{\Theta}+l(t)\gamma_{+}+l(t)\gamma_{0}\sin^2{\Theta} \right]^2}{1-l^2(t)}.
\label{eqT32}
\ee
The coherent part of the QFI with respect to time can be directly calculated from the definition \eqref{eqT23}, which gives
\be
\mathcal{F}_Q^{\text{C}}(t) = \omega^2 l^2(t) \sin^2{\Theta}=\omega^2\left(l^2(t)-n_3^2(t)\right),
\label{eqT33}
\ee
as a part of the quantum bound on the time-integrated entropy rate.

\begin{figure}[htbp]
\centering
\subfloat[Neumann: $\zeta=-1$]{\includegraphics[width=.33\textwidth]{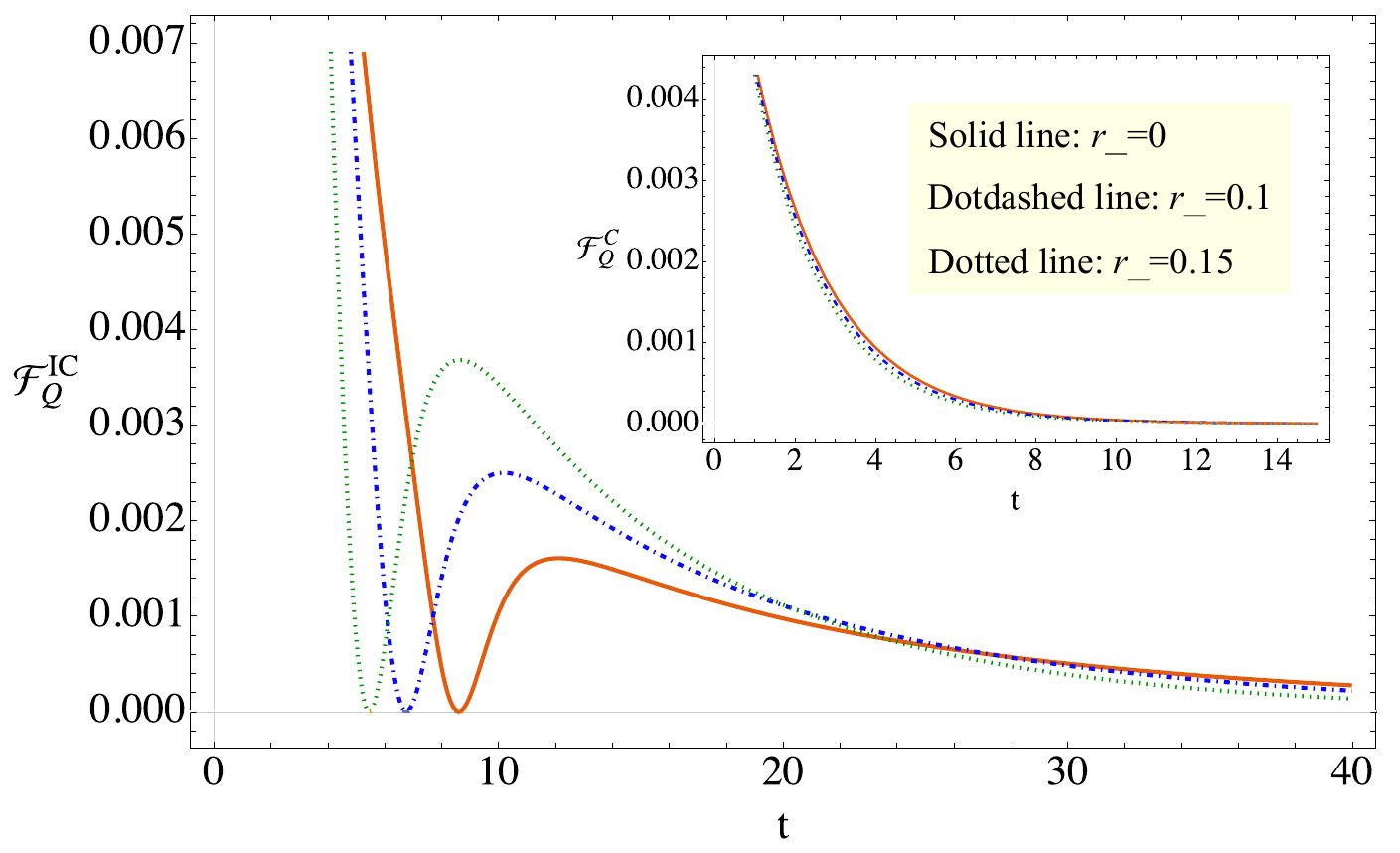}}
\subfloat[Transparent: $\zeta=0$]{\includegraphics[width=.33\textwidth]{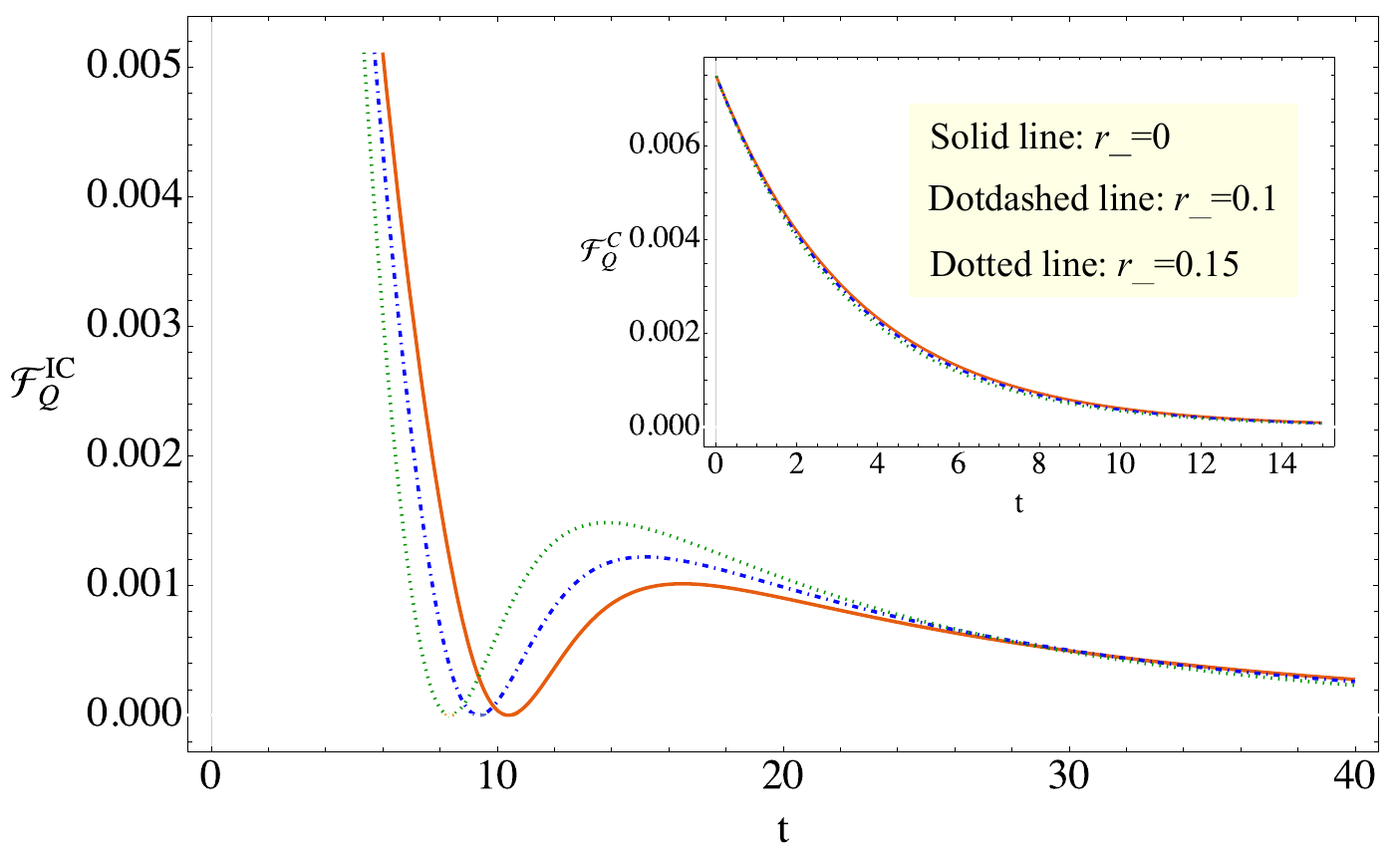}}
\subfloat[Dirichlet: $\zeta=1$]{\includegraphics[width=.33\textwidth]{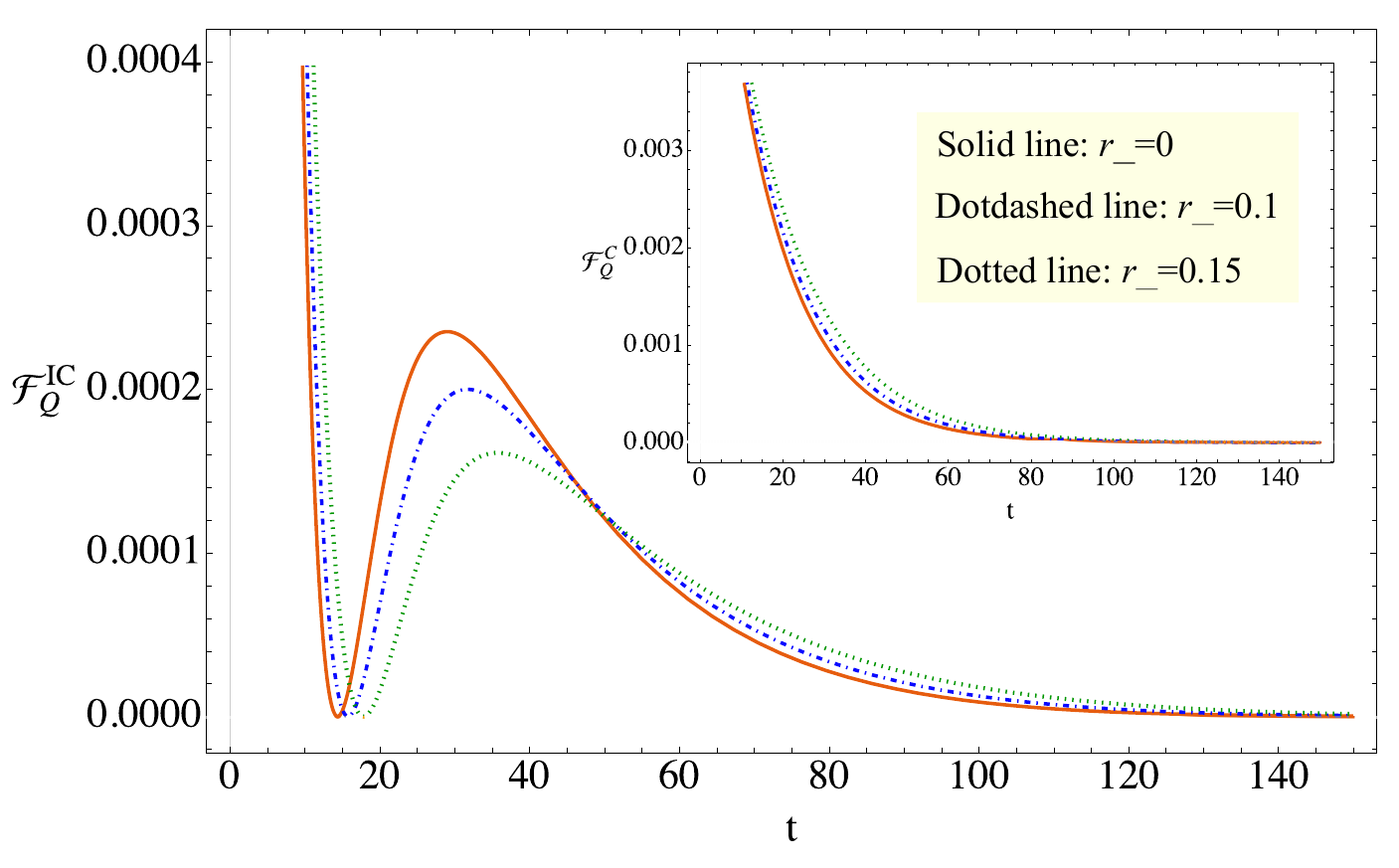}}
\caption{The QFI with respect to time as a function of BTZ black hole angular momentum, with background scalar chosen in (a) Neumann, (b) transparent, and (c) Dirichlet boundaries. The detector starts from a pure state with $l_0=1$ and $\theta_0=\pi/3$. The black hole is assumed to have $r_+=0.2$, and the effective Hawking temperature is set by $\beta\omega=2\pi$. The estimation is performed for an increasing inner horizon (red solid curves for $r_-=0$, blue dot-dashed curves for $r_-=0.1$, and green dotted curves for $r_-=0.15$), shows that the evolution of $\mathcal{F}_Q^{\text{IC}}$ is significantly impacted by the black hole rotation. In the insets, the monotonously varying coherent part $\mathcal{F}_Q^{\text{C}}$ is depicted.}
\label{fig9}
\end{figure}

We illustrate the incoherent and coherent parts of the QFI with respect to time as a function of black hole angular momentum in Fig.\ref{fig9}. We observe that the black hole's rotation significantly influences the non-monotonic evolution of $\mathcal{F}_Q^{\text{IC}}(t)$, which is more prominent than the subtle trace from the entropy flux or the entropy production rate shown in Fig.\ref{fig8}. We note that, as time passes, both $\mathcal{F}_Q^{\text{IC}}(t)$ and $\mathcal{F}_Q^{\text{C}}(t)$ approach zero, indicating a vanishing $\mathcal{F}_Q$ asymptotically. Recall that the QFI characterizes the evolution speed of the detector in quantum state space, $\mathcal{F}_Q\rar 0$ signifies an equilibrium of the open dynamics (i.e., a unique thermalization end) for the UDW detector. With the background field chosen with a Dirichlet boundary condition, we observe that the thermalization process of the detector is stretched to a timescale much larger than that of the Neumann or transparent boundary choices.

\begin{figure}[htbp]
\centering
\subfloat[Neumann: $\zeta=-1$]{\includegraphics[width=.33\textwidth]{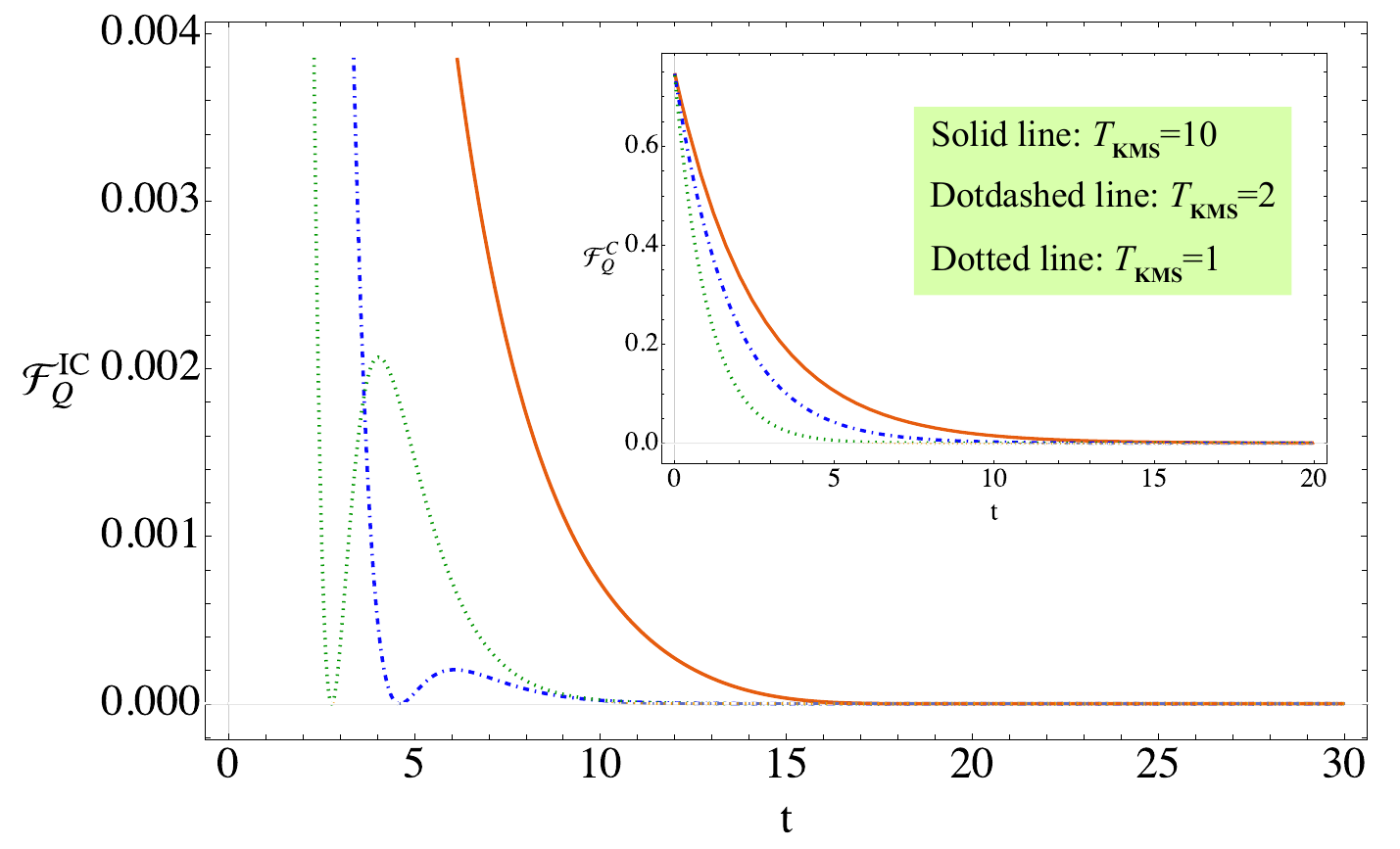}}
\subfloat[Transparent: $\zeta=0$]{\includegraphics[width=.33\textwidth]{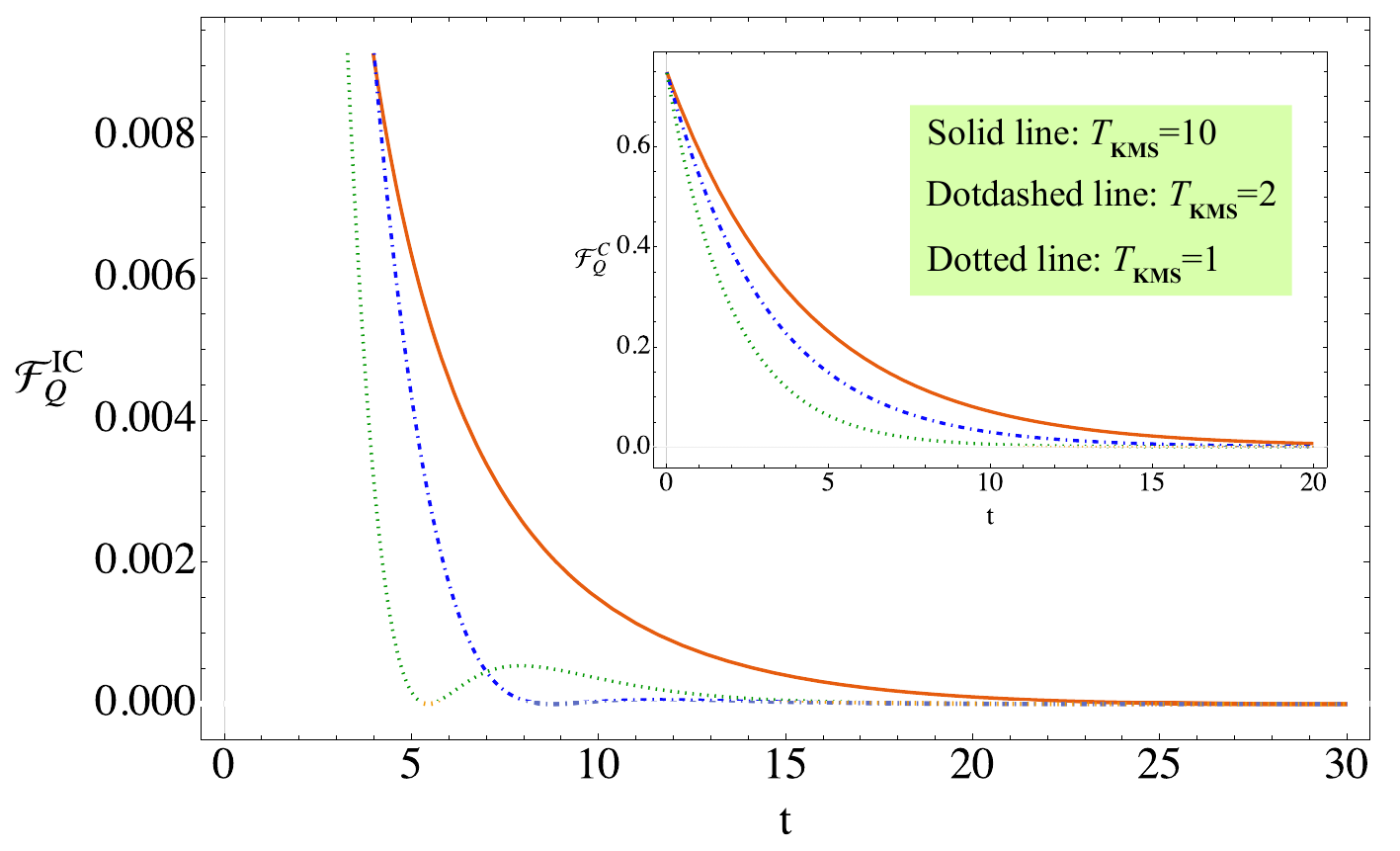}}
\subfloat[Dirichlet: $\zeta=1$]{\includegraphics[width=.33\textwidth]{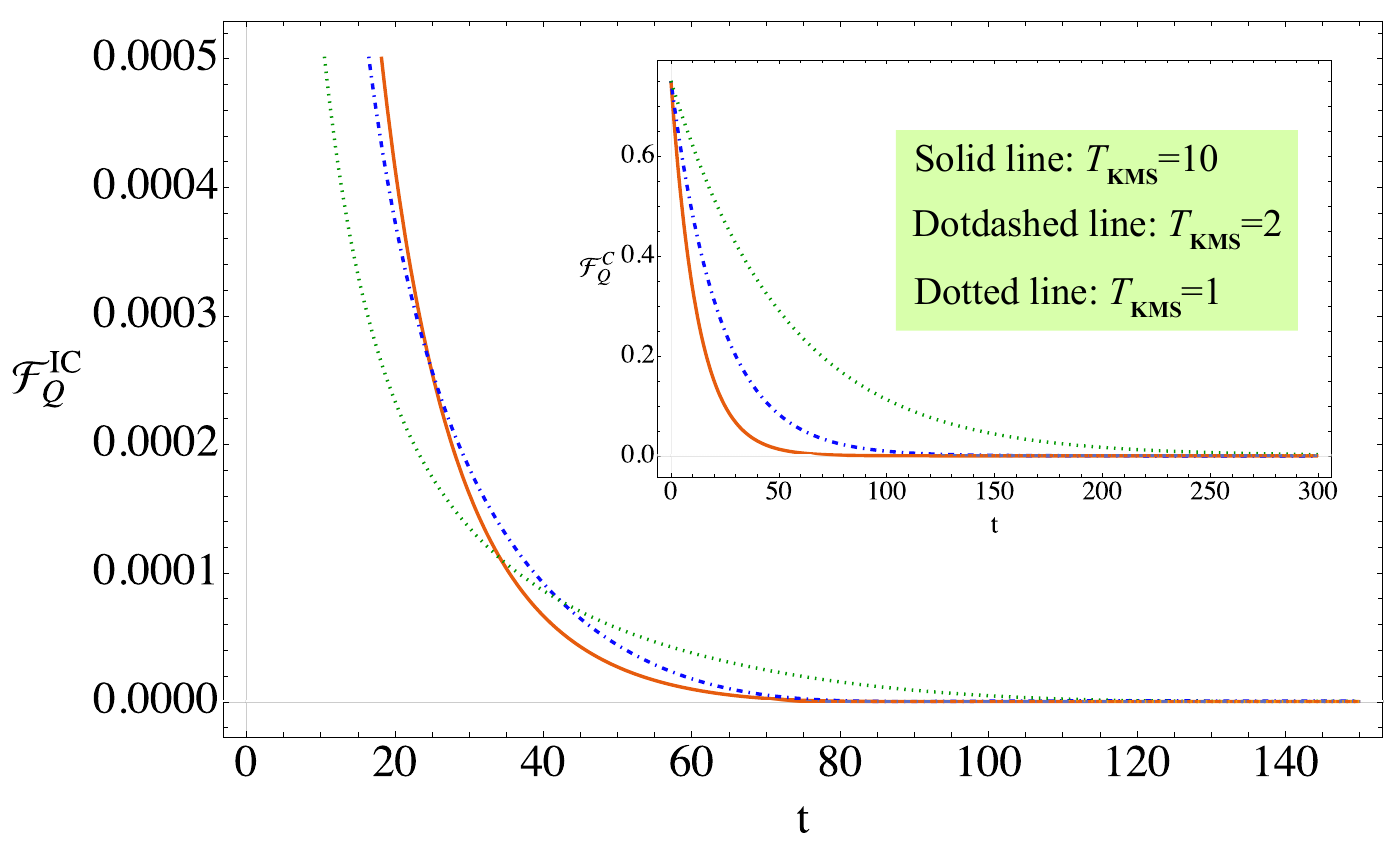}}
\caption{The QFI with respect to time as a function of the effective Hawking temperature $T_{\text{KMS}}$. The detector starts from a pure state with $l_0=1$ and $\theta=\pi/4$. The estimation is performed for the fixed BTZ geometry ($r_+=0.2$ and $r_-=0$) and a degrading effective temperature (red solid curves for $T_{\text{KMS}}=10$, blue dot-dashed curves for $T_{\text{KMS}}=2$, and green dotted curves for $T_{\text{KMS}}=1$). The scalar field is specified with (a) Neumann, (b) transparent, and (c) Dirichlet boundary conditions, respectively. In the insets, the monotonously varying coherent contribution $\mathcal{F}_Q^{\text{C}}$ is depicted.}
\label{fig10}
\end{figure}

In Fig.\ref{fig10}, we present the time evolution of the QFI with respect to time for the detector detecting the effective Hawking radiation at different temperatures. We notice an apparent change in the monotonicity of the incoherent part of the QFI, i.e., with higher $T_{\text{KMS}}$, the non-monotonicity of $\mathcal{F}_Q^{\text{IC}}$ is greatly suppressed. On the other hand, the coherent part of the QFI remains monotonic under Hawking radiation at any effective temperature. This is expected since $\mathcal{F}_Q^{\text{C}}$, given by \eqref{eqT33}, is measured by the difference between the length of the detector's Bloch vector $l(t)$ and its third component $n_3(t)$. It exhibits monotonic geometric evolution, as seen in Fig.\ref{fig3}, where the trajectory traced out by the Bloch vector is a spiral with a steadily decreasing radius. The different Hawking temperatures mainly influence the rate at which the radius contracts. Lastly, the decoherence time needed for the detector to thermalize is much longer with the scalar background under Dirichlet boundary condition (Fig.\ref{fig10}(c)), compared to Neumann (Fig.\ref{fig10}(a)) and transparent (Fig.\ref{fig10}(b)) scenarios. As previously mentioned, such incontrovertible discrimination in choosing Dirichlet boundary conditions over Neumann or transparent cases results from the special behavior of the transition rate \eqref{eq2.32}, which varies the slowest in the Dirichlet case \cite{sec1-33}.
  .


\section{Quantum thermal kinematics in BTZ spacetime}

\label{4}

So far, we have examined three quantum thermodynamic laws for the irreversible open process of the UDW detector in the BTZ spacetime. We observe that the black hole's angular momentum and Hawking radiation can significantly influence the time evolution of various quantum thermodynamic quantities, such as quantum heat, coherence, and entropy production rate. However, the transition rate \eqref{eq2.32} in BTZ exhibits a non-monotonic dependence on the effective temperature, making the pattern of influences for arbitrary values of $T_{\text{KMS}}$ complicated. This is in contrast to the cases of Hawking-Unruh bath in Rindler, Schwarzschild or de Sitter spacetime, where the transition rate function (as well as the Kossakowski coefficients) admits a simple form like $\mathcal{C}(\omega)\sim\left(e^{-\beta\omega}-1\right)^{-1}$. On the other hand, for the scalar background imposed a Dirichlet boundary, the detector's thermalization process is generally stretched to a very long timescale, compared to Neumann or transparent boundary choices. This phenomenon is rooted in the lowest varying behavior of the transition rate with $\zeta=1$, which remains intact regardless of Hawking radiation or the spinning degree.

\begin{figure}[htbp]
\centering
\subfloat[]{\includegraphics[width=.48\textwidth]{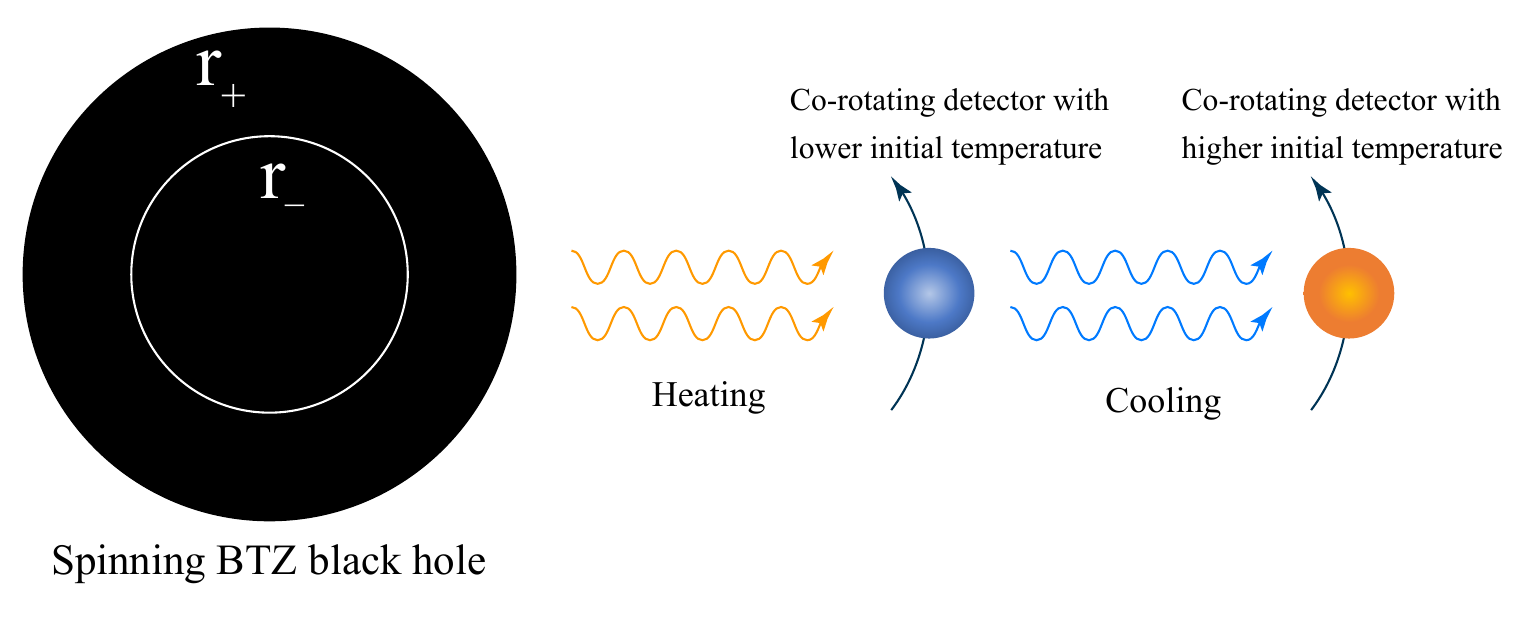}}~~~~
\subfloat[]{\includegraphics[width=.48\textwidth]{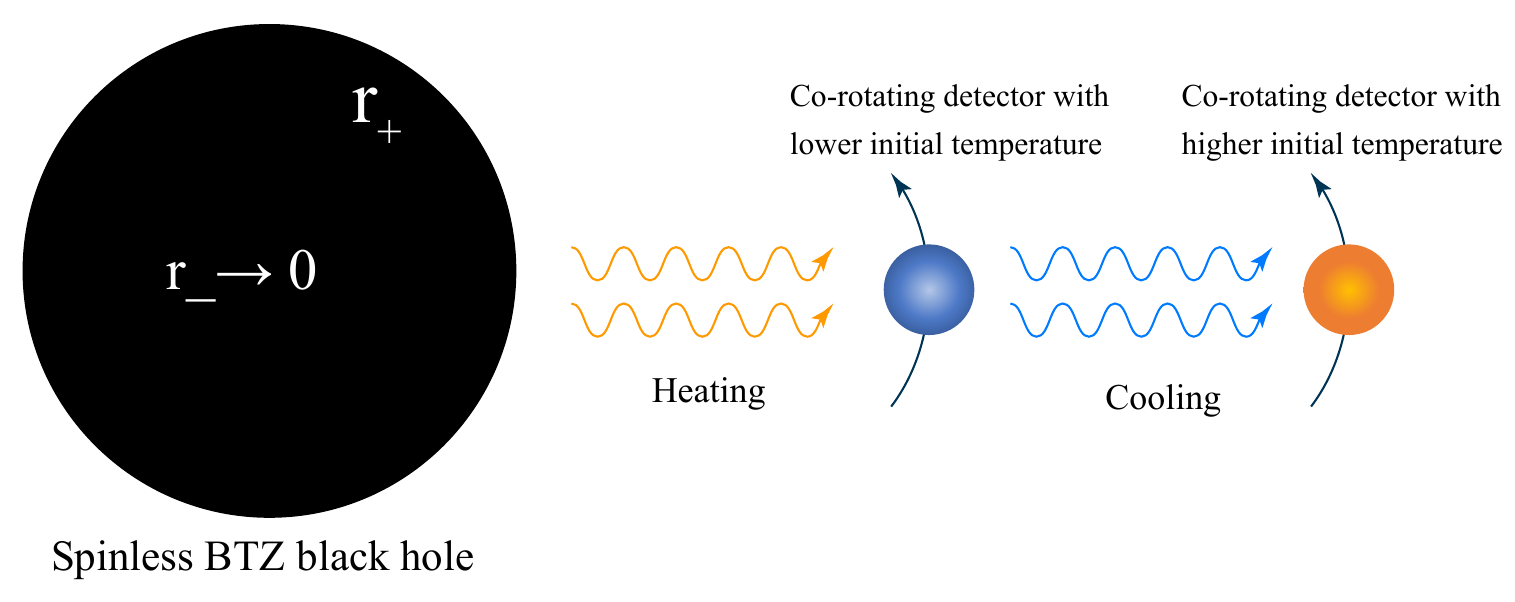}}
\caption{Asymmetry of detector thermalization in BTZ spacetime. (a) Outside a spinning BTZ black hole, two co-rotating detectors are located at different radial distances from the outer horizon. The detector (blue) closer to the horizon starts from a Gibbs state with a lower temperature $T_C$ than the local KMS temperature $T_{\text{KMS}}=T_H$ (orange arrowed wave lines); therefore, it undergoes a heating thermalization process. On the other hand, the detector (orange) starts from a Gibbs state with a higher temperature $T_H$ than the local KMS temperature $T_{\text{KMS}}=T_C$ (blue arrowed wave lines); therefore, it undergoes a cooling thermalization process. The same experiment is set for a spinless BTZ black hole in (b).}
\label{fig11}
\end{figure}

In this section, we explore the thermal kinematics \cite{QT13} of the UDW detector open process outside the BTZ black hole. We develop a specific protocol (see Fig.\ref{fig11}) for two UDW detectors to reach a fixed equilibrium determined by effective Hawking radiation at different locations. We place one detector closer to the horizon, starting from a Gibbs state with a lower temperature $T_C$ than the local Hawking temperature $T_{\text{KMS}}=T_H$, thereby undergoing a \emph{heating} thermalization process. Another detector, however, is placed at a position perceiving the Hawking radiation with effective temperature $T_{\text{KMS}}=T_C$. Starting from a Gibbs initial state with higher temperature $T_H$, the second detector undergoes a \emph{cooling} thermalization process. Using information geometry theory \cite{QT12}, we demonstrate an intriguing phenomenon that the heating and cooling between the same pair of temperatures, $T_C$ and $T_H$, are fundamentally asymmetric and evolve along distinct pathways. That is, the thermalization process toward equilibrium follows fundamentally different paths depending on whether the detector's temperature is rising or falling. The influence of black hole angular momentum on this asymmetry of the quantum (thermal) open process will also be examined.

To study thermal kinematics in the quantum regime, we need to develop several measures from information geometry to quantify the temporal variation of local flows in quantum state space. The first one is the Uhlmann fidelity between two states
\be
F\left(\rho_1, \rho_2\right):=\left( \operatorname{Tr}\left[\sqrt{\sqrt{\rho_1}\rho_2\sqrt{\rho_1}}\right] \right)^2,
\label{eqT34}
\ee
which measures how close two quantum states are in terms of their density matrices. The fidelity is symmetric and invariant under unitary operations and is bounded $0\leqslant F\leqslant 1$. However, for an open quantum system, the fidelity $F\left(\rho_{i}, \rho_{f}\right)$ cannot be symmetric between the initial and final states anymore, since the dissipative term in QMME \eqref{eq2.7-1} breaks the unitarity of the system evolution. Along the detector evolution path in quantum state space, we note that from the length \eqref{eqT20}, the instantaneous quantum "speed" of evolution can be identified as 
\be
v_{Q}:=ds/dt=\frac{1}{2}\sqrt{\mathcal{F}_Q(t)},
\label{eqT35}
\ee 
which is determined by the QFI with respect to time. For a non-equilibrium quantum process, comparing the velocity $v_{Q}$ of the process and its reverse, the appearing asymmetry then characterizes thermodynamic irreversibility \cite{QT13}.

We now apply the above measures to the open dynamics of a co-rotating detector in BTZ spacetime. We know that after a sufficiently long time, the detector will eventually be thermalized to an equilibrium \eqref{eq2.38}, which is a Gibbs final state
\be
\rho_f=\frac{1}{2}\left(1-\tanh\left(\frac{\omega}{2T}\right) \sigma_3\right),
\label{eqT36}
\ee
at effective temperature $T=\omega/\ln\left(\frac{1+\gamma}{1-\gamma}\right)$.

For an arbitrary detector state $\rho(t)=\frac{1}{2}(1+\boldsymbol{n}(t)\cdot \boldsymbol{\sigma})$, its fidelity to $\rho_f$ then is (see Appendix \ref{AppendixC}):
\be
F(\rho(t),\rho_f) =\frac{1}{2} \left\{ 1- n_3(t) \tanh\left(\frac{\omega}{2T}\right)+\sqrt{\left[1-l^2(t)\right]\sech^2\left(\frac{\omega}{2T}\right)} \right\}
\label{eqT37}
\ee
where $l(t)=\sqrt{\sum_{i=1}^3 n_i^2}$ is the length of the Bloch vector, given by \eqref{eq2.14} for BTZ spacetime. 

Specific to the protocol illustrated in Fig.\ref{fig11}, we choose the initial state for the closer detector (blue) as a Gibbs state with lower temperature as $T_C$. Then its state can be written as
\be
\rho_{i,\text{cold}}=\frac{1}{2}\left(1-\tanh\left(\frac{\omega}{2T_C}\right) \sigma_3\right).
\label{eqT38}
\ee 
After a sufficiently long time, the detector heats up to a thermalization end with effective temperature $T_H>T_C$. The fidelity \eqref{eqT37} during this process becomes
\be
F_{\text{heating}} =\frac{1}{2} \left\{ 1- n_{3,\text{heating}}(t) \tanh\left(\frac{\omega}{2T_H}\right)+\sqrt{\left[1-l_{\text{heating}}^2(t)\right]\sech^2\left(\frac{\omega}{2T_H}\right)} \right\}
\label{eqT39}
\ee
where $n_{3,\text{heating}}(t)$ and $l_{\text{heating}}$ are the solutions \eqref{eq2.13} and \eqref{eq2.14}, with the initial conditions determined by the initial cold state \eqref{eqT38} as $l_0=\tanh\left({\omega}/{2T_C}\right)$, $\theta_0=\pi$.

For the detector (orange) farther away in Fig.\ref{fig11}, we set its initial Gibbs state with a hotter effective temperature as
\be
\rho_{i,\text{hot}}=\frac{1}{2}\left(1-\tanh\left(\frac{\omega}{2T_H}\right) \sigma_3\right).
\label{eqT40}
\ee 
Further assuming the effective temperature at the detector's local position is $T_C$, then the thermalization process of the detector is identified as a cooling process. The fidelity \eqref{eqT37} during the process is 
\be
F_{\text{cooling}} =\frac{1}{2} \left\{ 1- n_{3,\text{cooling}}(t) \tanh\left(\frac{\omega}{2T_C}\right)+\sqrt{\left[1-l_{\text{cooling}}^2(t)\right]\sech^2\left(\frac{\omega}{2T_C}\right)} \right\}
\label{eqT41}
\ee
where $n_{3,\text{cooling}}(t)$ and $l_{\text{cooling}}$ are given by \eqref{eq2.13} and \eqref{eq2.14} with the initial conditions determined by the hot state \eqref{eqT40} as $l_0=\tanh\left({\omega}/{2T_H}\right)$, $\theta_0=\pi$. 

\begin{figure}[htbp]
\centering
\subfloat[Neumann: $\zeta=-1$]{\includegraphics[width=.33\textwidth]{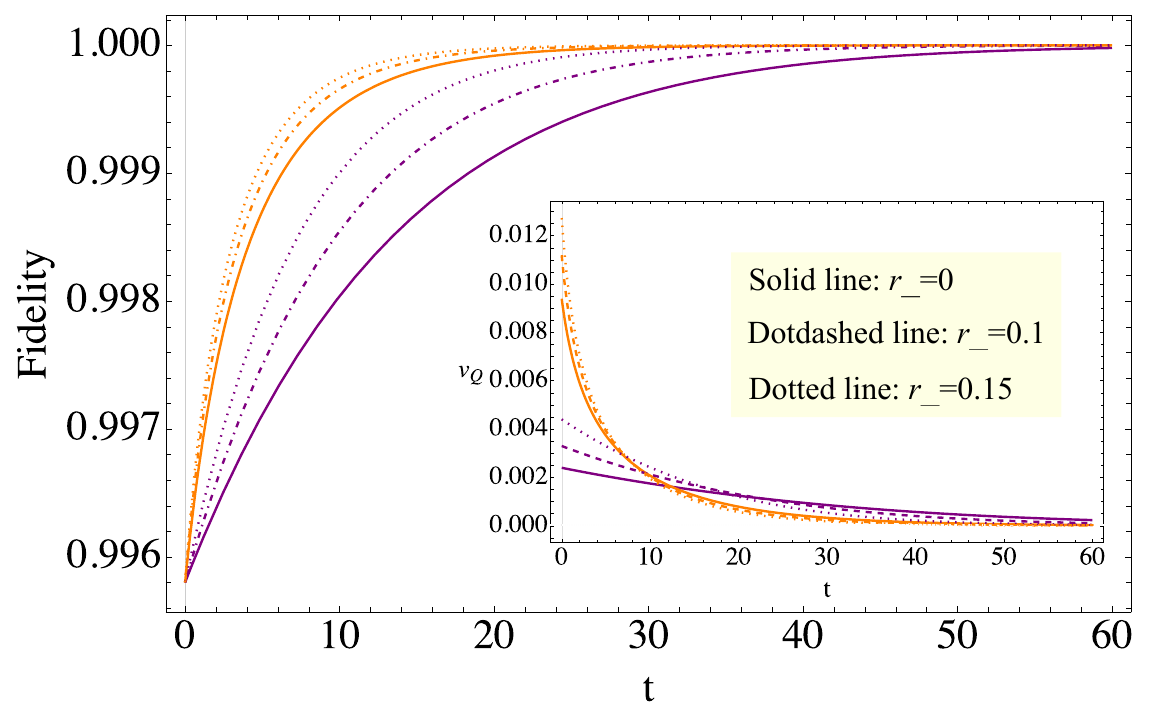}}
\subfloat[Transparent: $\zeta=0$]{\includegraphics[width=.33\textwidth]{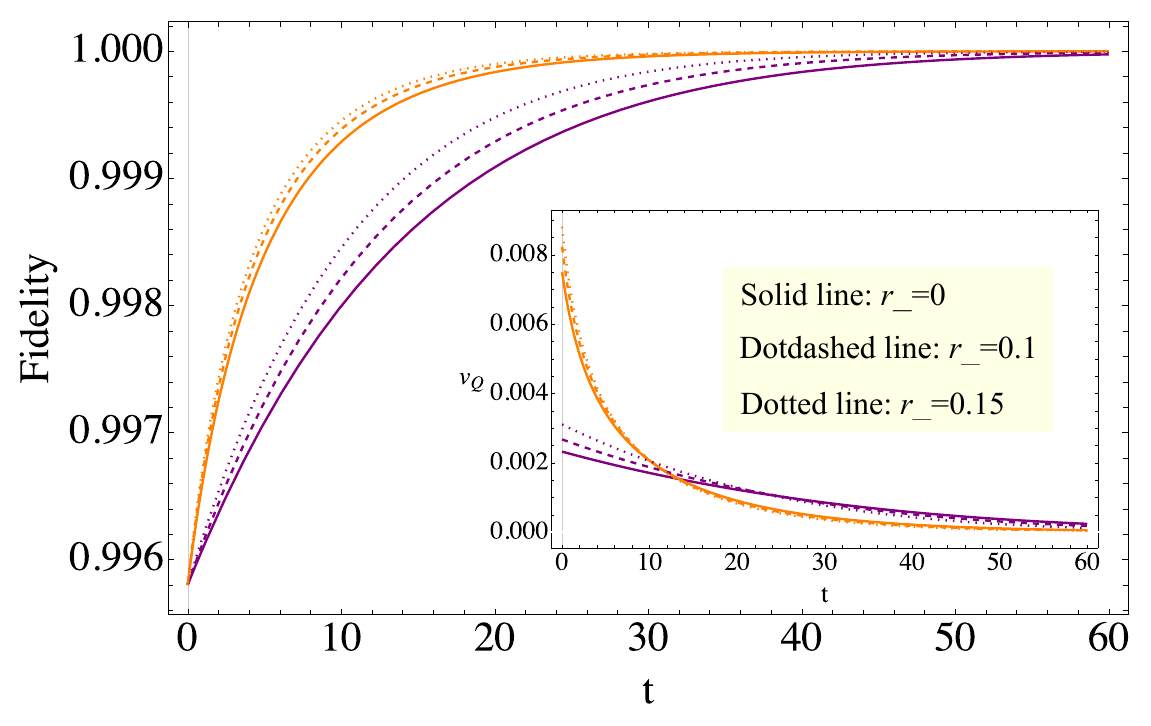}}
\subfloat[Dirichlet: $\zeta=1$]{\includegraphics[width=.33\textwidth]{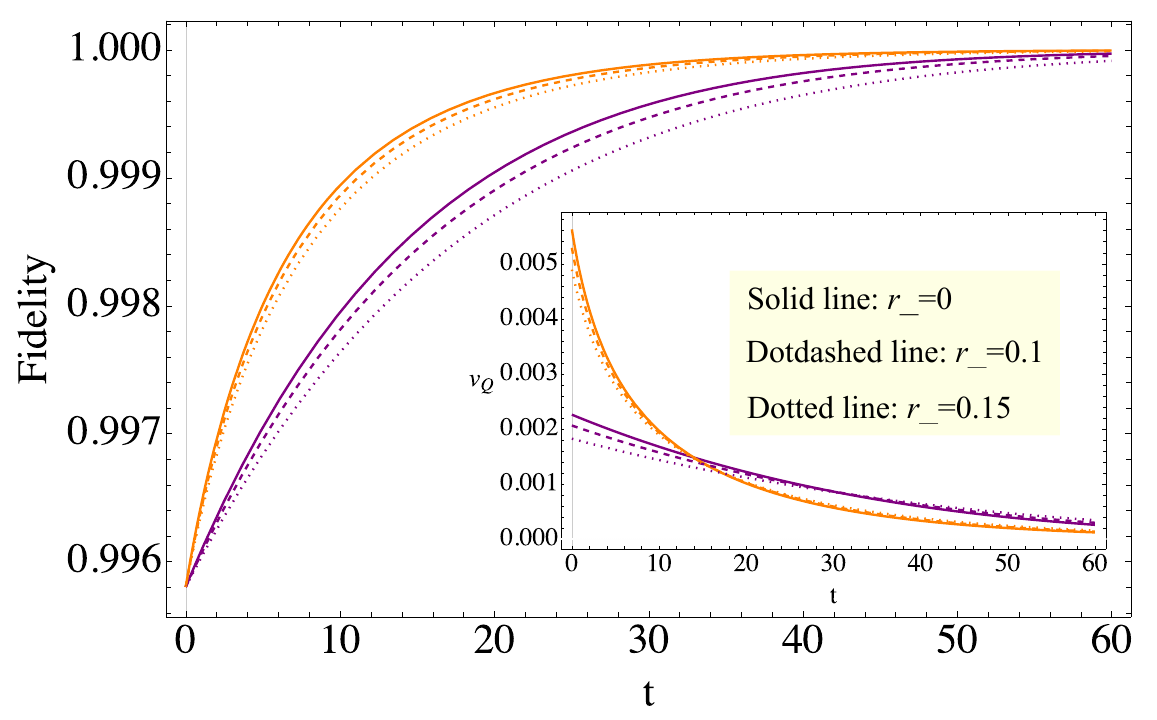}}
\caption{The influence of BTZ black hole rotation on the asymmetry of the detector's thermalization process. The fidelity and speed of the processes are estimated for a detector closer to the horizon, which heats up to a higher temperature $T_H=1/\sqrt{2}\pi$ (orange curves), and for a detector farther from the black hole, cooling down to a lower temperature $T_C=1/2\pi$ (purple curves). The estimation is performed for a fixed BTZ geometry ($r_+=0.2$) and an increasing inner horizon (solid curves for $r_-=0$, blue dot-dashed curves for $r_-=0.1$, and green dotted curves for $r_-=0.15$), shows that the evolution of ${F}(t)$ is significantly affected by black hole rotation. In the insets, the velocity $v_Q$ of detector thermalization for heating/cooling processes is depicted.}
\label{fig12}
\end{figure}

We illustrate in Fig.\ref{fig12}(a)(b)(c) the fidelity \eqref{eqT39} and \eqref{eqT41} during the heating/cooling processes of the co-rotating detector in the BTZ spacetime with different angular momentum, and various scalar background boundary choices. It is evident that, although the processes start and end with fixed initial and final states (with fixed $T_C$ and $T_H$), the heating and cooling protocols of the UDW detector are fundamentally asymmetric and follow distinct pathways, as shown by the corresponding time-evolving fidelity curves. In Fig.\ref{fig12}, we see that detector heating (orange curves) is always faster than its cooling (purple curves), indicated by $F_{\text{heating}}
\geqslant F_{\text{cooling}}$ for $0<t<\infty$. This phenomenon is analogous to the recently proposed quantum Mpemba effect \cite{QT14} for nonequilibrium systems. We also find that when the black hole has nonzero angular momentum, the evolution of the fidelities during heating/cooling protocols, yet the detector heating remains consistently faster than its cooling.

\begin{figure}[htbp]
\centering
\subfloat[]{\includegraphics[width=.42\textwidth]{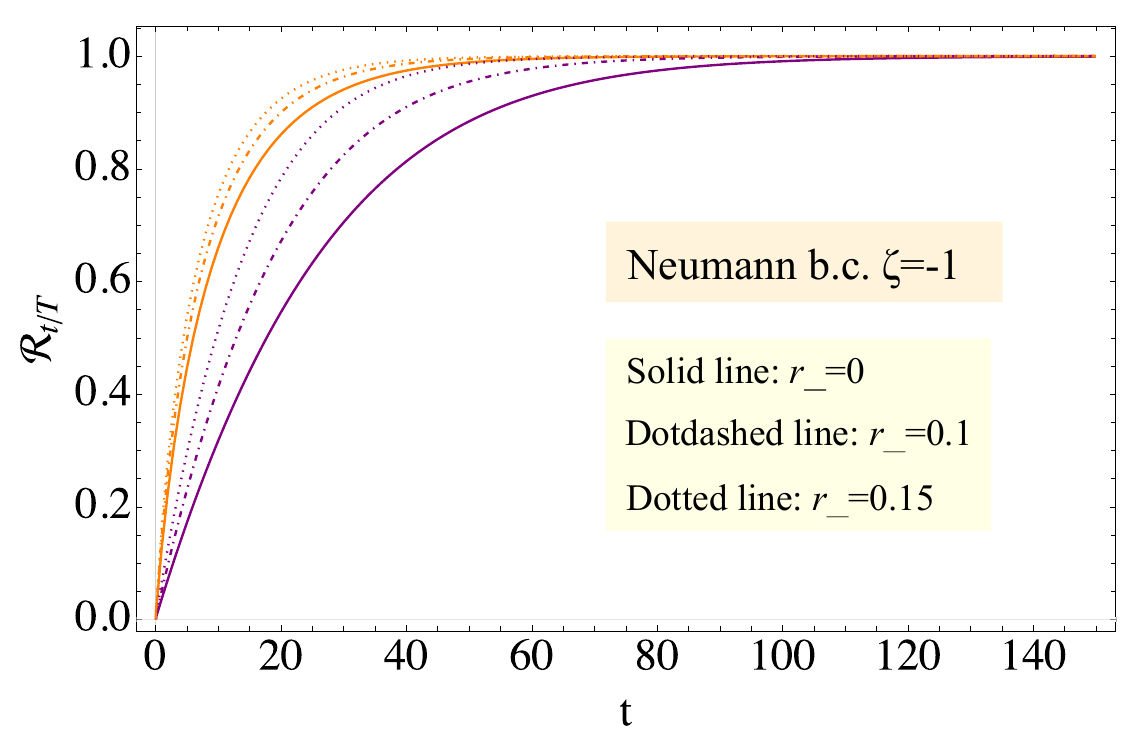}}~~~~~~~
\subfloat[]{\includegraphics[width=.42\textwidth]{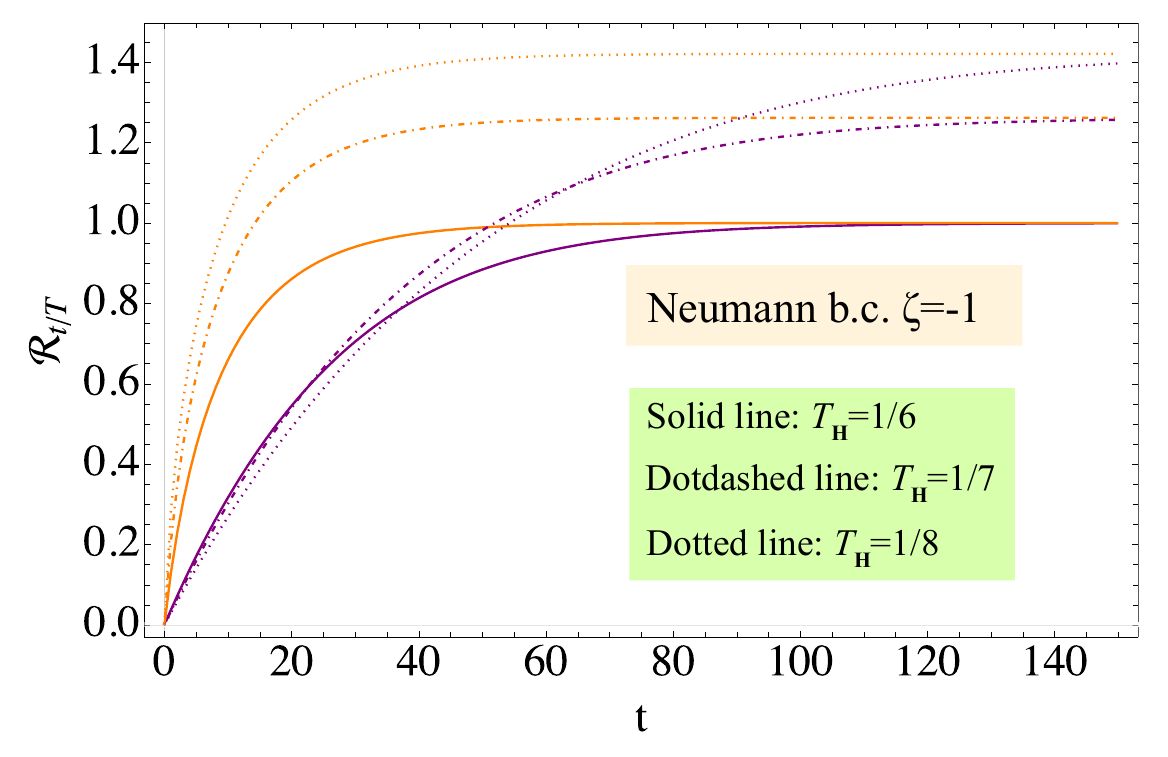}}
\caption{The quantum degree of completion $\mathcal{R}_{t/T}$ for the UDW detector thermalization. In (a), the estimation is performed for a detector outside a BTZ black hole with $r_+=0.2$ and angular momentum represented by $r_-=0, 0.1,0.15$. The heating/cooling protocol is conducted between a temperature pair $(T_H=1/\sqrt{2}\pi, T_C=1/{2}\pi)$, whose completion $\mathcal{R}_{t/T}$ reveals an intrinsic asymmetry between corresponding thermalization paths. In (b), the estimation is performed for BTZ geometry with $r_+=0.2$ and $r_-=0$, and temperature pairs $(T_H=1/\sqrt{2}\pi, T_C=1/6)$, $(T_H=1/\sqrt{2}\pi, T_C=1/7)$, and $(T_H=1/\sqrt{2}\pi, T_C=1/8)$, showing that the increasing temperature difference $\Delta T=T_H-T_C$ results in greater asymmetry in the quantum degree of completion $\mathcal{R}_{t/T}$ between heating/cooling protocols.}
\label{fig13}
\end{figure}

Using the results from incoherent \eqref{eqT32} and coherent \eqref{eqT33} contributions of the QFI, we also show in the insets of Fig.\ref{fig12} the quantum speed $v_Q$ given by \eqref{eqT35} for the heating/cooling processes of the UDW detector. We observe that, at early times, the evolution speed of detector heating exceeds that of cooling, and the nonvanishing black hole rotation affects this difference. However, at later times, the heating velocity surpasses cooling, which may raise the concern about if the condition $F_{\text{heating}}\geqslant F_{\text{cooling}}$ is sufficient to conclude that detector heating is faster than cooling. Since reaching a steady state during a dissipative process may take an infinite amount of time, we define the quantum degree of completion as the ratio between the length of the system's evolution path \eqref{eqT20} over different end times:
\be
\mathcal{R}_{t/T}:=\frac{\mathcal{L}\left(0, t\right)}{\mathcal{L}\left(0, T\right)}.
\ee
This is a monotonically increasing function bounded between 0 and 1, which allows us to refine our analysis of the detector's quantum thermal kinematics. We depict in Fig.\ref{fig13}(a) the quantum degree of completion for the detector's heating/cooling processes, with different values of black hole angular momentum. Clearly, the completion $\mathcal{R}_{t/T}$ for the heating process is always larger than that of the cooling protocol, confirming that heating is consistently faster than cooling, as seen by the comparison of the processes' fidelity.

\begin{figure}[htbp]
\centering
\subfloat[Neumann: $\zeta=-1$]{\includegraphics[width=.33\textwidth]{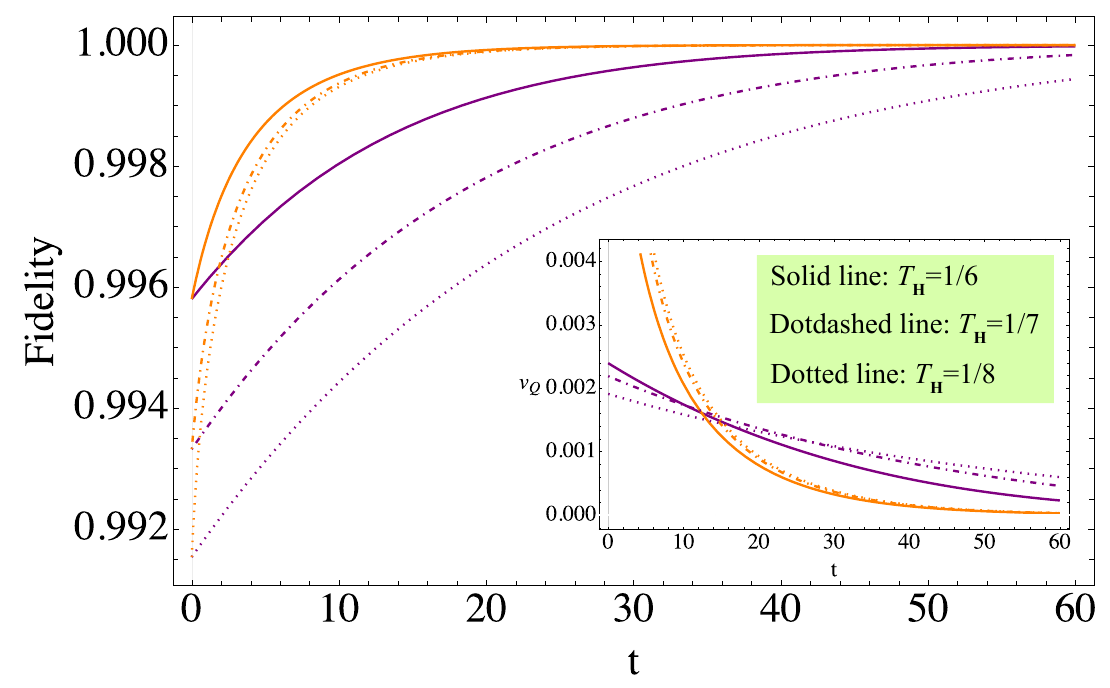}}
\subfloat[Transparent: $\zeta=0$]{\includegraphics[width=.33\textwidth]{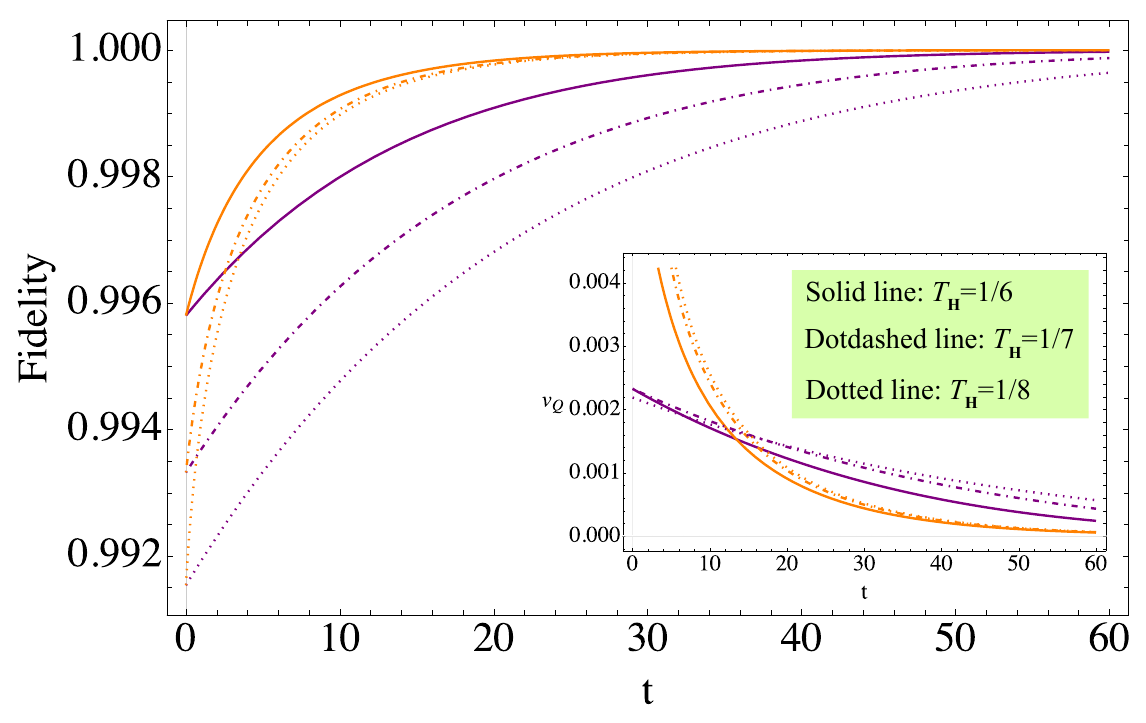}}
\subfloat[Dirichlet: $\zeta=1$]{\includegraphics[width=.33\textwidth]{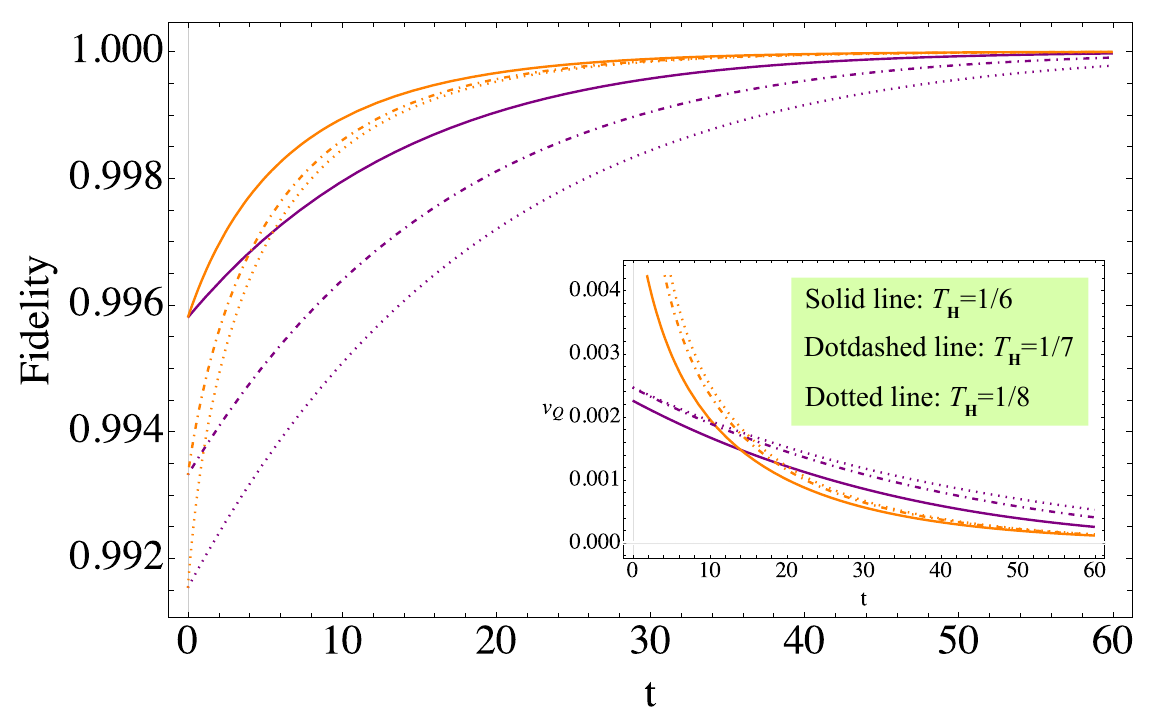}}
\caption{The influence of temperature difference $\Delta T=T_H-T_C$ on the asymmetry of detector thermalization. The BTZ geometry is fixed by $r_+=0.2$ and $r_-=0$. The fidelity and speed of processes are estimated for the detector closer to the horizon, heating from $T_C$ to $T_H$ (orange curves), and for the farther detector cooling from $T_H$ to $T_C$ (purple curves), respectively. The temperature pairs are set as $(T_H=1/\sqrt{2}\pi, T_C=1/6)$, $(T_H=1/\sqrt{2}\pi, T_C=1/7)$, and $(T_H=1/\sqrt{2}\pi, T_C=1/8)$. The increasing temperature difference $\Delta T$ results in greater asymmetry between heating and cooling processes. In the insets, the quantum speed $v_Q$ of detector thermalization s depicted.}
\label{fig14}
\end{figure}

We finally explore how the temperature difference $\Delta T=T_H-T_C$ affects the asymmetry of detector heating/cooling processes. In Fig.\ref{fig14}, which considers explicitly a nonrotating BTZ geometry (i.e., $r_+=0.2$ and $r_-=0$), we plot the fidelity and quantum speed $v_Q$ of the heating/cooling protocols for the UDW detector as the temperature difference $\Delta T$ increases. First, for any $\Delta T$ or scalar boundary choices, we see $F_{\text{heating}}\geqslant F_{\text{cooling}}$ always holds, indicating that the detector heats up faster than it cools down during its thermalization. Second, we find that for larger values of $\Delta T$, the difference between the heating and cooling paths toward the steady state becomes more pronounced. This is reasonable since, with larger $\Delta T$, the initial state of the detector $\rho_i$, appears more distinct from its thermalized end state $\rho_f$.

In the insets of Fig.\ref{fig14}, the quantum speed $v_Q$ is depicted for various $\Delta T$. We find that, at early times, the heating speed of the detector exceeds that of cooling, and is surpassed by the cooling speed at some later time. Nevertheless, as shown in Fig.\ref{fig13}(b), the quantum degree of completion $\mathcal{R}_{t/T}$ of detector heating is always larger than that of the cooling protocol, confirming that heating is consistently faster than cooling for any temperature pairs with difference $\Delta T$.

\section{Summary and discussion}

In this paper, we explore the thermalization process of a UDW detector outside a BTZ black hole from the perspective of quantum thermodynamics. In particular, we derive the complete dynamics of the detector within the framework of an open quantum system, with the detector response to the scalar background in rotating BTZ being analytically calculated. We establish three quantum thermodynamic laws for the co-rotating detector thermalization process. We demonstrate the quantum Zeroth Law as a time evolution of the detector's quantum relative entropy (QRE), which reaches its unique thermalization end once the QRE vanishes. The quantum First Law for the open system decomposes the change rate of the system's internal energy into the time derivatives of quantum work, heat, and coherence, all of which serve as refined process functions to distinguish the thermalization paths in the detector's Hilbert space. The Second Law, expressed in the context of information geometry, links the entropy production rate of a detector to the widely known quantum Fisher information (QFI). Finally, we explore the thermal kinematics of the UDW detector. Using information geometry theory, we demonstrate that the thermalization process of a UDW detector toward equilibrium is fundamentally asymmetric, as it follows different paths depending on whether the detector temperature is rising or falling. In particular, we find that the heating protocol for a UDW detector is always faster than the cooling one. 

Throughout our analysis, we find that the BTZ black hole's angular momentum, the effective Hawking radiation perceived by the detector, as well as the infinity boundary conditions on the scalar background, significantly modify the quantum thermodynamic quantities as feature functions of the detector's open dynamics. However, due to the highly non-monotonic behavior of the detector response rate in BTZ geometry, unlike in static black hole or dynamic de Sitter scenarios, complicated influence patterns have been observed. Nevertheless, such rich modifications on the open dynamics of a local quantum system may have a positive side. For example, it could be compelling to further explore the quantum thermal machine in BTZ spacetime and anticipate a promising efficiency than previous studies in static spacetimes.


\section*{Acknowledgement}
This work is supported by the National Natural Science Foundation of China (Nos. 12475061, 12075178) and the Shaanxi Fundamental Science Research Project for Mathematics and Physics (No. 23JSY006), and the Innovation Program for Quantum Science and Technology (No. 2021ZD0302400). 

\appendix
\section{Derivation of analytical response $\mathcal{C}(\omega)$ in \eqref{eq2.32}}
\label{appendixA}

After applying the sharp switching and taking the switch-on to be in the asymptotic past, the response $\mathcal{C}(\omega)$ defined in \eqref{eq2.11} is reduced into the form \cite{sec1-33}:
\begin{equation}
\begin{aligned}
\mathcal{C}(\omega) =\frac{1}{4}+\frac{1}{4 \pi \sqrt{\alpha(r)-1}} & \sum_{n=-\infty}^{\infty} \int_0^{\infty} \mathrm{d}s \operatorname{Re}\left[\mathrm { e } ^ {  i \omega \ell s } \left(\frac{1}{\sqrt{K_n-\sinh ^2\left(\Xi s+n \pi r_{-} / \ell\right)}}\right.\right. \\
& \left.\left.-\frac{\zeta}{\sqrt{Q_n-\sinh ^2\left(\Xi s+n \pi r_{-} / \ell\right)}}\right)\right].
\label{app1}
\end{aligned}
\end{equation}
Recall that $\zeta=-1, 0, 1$ denotes the Neumann, transparent, or Dirichlet boundary conditions imposed on the field at infinity, and new functions are introduced as follows: 
\be
\begin{aligned}
K_n &= (1-\alpha^{-1})^{-1} \sinh^2\left(\frac{n \pi r_{+}}{l}\right), \\ 
Q_n &= (\alpha-1)^{-1}\left[\alpha \sinh^2\left(\frac{n \pi r_{+}}{l}\right)+1\right], \\
\Xi &= (2\sqrt{\alpha-1})^{-1}.
\label{app2}
\end{aligned}
\ee
By deforming the integration contour appropriately, the integral \eqref{app1} becomes:
\begin{equation}
\begin{aligned}
\mathcal{C}(\omega) = & \frac{\mathrm{e}^{\beta \omega / 2}}{2 \pi} \sum_{n=-\infty}^{\infty} \cos \left(\frac{n \beta \omega r_{-}}{\ell}\right) \int_0^{\infty} \mathrm{d} y \cos \left(\frac{y \beta \omega}{\pi}\right) \\
&~~~~~~~~~~~~~~~~~~~~~~~~ \times\left(\frac{1}{\sqrt{K_n+\cosh ^2 y}}-\frac{\zeta}{\sqrt{Q_n+\cosh ^2 y}}\right) .
\label{app3}
\end{aligned}
\end{equation}

We observe that the key part in the integrand of \eqref{app3} can be analytically calculated once the following integral formula is proved:
\be
\int_0^{\infty}{dx \frac{\cos{(a x)}}{\sqrt{b+\cosh^2{x}}}} = \frac{\pi}{2\cosh{\left( \frac{a\pi}{2} \right)}} P_{-\frac{1}{2}+\frac{i a}{2}}(2b+1), ~~ a\in \mathbb{R}, b>0  \label{app4},
\ee
where $P_{\nu}$  is the associated Legendre function of the first kind. Substituting this formula into \eqref{app3}, we immediately derive \eqref{eq2.32} in Section \ref{Fourier-correlator}.

\begin{figure}[hbtp]
\centering  
\includegraphics[width=0.8\textwidth]{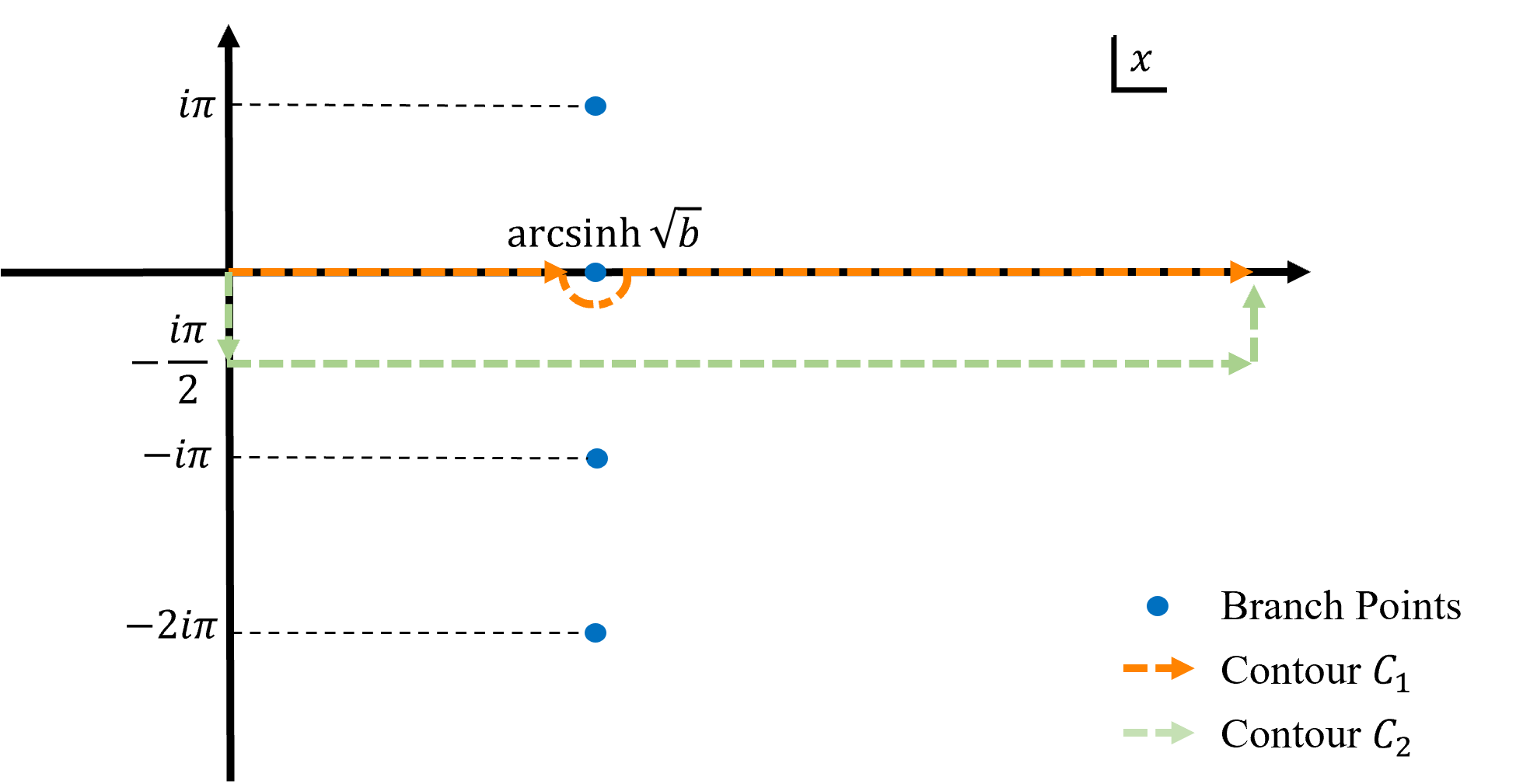}
\caption{Deforming the integral contour from $C_1$ to $C_2$. The integrand has the branch points $\arcsinh{b}+i\pi n$ ($n\in \mathbb{Z}$).} \label{fig15}
\end{figure}

To prove (\ref{app4}), let
\be
I(a, b) := \operatorname{Re}\left[ \int_0^{\infty} \frac{\mathrm{e}^{i a x} \mathrm{~d} x}{\sqrt{b-\sinh ^2 x}} \right], ~~ a\in \mathbb{R}, b>0 .  \label{app5}
\ee
By deforming the integral contour of $I(a,b)$ into $C_1$ to $C_2$ shown in Fig.\ref{fig15}, we obtain:
\begin{equation}
\begin{aligned}
I(a, b) & = \operatorname{Re}\left[ \int_{C_2} \mathrm{d} x \frac{\mathrm{e}^{i a x} }{\sqrt{b-\sinh ^2 x}} \right] \\
& = \operatorname{Re}\left[ \int_0^{-\frac{\pi}{2}} i \mathrm{d} y \frac{\mathrm{e}^{-a y} }{\sqrt{b+\sin ^2 y}} \right] + \operatorname{Re}\left[ \int_0^{\infty} \mathrm{d} y \frac{\mathrm{e}^{i a (y-i\pi/2)} }{\sqrt{b-\sinh ^2 (y-i\pi/2)}} \right]  \\
& = \mathrm{e}^{a \pi/2} \int_0^{\infty} \mathrm{d} y \frac{\cos{(a y)}}{\sqrt{b+\cosh ^2 y}}  \label{app6}
\end{aligned}
\end{equation}
where the first term in the second line is zero, as the integral is pure imaginary. 

On the other hand, we note the associated Legendre function (for $\alpha>0$, $\operatorname{Re}[\mu]<\frac{1}{2}$, $\operatorname{Re}[\nu+\mu]>-1$) as:
\be
\begin{aligned}
& P_\nu^\mu(\cosh \alpha)=\frac{\sqrt{2} \sinh ^\mu \alpha}{\sqrt{\pi} \Gamma\left(\frac{1}{2}-\mu\right)} \int_0^\alpha \frac{\cosh \left[\left(\nu+\frac{1}{2}\right) x\right] d x}{(\cosh \alpha-\cosh x)^{\mu+1 / 2}}, \\
& Q_\nu^\mu(\cosh \alpha)=\sqrt{\frac{\pi}{2}} \frac{e^{\mu \pi i} \sinh ^\mu \alpha}{\Gamma\left(\frac{1}{2}-\mu\right)} \int_\alpha^{\infty} \frac{e^{-\left(\nu+\frac{1}{2}\right) x} d x}{(\cosh x-\cosh \alpha)^{\mu+1 / 2}}, \label{Eq.(A.9)}
\end{aligned}
\ee
and the following identities:
\be
\begin{aligned}
Q_{-\nu-1} & =Q_\nu-\pi \cot (\pi \nu) P_\nu, \quad \text { for } \sin (\pi \nu) \neq 0, \\
\operatorname{Im}\left[Q_{-\frac{1}{2}+i \lambda}\right] & =-\operatorname{Im}\left[Q_{-\frac{1}{2}-i \lambda}\right]  .\label{Eq.(A.10)}
\end{aligned}
\ee
Choosing $\mu=0, \nu=-\frac{1}{2}+\frac{i a}{2}$, by using \eqref{Eq.(A.9)} and \eqref{Eq.(A.10)}, we have:
\be
\begin{aligned}
 P_{-\frac{1}{2}+\frac{i a}{2}}(\cosh \alpha)&=\frac{\sqrt{2}}{\pi} \int_0^\alpha d x \frac{\cos{\frac{a x}{2}} }{\sqrt{\cosh \alpha-\cosh x}}, \\
 Q_{-\frac{1}{2}-\frac{i a}{2}}(\cosh \alpha)&=\frac{1}{\sqrt{2}} \int_\alpha^{\infty} d x \frac{e^{\frac{i a}{2} x} }{\sqrt{\cosh x-\cosh \alpha}}, \\
 \operatorname{Im}\left[Q_{-\frac{1}{2}-\frac{i a}{2}}\right] &= \frac{\pi}{2} \tanh{\left( \frac{a\pi}{2} \right)} \operatorname{Re}\left[ P_{-\frac{1}{2}+\frac{i a}{2}} \right]= \frac{\pi}{2} \tanh{\left( \frac{a\pi}{2} \right)}  P_{-\frac{1}{2}+\frac{i a}{2}}  \label{Eq.(A.11)}
\end{aligned}
\ee
where we neglect the upper index $\mu=0$ for simplicity. 

Corresponding to the way we deforming the contour in Fig.\ref{contour}, similar to the integral contour in the last folia of Riemann surface of $\sqrt{x}$, we have $(\cosh \alpha-\cosh x)^{-1/2} = \left[e^{-i\pi}(\cosh x-\cosh \alpha)\right]^{-1/2} = i (\cosh x-\cosh \alpha)^{-1/2}$ for $x>\alpha$. Then, the integral $I(a,b)$ defined in \eqref{app5} can be recast as:
\begin{equation}
I(a, b)  = \operatorname{Re}\left[ \int_0^{\infty} \mathrm{d} x \frac{\mathrm{e}^{i a x} }{\sqrt{b-\sinh ^2 x}} \right]  = \sqrt{2} \operatorname{Re}\left[ \int_0^{\infty} \mathrm{d} x\frac{\mathrm{e}^{i a x} }{\sqrt{(2b+1)-\cosh(2x)}} \right]
\ee
Denoting $2b+1:=\cosh \alpha$, we further deduce $I(a,b)$ by \eqref{Eq.(A.11)} as:
\begin{equation}
\begin{aligned}
I(a, b) & = \operatorname{Re}\left[ \int_0^{\infty} \mathrm{d} x\frac{\mathrm{e}^{\frac{i a}{2}x} }{\sqrt{\cosh \alpha-\cosh x}} \right] \\
& = \frac{1}{\sqrt{2}} \int_0^{\alpha} \mathrm{d} x\frac{\cos{\frac{a x}{2}} }{\sqrt{\cosh \alpha-\cosh x}} + \frac{1}{\sqrt{2}}\operatorname{Im}\left[ \int_{\alpha}^{\infty} \mathrm{d} x\frac{\mathrm{e}^{\frac{i a}{2}x} }{\sqrt{\cosh x-\cosh \alpha}} \right] \\
& = \frac{\pi}{2} P_{-\frac{1}{2}+\frac{i a}{2}}(\cosh \alpha) + \operatorname{Im}\left[Q_{-\frac{1}{2}-\frac{i a}{2}}(\cosh \alpha)\right] \\
& = \frac{\mathrm{e}^{{a\pi}/{2}} \pi}{2\cosh{\left( {a\pi}/{2} \right)}}  P_{-\frac{1}{2}+\frac{i a}{2}}(\cosh \alpha) \label{Eq(A.12)}
\end{aligned}
\end{equation}
After combining \eqref{app6}) and \eqref{Eq(A.12)}, we obtain \eqref{app3}.

\section{Derivation of quantum heat, work and coherence in \eqref{eq2.43}}
\label{AppendixB}

We start from a rewrite of QMME \eqref{eq2.7-1} into an equation of Bloch vector. Assuming a more general effective Hamiltonian with the Lamb shift as $H_{\text{LS}}=\Omega~ \boldsymbol{m}\cdot\boldsymbol{\sigma}/2$, using the identity
\be
\left[\boldsymbol{a}\cdot\boldsymbol{\sigma},\boldsymbol{b}\cdot\boldsymbol{\sigma}\right]=2i\left(\boldsymbol{a}\times\boldsymbol{b}\right)\cdot\boldsymbol{\sigma},
\ee
the first term of QMME \eqref{eq2.7-1} can be written as
\be
-i\left[H_{\text{LS}},\rho\right]=\frac{\Omega}{2}\left(\boldsymbol{m}\times\boldsymbol{n}\right)\cdot\boldsymbol{\sigma},
\label{eq.B2}
\ee
where $\boldsymbol{n}(t)$ is the Bloch vector of the detector. Substituting the decomposition \eqref{eq2.9} of Kossakowski matrix into the dissipator \eqref{eq2.8}, in terms of the Bloch vectior $\boldsymbol{n}(t)$, we obtain
\begin{equation}
\mathcal{L}[\rho] = \left[ \gamma_0 \left(\boldsymbol{m}\cdot\boldsymbol{n}\right) - \gamma_{-} \right] \boldsymbol{m}\cdot\boldsymbol{\sigma} - \left(\gamma_{+} + \gamma_0\right) \boldsymbol{n}\cdot \boldsymbol{\sigma}. \label{eq.B3}
\end{equation}
Combining \eqref{eq.B2} and \eqref{eq.B3}, the QMME \eqref{eq2.7-1} can be rewritten as an equation of the Bloch vector of the detector:
\begin{equation}
\frac{d\boldsymbol{n}}{dt} = \Omega\left(\boldsymbol{m} \times \boldsymbol{n}\right) + 2g^2 \left\{
\left[ \gamma_0 \left(\boldsymbol{m}\cdot\boldsymbol{n}\right) - \gamma_{-} \right] \boldsymbol{m} - \left(\gamma_{+} + \gamma_0\right) \boldsymbol{n}
\right\}. \label{eq.B4}
\end{equation}

We are now ready to calculate the quantum thermal quantities in the First Law. By definition, the internal energy is given by
\begin{equation}
U_t = \operatorname{Tr}(H_{\text{eff}} \rho) = \frac{\Omega}{2} \boldsymbol{m} \cdot \boldsymbol{n}.
\end{equation}
Using the change rate of Bloch vector \eqref{eq.B4}, the time derivative of detector internal energy is:
\begin{equation}
\frac{dU_t}{dt} = \frac{\Omega}{2} \boldsymbol{m} \cdot \frac{d\boldsymbol{n}}{dt} = -g^2 \Omega \left( \gamma_{-} + \gamma_{+} \, \boldsymbol{m} \cdot \boldsymbol{n} \right).
\end{equation}
For the time-independent detector, its power term vanishes $\dot{\mathds{W}}_t = 0$ by definition. For the UDW detector, the length of its Bloch vector \eqref{eq2.14} is time-dependent. Therefore, the heat flow $\dot{\mathds{Q}}$ of the detector is given by:
\begin{equation}
\begin{aligned}
\dot{\mathds{Q}} := \frac{\Omega}{2} \cos{\Theta} \, \frac{dl(t)}{dt} 
&= \frac{\Omega}{2} \frac{ \boldsymbol{m} \cdot \boldsymbol{n} }{ l^2 } \left( \boldsymbol{n} \cdot \frac{d\boldsymbol{n}}{dt} \right) \\
&= -g^2 \Omega \left\{ \gamma_{-} + \gamma_{+} (\boldsymbol{m} \cdot \boldsymbol{n}) + \left[ \gamma_0 (\boldsymbol{m} \cdot \boldsymbol{n}) - \gamma_{-} \right] \sin^2 \Theta \right\}.
\end{aligned}
\end{equation}
Finally, the coherence-induced energy change is defined by
\begin{equation}
\dot{\mathds{C}} = \dot{U}_t - \dot{\mathds{Q}} = g^2 \Omega \left[ \gamma_0 (\boldsymbol{m} \cdot \boldsymbol{n}) - \gamma_{-} \right] \sin^2 \Theta.
\end{equation}
For the detector model in Section \ref{2.1}, we set $\boldsymbol{m}=(0,0,1)$ and obtain the result \eqref{eq2.43}.

\section{Derivation of quantum fidelity \eqref{eqT37}}
\label{AppendixC}

Let $\rho(t)$ be an arbitrary qubit state with Bloch vector length $l(t)$. In a diagonal form, the density matrix can be written as $\rho = \rho_{+} \ket{+}\bra{+} + \rho_{-} \ket{-}\bra{-}$, where $\rho_{\pm} = \frac{1}{2}(1 \pm l)$ and eigenbasis $\ket{\pm}$ have been given in \eqref{eq-IG5}.

In the eigenbasis $\{\ket{+}, \ket{-}\}$, the thermalized final state \eqref{eqT36} takes the form:
\begin{equation}
\rho_f =
\frac{1}{2}
\left(\begin{array}{cc}
\displaystyle 1 - \tanh\left(\frac{\omega}{2T}\right) \cos \Theta &~~ \displaystyle e^{i \phi_t} \tanh\left(\frac{\omega}{2T}\right) \sin \Theta \\
&\\
\displaystyle e^{-i \phi_t} \tanh\left(\frac{\omega}{2T}\right) \sin \Theta &~~ \displaystyle  1 + \tanh\left(\frac{\omega}{2T}\right) \cos \Theta
\end{array}\right)
\end{equation}
where $\cos{\Theta} = {n_3}/l(t)$ and $\varphi_t$ is a global phase. To calculate the fidelity \eqref{eqT34} between $\rho(t)$ and $\rho_f$, the key step is to find the eigenvalues of the matrix:
\begin{equation}
\sqrt{\rho}\rho_f \sqrt{\rho} =
\frac{1}{4}
\left(\begin{array}{cc}
\displaystyle 
(1+l)(1 - \tanh\left(\frac{\omega}{2T}\right)\cos \Theta) &~~\displaystyle \sqrt{1 - l^2} \tanh\left(\frac{\omega}{2T} \right)\sin \Theta e^{i \phi_t} \\
&\\
\displaystyle \sqrt{1 - l^2}\tanh\left(\frac{\omega}{2T}\right) \sin \Theta e^{-i \phi_t} & ~~\displaystyle(1 - l)(1 + \tanh\left(\frac{\omega}{2T}\right) \cos \Theta)
\end{array}\right)
\end{equation}
which are:
\begin{equation}
\begin{aligned}
\lambda_{\pm} = &\frac{1}{4} \left(1 - l(t) \tanh\left(\frac{\omega}{2T}\right) \cos \Theta \right)\\
&\pm \frac{1}{4} \sqrt{ \left[1 - l(t) \tanh\left(\frac{\omega}{2T}\right) \cos \Theta \right]^2 - \left(1 - l^2 \right)\left[1 - \tanh^2\left(\frac{\omega}{2T}\right)\right]}.
\end{aligned}
\end{equation}
Then, the fidelity can be obtained as:
\begin{equation}
\begin{aligned}
F(\rho, \rho_f) &=\left[\sum_j \sqrt{\lambda_j}\right]^2\\
&=\frac{1}{2} \left( 1 - l \tanh\left(\frac{\omega}{2T}\right)\cos \Theta + \sqrt{(1 - l^2)\left[1 - \tanh^2\left(\frac{\omega}{2T}\right)\right]} \right),
\label{C4}
\end{aligned}
\end{equation}
which is exactly \eqref{eqT37} using $\cos{\Theta} = {n_3}/l(t)$. 

As a consistency check, we consider the case that $\rho$ approaches a thermal state. This means $l \to \tanh\left({\omega}/{2T}\right)$ and $\cos \Theta \to -1$. Substituting these asymptotic values into \eqref{C4}, we find $F(\rho, \rho_f) \to 1$, as expected.

\end{document}